\documentclass[onecolumn,journal,noadjust]{IEEEtran}
\usepackage{cite}

\ifCLASSINFOpdf
\else
\fi
\usepackage{amssymb}
\usepackage{dsfont}
\usepackage{float}
\usepackage{graphicx}
\newtheorem{lemma}{Lemma}
\newtheorem{example}{Example}
\newtheorem{definition}{Definition}
\newtheorem{proposition}{Proposition}
\newtheorem{theorem}{Theorem}
\newtheorem{remark}{Remark}
\newtheorem{corollary}{Corollary}
\usepackage{xcolor}
\usepackage{mathrsfs}

\usepackage [autostyle, english = american]{csquotes}
\MakeOuterQuote{"}

\usepackage{amsmath}
\DeclareMathOperator*{\argmax}{arg\,max}

\begin{document}
%

\title{Channel Coding with Mean and Variance \\Cost Constraints}
%
%
%

\author{Adeel Mahmood and Aaron B.~Wagner\\
School of Electrical and Computer Engineering, Cornell University}

\maketitle

\begin{abstract}
We consider channel coding for discrete memoryless channels (DMCs) with a novel cost constraint that constrains both the mean and the variance of the cost of the codewords. We show that the maximum (asymptotically) achievable rate under the new cost formulation is equal to the capacity-cost function; in particular, the strong converse holds. We further characterize the optimal second-order coding rate of these cost-constrained codes; in particular, the optimal second-order coding rate is finite. We then show that the second-order coding performance is strictly
improved with feedback using a new variation of timid/bold coding, significantly
broadening the applicability of timid/bold coding schemes from
unconstrained
compound-dispersion channels to all cost-constrained channels. Equivalent results on the minimum average probability of error are also given.        
\end{abstract}

\begin{IEEEkeywords}
Channel coding, feedback communications, second-order coding rate, stochastic control.
\end{IEEEkeywords}

%

\IEEEpeerreviewmaketitle

\section{Introduction}

Channel coding is a fundamental problem in information theory, where the goal is to reliably transmit information over a noisy channel. Information transmission with arbitrarily small error probability is possible at all rates below what is called the capacity $C$ of the channel, if the number $n$ of channel uses (also called blocklength) is allowed to go to infinity \cite{korner1}. A channel satisfies a weak converse if for all rates above capacity, the error probability $\epsilon$ cannot be made arbitrarily small. A channel satisfies the strong converse if for all rates above capacity, the error probability $\epsilon \to 1$ as $n \to \infty$.

In many practical scenarios, the channel input is subject to some cost constraints which limit the amount of resources that can be used for transmission. Such constraints may arise out of concern for interference with other terminals or, especially in the case of mobile
devices, power consumption. With a cost constraint present, the role of capacity is replaced by the capacity-cost function \cite[Theorem 6.11]{korner1}. The coding performance however is sensitive to the formulation of the cost constraint. For instance, whether the strong converse holds
depends not only on the channel but also on how the cost constraint is imposed. We focus on discrete memoryless channels (DMCs) with a cost function denoted by $c(\cdot)$. One common cost constraint called the almost-sure (a.s.) cost constraint \cite{5290292,6767457} bounds the time-average cost of the channel input $X^n$ over all messages, realizations
of any side randomness, channel noise (if there is feedback), etc.:  
\begin{align}
    \label{eq:constraint:as}
    \frac{1}{n} \sum_{i = 1}^n c(X_i) \le \Gamma \quad \text{a.s.}
\end{align}
On the other hand, the expected cost constraint bounds the sum-cost in the average sense:
\begin{align}
    \label{eq:constraint:mean}
    \frac{1}{n} \sum_{i = 1}^n \mathbb{E}[c(X_i)] \le \Gamma.
\end{align}
These two cost constraints are also called short-term and long-term power constraints, respectively, in certain contexts \cite{7156144}. With an almost-sure cost constraint, the strong converse holds \cite[Theorem 6.11]{korner1}. With an expected cost constraint, the strong converse ceases to hold \cite[Theorem 77]{Polyanskiy2010} because the constraint
admits signaling schemes that use power in a 
highly non-ergodic fashion.\footnote{We refer to $c(X_i)$ as
\emph{power} although it could represent other resources.} Accordingly,
past work on second-order coding rates with cost constraints (\cite{5290292, 7055296}) has focused on the
almost-sure constraint, namely~(\ref{eq:constraint:as}). The second-order coding rate (\cite{9099482}, \cite{5290292}, \cite{strassen}, \cite{7447062}, \cite{6816070}) quantifies the $O(n^{-1/2})$ convergence to the capacity-cost function (or to the capacity in the unconstrained case). Under the a.s. cost
formulation, the optimal second-order coding rate (SOCR) is known~\cite[Theorem 3]{5290292}.  

One of the lessons of information-theoretic studies of channel
coding is that the various codewords should appear to be selected randomly 
and independent and identically distributed (i.i.d.) according to $P^*$, where 
\begin{equation}
    P^* = \argmax_{P: \sum_x P(x) c(x) \le \Gamma} I(P,W)
\end{equation}
is a capacity-cost-achieving input distribution (at cost
level $\Gamma$) for a discrete memoryless channel $W(\cdot|\cdot)$. The idea of generating codewords in an i.i.d. fashion
is so natural and ubiquitous that it is notable that it is 
actually impermissible under~(\ref{eq:constraint:as}).  

More seriously, one incurs a performance loss by prohibiting the use
of i.i.d.-generated codewords in second-order coding rate (SOCR) analyses. Consider the problem with ideal feedback from the output of the
channel to the encoder. 
For channels without cost constraints,
it is known that feedback can improve the second-order coding
rate for compound-dispersion channels~\cite{9099482}. Specifically, 
suppose
a channel $W(\cdot|\cdot)$ has two capacity-achieving input
distributions $P_1^*$ and $P_2^*$ such that
\begin{equation}
    \text{Var}_{P_1^* \circ W}\left( \log \frac{W(Y|X)}{P_1^*W(Y)}\right)
    < 
    \text{Var}_{P_2^* \circ W}\left( \log \frac{W(Y|X)}{P_2^*W(Y)}\right),
    \label{eq:compound}
\end{equation}
where $P_i^* \circ W$ denotes the joint distribution over inputs
and outputs induced by the distributions $P_i^*$ and $W(\cdot|\cdot)$,
and $P_i^* W$ denotes the marginal distribution of the induced
output. While codewords drawn from $P_1^*$ and $P_2^*$ have the
same mean information-carrying ability
by virtue of $P_1^*$ and $P_2^*$ both being capacity-achieving, codewords drawn from $P_2^*$ are more variable as a consequence of (\ref{eq:compound}).
Thus,
the encoder can employ codewords drawn from $P_1^*$ (``timid'')
so long as the transmission is proceeding well and from
$P_2^*$ (``bold'') if an error appears likely.
This is referred to as
\emph{timid/bold} coding.

One limitation of the above idea is that the channel
must be compound-dispersion.
In particular, the capacity-achieving input
distribution for $W(\cdot|\cdot)$ cannot be
unique. In fact, for simple-dispersion channels for which $(\ref{eq:compound})$ does not hold, feedback does not improve the second-order coding
rate \cite[Theorem 3]{9099482}.

The recent work \cite{adeeltimid} studied the feedback improvement of SOCR with a cost constraint. The cost constraint in \cite{adeeltimid} was intermediate between an almost-sure cost constraint and an expected cost constraint. With the intermediate cost constraint, \cite{adeeltimid} showed that the timid/bold feedback scheme can improve the SOCR for DMCs even if the capacity-cost-achieving distribution is unique, thus broadening the scope of timid/bold coding beyond \cite{9099482}. Specifically, let $P^*$ denote a
capacity-cost-achieving input distribution for
the DMC $W(\cdot|\cdot)$, which might well be unique. By the law of total variance,
the $n$-length form of the variance in 
(\ref{eq:compound}) can be written as
\begin{equation}
    \text{Var}\left( \log \frac{W(Y^n|X^n)}{P^*W(Y^n)}\right)
    =
    \mathbb{E} \left [\text{Var}\left( \log \frac{W(Y^n|X^n)}{P^*W(Y^n)} \middle | X^n  \right) \right]  
          + \text{Var}\left( \mathbb{E}\left [  \log \frac{W(Y^n|X^n)}{P^*W(Y^n)} \middle | X^n \right] \right).
          \label{eq:nlength}
\end{equation}
If the channel input $X^n$ is constant-composition, i.e., drawn uniformly from a fixed type class associated
with a distribution that is close to $P^*$, then the
quantity
\begin{equation}
     \mathbb{E}\left [  \log \frac{W(Y^n|X^n)}{P^*W(Y^n)} \middle | X^n \right] \label{blindarrow}
\end{equation}
is a.s.\ constant and the second term in (\ref{eq:nlength})
is zero.
In contrast, if $X^n$ is i.i.d. according to $P^*$, then the second
term in (\ref{eq:nlength}) is order-$n$ (see \cite[Lemma 2]{adeeltimid}). The first term, in contrast, is approximately
the same between the two cases. Thus both timid and bold
signaling mechanisms can be created from $P^*$ alone, depending
on whether one uses constant-composition or i.i.d.\ codewords. Yet, this observation cannot be applied under the prevailing a.s. cost
formulation for second-order rate analysis because i.i.d.\ codewords 
are impermissible
under the a.s. constraint in (\ref{eq:constraint:as}). 

By formulating an intermediate cost constraint that allows both i.i.d. and constant-composition channel inputs, \cite{adeeltimid} showed a strict improvement of SOCR with feedback. However, relaxing the a.s. cost constraint to permit i.i.d. codewords should be approached cautiously, as the expected cost constraint from $(\ref{eq:constraint:mean})$, which similarly accommodates i.i.d. codewords, also admits signaling schemes characterized by a highly non-ergodic power usage, leading to the absence of a strong converse. 

In this paper, we introduce a new $(\Gamma, V)$ cost constraint which constrains both the mean and the variance of the codewords: 
\begin{align}
    \mathbb{E}\left[\sum_{i = 1}^n c(X_i)\right] & \le n\Gamma \\
    \text{Var}\left(\sum_{i = 1}^n c(X_i)\right) & \le nV. \label{eq:intro:var}
\end{align}
The $(\Gamma, V)$ cost constraint is a natural strengthening of the expected cost constraint via a second-moment constraint. The idea is that we want power to be consumed at 
a limited rate but also in a predictable fashion, both
to ensure a gradual consumption of energy and so
that the transmitted signal is sufficiently ergodic \cite[(5)]{mahmood2024improvedchannelcodingperformance}
that it can be treated as noise by other terminals (which mitigates the negative impact of interference). Note that (\ref{eq:intro:var}) ensures that $c(X_i)$
satisfies the weak law of large numbers as $n \rightarrow
\infty$.

With an additional variance constraint in $(\Gamma, V)$ channel codes for DMCs, the strong converse holds. Our new cost constraint also admits a finite second-order converse,  i.e., the maximum achievable second-order coding rate is finite. We give matching achievability and converse results characterizing the optimal SOCR in terms of a function of $\Gamma$ and $V$, which takes the form 
\begin{align}
    \inf_{\Pi} \mathbb{E}\left [ \Phi(\Pi) \right ], \label{fvcx}
\end{align}
where $\Phi(\cdot)$ is the standard Gaussian CDF and the infimum is over all random variables $\Pi$ with an appropriately constrained expectation and variance. We characterize the solution to the optimization problem in $(\ref{fvcx})$ as well as the properties of a function $\mathcal{K}(r, V)$ which is equal to $(\ref{fvcx})$ with the expectation and variance constrained by $r$ and $V$, respectively. For $V > 0$, it is shown in \cite[Theorem 1]{mahmood2024improvedchannelcodingperformance} that the optimal SOCR under the $(\Gamma, V)$ cost constraint is strictly higher than the optimal SOCR under the a.s. cost constraint. Similar results for Gaussian channels under the $(\Gamma, V)$ cost constraint are proven in \cite{gaussianadeel}. We also note that a variation of the optimization problem in $(\ref{fvcx})$ has appeared in \cite{7156144}.  

After establishing the optimal second-order coding performance under the $(\Gamma, V)$ constraint, we show that this performance
is strictly improved with feedback (Theorem \ref{Feedback_Achievability_Theorem}). Our feedback scheme is a new variant of timid/bold coding where feedback improvement requires neither multiple capacity-cost-achieving distributions as in \cite{9099482} nor i.i.d. $P^*$ codewords as in \cite{adeeltimid}. Note that i.i.d. $P^*$ codewords are admissible under the $(\Gamma, V)$ cost constraint for $V \geq \text{Var}_{P^*}(c(X))$ so that the feedback scheme described in $(\ref{eq:nlength})$ and $(\ref{blindarrow})$ can be used. However, our feedback scheme uses a more general approach, i.e., we show that feedback gives a strict SOCR improvement for all values of $V > 0$, even for $V < \text{Var}_{P^*}(c(X))$ where i.i.d. $P^*$ codewords are not permissible. Combining Theorem \ref{Feedback_Achievability_Theorem} with \cite[Theorem 2]{mahmood2024improvedchannelcodingperformance}, it follows that the aforementioned feedback improvement is possible if and only if $V > 0$, for DMCs with a unique capacity-cost-achieving distribution. Therefore, a more foundational advantage of the $(\Gamma, V)$ cost constraint is allowing a nonzero variance of the cost of the channel input around the cost point $\Gamma$, while still sufficiently regulating power consumption to ensure a finite second-order coding rate.

\section{Preliminaries \label{prelims}}

Let $\mathcal{A}$ and $\mathcal{B}$ be finite input and output alphabets, respectively, of a DMC. A sequence $(x_1, \ldots, x_n) \in \mathcal{A}^n$ of length $n$ will be written as $x^n$, the first $k$ components ($1 \leq k \leq n$) of $x^n$ will be denoted by $x^k$ and the $k_1^{\text{th}}$ to $k_2^{\text{th}}$ components (both inclusive) will be denoted by $x^{k_1:k_2}$. We will write $\log$ to denote logarithm to the base $e$ and $\exp(x)$ to denote $e$ to the power of $x$. Let $\mathcal{P}(\mathcal{A})$ be the set of probability distributions on $\mathcal{A}$ and let $\mathcal{P}(\mathcal{B} | \mathcal{A})$ denote the set of stochastic matrices from $\mathcal{A}$ to $\mathcal{B}$. We will use $W \in \mathcal{P}(\mathcal{B}| \mathcal{A})$ to denote the DMC. We will write $X^n$ to denote a random channel input of blocklength $n$ and $Y^n$ to denote the channel output. Let $\mathcal{P}_n(\mathcal{A})$ be the set of $n$-types on $\mathcal{A}$. For a given $t \in \mathcal{P}_n(\mathcal{A})$, $T^n_{\mathcal{A}}(t)$ denotes the type class, i.e., the set of sequences $x^n \in \mathcal{A}^n$ with empirical distribution equal to $t$. $\phi(\cdot)$ and $\Phi(\cdot)$ denote the PDF and CDF of the standard Gaussian random variable, respectively. We will use $s(P)$ to denote the size of the support of a distribution $P \in \mathcal{P}(\mathcal{A})$. For a random variable $X$, $||X||_\infty$ will denote its essential supremum.

For a given $P \in \mathcal{P}(\mathcal{A})$, $P \circ W$ denotes the joint distribution on $\mathcal{A} \times \mathcal{B}$ induced by $P$ and $W$, and $PW$ denotes the corresponding marginal distribution on $\mathcal{B}$, i.e., for any $a \in \mathcal{A}$, $b \in \mathcal{B}$,
\begin{align*}
    (P \circ W)(a, b) &:= P(a) W(b|a),\\
    PW(b) &:= \sum_{a \in \mathcal{A}}(P \circ W)(a, b).
\end{align*}
The cost function is denoted by $c(\cdot)$ where $c: \mathcal{A} \to [0, c_{\max}]$ and $c_{\max} > 0$ is a constant. Let $\Gamma_0 = \min_{a \in \mathcal{A}} c(a)$. For $\Gamma > \Gamma_0$, the capacity-cost function is defined as 
\begin{align}
    C(\Gamma ) &= \max_{\substack{P \in \mathcal{P}(\mathcal{A})\\
    c(P) \leq \Gamma} } I(P, W), \label{main_form}
\end{align}
where $c(P) := \sum_{a \in \mathcal{A}} P(a) c(a)$. For a given $x^n \in \mathcal{A}^n$, we define 
\begin{align*}
    c(x^n) := \frac{1}{n} \sum_{i=1}^n c(x_i).
\end{align*}
By \cite[Theorem 6.11]{korner1}, $C(\Gamma)$ is a non-decreasing, concave function of $\Gamma$ for $\Gamma \geq \Gamma_0$. Let $\Gamma^*$ denote the smallest $\Gamma$ such that the capacity-cost function  $C(\Gamma)$ is equal to the unconstrained capacity $C$, i.e., $C(\Gamma) = C = \max_{P \in \mathcal{P}(\mathcal{A})} I(P, W)$. We assume $\Gamma^* > \Gamma_0$ and $\Gamma \in (\Gamma_0, \Gamma^*)$ throughout the paper. The function $C(\Gamma)$ is strictly increasing and differentiable \cite[Problem 8.4]{korner1} in the interval $(\Gamma_0, \Gamma^*)$. Therefore, $C(\Gamma)$ is continuously differentiable \cite[Corollary 25.5.1]{rockafellar1997convex}. From \cite[Theorem 1]{7055296} (see also \cite[Lemma 8.1]{korner1} and \cite[Theorem 8.4]{korner1}), we have the following proposition:
\begin{proposition}
Let $P^*$ be a solution to $(\ref{main_form})$ and $Q^* = P^*W$. For $\Gamma \in (\Gamma_0, \Gamma^*)$, we have $c(P^*) = \Gamma$ and for every $a \in \mathcal{A}$, 
    \begin{align}
        \sum_{b \in \mathcal{B}} W(b|a) \log \frac{W(b|a)}{Q^*(b)} \begin{cases}
            = C(\Gamma) - C'(\Gamma)(\Gamma - c(a)) & \text{ if } P^*(a) > 0\\
            \leq C(\Gamma) - C'(\Gamma)(\Gamma - c(a)) & \text{ otherwise},
        \end{cases} \label{goldenproperty}
    \end{align}
    where $C'(\Gamma) > 0$.  
    \label{getout}
\end{proposition}
The result in $(\ref{goldenproperty})$ is a generalization of the unconstrained case which is stated in the following proposition (see \cite[Theorem 4.5.1]{gallager1968}):
\begin{proposition}
    Let $P^* = \argmax_{P \in \mathcal{P}(\mathcal{A})} I(P, W)$ and $Q^* = P^*W$. Then for every $a \in \mathcal{A}$,  
    \begin{align}
        \sum_{b \in \mathcal{B}} W(b|a) \log \frac{W(b|a)}{Q^*(b)} \begin{cases}
            = C & \text{ if } P^*(a) > 0\\
            \leq C & \text{ otherwise}.
        \end{cases} \label{gol2denproperty}
    \end{align}
    \label{ihavenoidea}
\end{proposition}

Throughout the paper, we will assume that the capacity-cost-achieving distribution for cost $\Gamma$ is unique. We will use $P^*$ to denote the unique solution to $(\ref{main_form})$. For the application to feedback communication, 
this is the
most interesting case, since if $P^*$ is not unique, the timid/bold scheme
of~\cite{9099482} is already applicable. This assumption also has precedent
in the literature (e.g.,~\cite{7055296}), because it affords certain technical
simplifications (e.g., Lemma~\ref{lemcon3t} in 
Appendix~\ref{lemcon3t_proof}). Note that we do not assume 
uniqueness for costs $\Gamma' \neq \Gamma$.

The following definitions will remain in effect throughout the paper:  
\begin{align}
\begin{split}
    Q^* &:= P^* W,\\
    \nu_{a} &:= \text{Var}\left( \log \frac{W(Y|a)}{Q^*(Y)} \right),\quad  \text{ where } Y \sim W(\cdot|a),\\
    \nu_{\min} &:= \min_{a \in \mathcal{A}} \nu_a,\\
    \nu_{\max} &:= \max_{a \in \mathcal{A}} \nu_a,\\
    i(a, b) &:= \log \frac{W(b|a)}{Q^*(b)},\\ 
    i_{\max} &:= \max_{a \in \mathcal{A}, b \in \mathcal{B} : W(b|a) > 0} | i(a, b)  |,\\
    V(\Gamma) &:= \sum_{a \in \mathcal{A}} P^*(a) \nu_{a},\\
    [P^*]_n &:= \arg\min_{\substack{t \in \mathcal{P}_n(\mathcal{A}):\\
    c(t) \leq \Gamma}} ||t - P^*||_1,
    \end{split}
    \label{basicdefs}
\end{align}
where $||\cdot||_1$ denotes the $l^1$ norm. Note that the output distribution $Q^*$ is always unique, and without loss of generality, $Q^*$ can be assumed to satisfy $Q^*(b) > 0$ for all $b \in \mathcal{B}$ \cite[Corollaries 1 and 2 to Theorem 4.5.1]{gallager1968}. 

A channel input $X^n \sim \text{Unif}(T^n_{\mathcal{A}}(t))$ drawn uniformly from a type class $t$ is called a constant-composition (cc) input. We will denote by $Q^{cc}$ the output distribution induced by the input $X^n \sim \text{Unif}(T^n_{\mathcal{A}}([P^*]_n))$ through the DMC $W$. For $X^n \sim \text{Unif}(T^n_{\mathcal{A}}([P^*]_n))$ and $Y^n \sim W(\cdot|X^n)$, we define   
\begin{align}
\begin{split}
    [C(\Gamma)]_n &:= \frac{1}{n} \mathbb{E}\left [\sum_{i=1}^n \log \frac{W(Y_i|X_i)}{Q^*(Y_i)}\right ]\\
    [V(\Gamma)]_n &:= \sum_{a \in \mathcal{A}} [P^*]_n(a) \nu_{a}\\
    Q^{cc}(y^n) &:= \sum_{x^n \in T^n_{\mathcal{A}}([P^*]_n)} \frac{1}{|T^n_\mathcal{A}([P^*]_n)|} \prod_{i=1}^n W(y_i|x_i).
    \end{split}
    \label{ccquantities}
\end{align}
The discrepancy between the above three quantities and $C(\Gamma), V(\Gamma)$ and $Q^*$ is quantified by Lemma \ref{combinedlemmas} in Section \ref{suppsection}.

With a blocklength $n$ and a fixed rate $R > 0$, let $\mathcal{M}_R = \{1, \ldots, \lceil \exp(nR) \rceil \}$ denote the message set. Let $M \in \mathcal{M}_R$ denote the random message drawn uniformly from the message set.

\begin{definition}
An $(n, R)$ code for a DMC consists of an encoder $f$ which, for each message $m \in \mathcal{M}_R$, chooses an input $X^n = f(m) \in \mathcal{A}^n$, and a decoder $g$ which maps the output $Y^n$ to $\hat{m} \in \mathcal{M}_R$. The code $(f,g)$ is random if $f$ or $g$ is random. 
\label{defwocostwofeedback}
\end{definition}

\begin{definition}
An $(n, R)$ code with ideal feedback for a DMC consists of an encoder $f$ which, at each time instant $k$ ($1 \leq k \leq n$) and for each message $m \in \mathcal{M}_R$, chooses an input $x_k = f(m, x^{k-1}, y^{k-1}) \in \mathcal{A}$, and a decoder $g$ which maps the output $y^n$ to $\hat{m} \in \mathcal{M}_R$. The code $(f,g)$ is random if $f$ or $g$ is random.   
\label{defwocostwfeedback}
\end{definition}

As noted in the introduction, we consider a cost 
constraint that restricts both the mean  and the variance of the codewords by some $\Gamma \in (\Gamma_0, \Gamma^*)$ and $V > 0$, respectively. We consider the following
formulation.

\begin{definition}
An $(n, R, \Gamma, V)$ code for a DMC is an $(n, R)$ code such that $\mathbb{E}\left [ \sum_{i=1}^n c(X_i) \right] \leq n \Gamma$ and $\text{Var}\left(\sum_{i=1}^n c(X_i) \right) \leq n V$, where the message $M \sim \text{Unif}(\mathcal{M}_R)$ has a uniform distribution over the message set $\mathcal{M}_R$. 
\label{defwcostwofeedback}
\end{definition}

\begin{definition}
    An $(n, R, \Gamma, V)$ code with ideal feedback for a DMC is an $(n,R)$ code with ideal feedback such that $\mathbb{E}\left[\sum_{i=1}^n c(X_i)\right] \leq n\Gamma$ and $\text{Var}\left(\sum_{i=1}^n c(X_i) \right) \leq n V$, where the message $M \sim \text{Unif}(\mathcal{M}_R)$ has a uniform distribution over the message set $\mathcal{M}_R$. 
    \label{defwcostwfeedback}
\end{definition}

We consider optimum coding performance for $(n, R,\Gamma, V)$ codes defined in Definitions 
\ref{defwcostwofeedback} and \ref{defwcostwfeedback}.
Given $\epsilon \in (0, 1)$, define 
\begin{align*}
    M^*_{\text{fb}}(n, \epsilon,  \Gamma, V) := \max \{ \lceil \exp(nR) \rceil : \bar{P}_{\text{e,fb}}(n,R, \Gamma, V) \leq \epsilon   \},
\end{align*}
where $\bar{P}_{\text{e,fb}}(n,R,\Gamma, V)$ denotes the minimum average error probability attainable by any $(n,R, \Gamma, V)$ code with feedback. Similarly, define \begin{align*}
    M^*(n, \epsilon,  \Gamma, V) := \max \{ \lceil \exp(nR) \rceil : \bar{P}_{\text{e}}(n,R, \Gamma, V) \leq \epsilon   \},
\end{align*}  
where $\bar{P}_{\text{e}}(n,R,\Gamma, V)$ denotes the minimum average error probability attainable by any $(n,R, \Gamma, V)$ code without feedback.  

The second-order coding rate (SOCR) is defined as 
\begin{align}
    \liminf_{n \to \infty} \frac{\log M^*(n, \epsilon, \Gamma, V) - n C(\Gamma)}{\sqrt{n}}. \label{oursocrhai}
\end{align}
For $(n, R,\Gamma, V)$ codes for DMCs defined in Definition \ref{defwcostwofeedback}, the following (achievability) bound on SOCR can be obtained by using the coding scheme from \cite[Theorem 3]{5290292}:
\begin{align}
    \liminf_{n \to \infty} \frac{\log M^*(n, \epsilon, \Gamma, V) - nC(\Gamma)}{\sqrt{n}} \geq \sqrt{V(\Gamma)} \Phi^{-1}(\epsilon) \label{improvedkaro}
\end{align}
for $\epsilon \in (0, 1)$, where the right-hand side of $(\ref{improvedkaro})$ is the optimal SOCR associated with the almost-sure cost constraint,
assuming a unique capacity-cost-achieving distribution $P^*$ at cost $\Gamma$. The aforementioned coding scheme, called a constant-composition code, consists of generating random codewords uniformly distributed over a fixed type class of a type which is closest to $P^*$ and which satisfies the cost constraint almost surely. With a fixed type, the constant-composition code trivially satisfies the $(\Gamma, V)$ cost constraint for all $V > 0$. However, while constant-composition codes hit the optimal second-order coding rate with an almost-sure cost constraint, such codes are not necessarily optimal with the new $(\Gamma, V)$ cost constraint introduced in this paper.

\begin{definition}
    A controller is a function $F : (\mathcal{A} \times \mathcal{B})^* \to \mathcal{P}(\mathcal{A})$. 
\end{definition}
Recall that for any set $\mathcal{X}$, $\mathcal{X}^*$ denotes the set of all finite sequences (including the empty sequence) that can be formed from elements of $\mathcal{X}$. We shall sometimes write $F(\cdot| x^k, y^k)$ for $F(x^k, y^k)(\cdot)$. The design of random feedback codes $(f,g)$ can be directly related to the design of controllers \cite{9099482}. Specifically, for each message $m \in \mathcal{M}_R$, the channel input at time $k$ when the past inputs are $x^{k-1}$ and the past outputs are $y^{k-1}$ is chosen randomly according to $F(\cdot| x^{k-1}, y^{k-1})$. In other words, $f(m, x^{k-1}, y^{k-1})$ has the distribution $F(x^{k-1}, y^{k-1})$. Given $y^n$, the decoder $g$ selects the message $m$ with the lowest index that achieves the maximum over $m$ of 
\begin{align*}
   \prod_{i=1}^n W\left (y_i | f(m, x^{i-1}, y^{i-1}) \right ). 
\end{align*}

Any controller $F$ can be used to construct an $(n, R)$ code with ideal feedback using the aforementioned construction. Moreover, any controller $F$ satisfying
\begin{align}
\begin{split}
    \mathbb{E}\left[\sum_{i=1}^n c(X_i)\right] &\leq n\Gamma\\
    \text{Var}\left(\sum_{i=1}^n c(X_i) \right) &\leq nV 
    \end{split}
    \label{faby}
\end{align}
can be used to construct an $(n, R, \Gamma, V)$ code with ideal feedback.

Given a random $(n ,R, \Gamma, V)$ feedback code based on a controller $F$, the following lemma, which is \cite[Lemma 14]{9099482} particularized to the class of controllers satisfying $(\ref{faby})$, gives an upper bound on the average error probability of the code in terms of the distribution of $(X^n, Y^n)$ induced by the controller.

\begin{lemma}[{\cite[Lemma 14]{9099482}}]
    For any $\Gamma \in (\Gamma_0, \Gamma^*)$ and $V > 0$, a controller $F$ satisfying $(\ref{faby})$, and any $n$, $\theta$ and $R$,
    \begin{align}
        \bar{P}_{\text{e,fb}}(n,R,  \Gamma, V) \leq (F \circ W) \left(\frac{1}{n} \log \frac{W(Y^n|X^n)}{FW(Y^n)} \leq R + \theta \right) + e^{-n \theta}, \label{patanahimujhe}
    \end{align}
    where $(X^n, Y^n)$ have the joint distribution specified by 
    \begin{align*}
        (F \circ W)(x^n, y^n) &= \prod_{k=1}^n F(x_k|x^{k-1}, y^{k-1}) W(y_k|x_k),
    \end{align*}
    and $FW$ denotes the marginal distribution of $Y^n$. Furthermore, if for some $\alpha$ and $\epsilon$,
    \begin{align}
        \limsup_{n \to \infty} \, (F \circ W)\left(\frac{1}{n} \log \frac{W(Y^n|X^n)}{FW(Y^n)} \leq C(\Gamma) + \frac{\alpha}{\sqrt{n}} \right) < \epsilon, \label{ghi}
    \end{align}
    then the controller $F$ gives rise to an achievable second-order coding rate of $\alpha$, i.e.,  
    \begin{align}
        \liminf_{n \to \infty} \frac{\log M^*_{\text{fb}}(n, \epsilon, \Gamma, V) - nC(\Gamma)}{\sqrt{n}} \geq \alpha. \label{ghi2}
    \end{align}
    \label{aaron'slemma}
Similar results to $(\ref{patanahimujhe})$, $(\ref{ghi})$ and $(\ref{ghi2})$ hold for $\bar{P}_{\text{e}}(n, R, \Gamma, V)$ and $M^*(n, \epsilon, \Gamma, V)$ in the non-feedback case by replacing controllers $F$ by distributions $P \in \mathcal{P}(\mathcal{A}^n)$. Hence, in the non-feedback case, the joint distribution of $(X^n, Y^n)$ is specified by $(P \circ W)(x^n, y^n) = \prod_{k=1}^n P(x_k|x^{k-1}) W(y_k|x_k)$ for $P \in \mathcal{P}(\mathcal{A}^n)$. 
    
\end{lemma}

\begin{remark}
Lemma \ref{aaron'slemma} is a starting point to prove achievability results both with feedback (Theorem \ref{Feedback_Achievability_Theorem}) and without feedback (Theorem \ref{Achievability_Theorem}).   
\end{remark}

The following discussion illustrates how a controller can be used to generate a constant-composition channel input $X^n$. For any given type $t \in \mathcal{P}_n(\mathcal{A})$, let $\{T^n_{\mathcal{A}}(t)\}$ denote the multiset of size $n$ in which each $a \in \mathcal{A}$ has multiplicity equal to $n \cdot t(a)$. Let $\text{Unif}\left ( \{T^n_{\mathcal{A}}(t) \}\right ) \in \mathcal{P}(\mathcal{A})$ denote the uniform distribution over the multiset $\{T^n_{\mathcal{A}}(t) \}$; in other words, $\text{Unif}\left ( \{T^n_{\mathcal{A}}(t) \}\right ) = t$. Furthermore, for any sequence $x^k \in \mathcal{A}^k$,
\begin{align*}
    \text{Unif}\left (\{T^n_{\mathcal{A}}(t) \} - x^k\right )
\end{align*}
denotes the uniform distribution over the multiset $\{T^n_{\mathcal{A}}(t) \}$ with elements in the sequence $x^k$ removed.  
\begin{example}
A controller $F$ specified by 
\begin{align*}
    F(x^{k-1}, y^{k-1}) &= \text{Unif}\left (\{T^n_{\mathcal{A}}(t) \} - x^{k-1}\right ) 
\end{align*}   
for $k = 1, \ldots, n$ gives rise to a constant-composition code from the type class $t$, i.e., the total channel input $X^n$ is a random sequence with uniform distribution over the type class $T^n_{\mathcal{A}}(t)$, denoted\footnote{Note that in our notation, $\text{Unif}(T^n_\mathcal{A}(t)) \in \mathcal{P}(\mathcal{A}^n)$ whereas $\text{Unif}(\{T^n_\mathcal{A}(t)\}) \in \mathcal{P}(\mathcal{A})$.} as $X^n \sim \text{Unif}(T^n_\mathcal{A}(t))$. Note that this is an example of a controller without feedback.
\end{example}

Lemma \ref{mostgenconv} given below serves as a starting point to prove converse results. Different variants of the converse in Lemma \ref{mostgenconv} can be found in \cite[Theorem 27]{5452208}, \cite[(42)]{8012458} and \cite[Lemma 15]{9099482}, showing the wide applicability of the result. Motivated by this, we state Lemma \ref{mostgenconv} in a more general form than is needed for Theorem \ref{Converse_Theorem}.

\begin{lemma}
Consider a channel $W$ with cost constraint $(\Gamma,V) \in (\Gamma_0, \Gamma^*) \times (0, \infty)$, where $(\Gamma, V)$ maps to some subset $\mathcal{P}_{\Gamma,V}(\mathcal{A}^n) \subset \mathcal{P}(\mathcal{A}^n)$ of distributions. Consider a random non-feedback $(n,R)$ code with minimum average error probability at most $\epsilon \in (0, 1)$ such that the codewords are distributed according to some $\overline{P} \in \mathcal{P}_{\Gamma,V}(\mathcal{A}^n)$. Then for every $n, \rho > 0$ and $\epsilon \in (0, 1)$,   
\begin{align}
    \log \lceil \exp(nR) \rceil &\leq \log \rho - \log \left [ \left( 1 - \epsilon - \sup_{\overline{P} \in \mathcal{P}_{\Gamma,V}(\mathcal{A}^n)}\, \inf_{q \in \mathcal{P}(\mathcal{B}^n)} (\overline{P} \circ W) \left( \frac{W(Y^n|X^n)}{q(Y^n)} > \rho \right)\right)^+\right]. \label{qandp}
\end{align}

\label{mostgenconv}
\end{lemma}

\textit{Proof:} The proof is similar to \cite[Lemma 15]{9099482} and is given in Appendix \ref{mostgenconvproof} for convenience. 

\begin{remark}
Lemma \ref{mostgenconv} formulates the cost constraint as a set of allowable distributions of the channel input, and this set can be parametrized by an arbitrary number of constraint parameters such as mean, variance and higher-order moments. In our case, we take the cost-constrained set $\mathcal{P}_{\Gamma, V}(\mathcal{A}^n) \subset \mathcal{P}(\mathcal{A}^n)$ to be equal to the set of distributions $\overline{P}$ such that the channel input $X^n \sim \overline{P}$ satisfies $\mathbb{E}\left [ \sum_{i=1}^n c(X_i) \right ] \leq n \Gamma$ and $\text{Var}\left(\sum_{i=1}^n c(X_i) \right) \leq n V$. Furthermore, a feedback version of Lemma \ref{mostgenconv} can be proved and stated by replacing $\overline{P}$ in $(\ref{qandp})$ by controllers $F$ such that the marginal distribution of $X^n$ induced by $F \circ W$ lies in $\mathcal{P}_{\Gamma, V}(\mathcal{A}^n)$.   
\end{remark}

\section{Main Results}

\begin{definition}
    The function $\mathcal{K} : \mathbb{R} \times (0, \infty) \to (0, 1)$ is defined as 
    \begin{align}
        \mathcal{K}(r, V) := \inf_{\Pi} \mathbb{E}[\Phi(\Pi)], \label{jzz}
    \end{align}
    where the infimum is over all random variables $\Pi$ satisfying $\mathbb{E}[\Pi] \geq r$ and $\text{Var}(\Pi) \leq V$. \label{Kdef}
\end{definition}

The properties of the function $\mathcal{K}(r, V)$ are given in the following lemma. 

\begin{lemma}
    The function $\mathcal{K}(r, V)$ satisfies the following three properties: 
    \begin{enumerate}
        \item The infimum in $(\ref{jzz})$ is a minimum, and there exists a minimizer which is a discrete probability distribution with at most 3 point masses;
    \item $\mathcal{K}(r, V)$ is a strictly increasing function w.r.t. $r$ (for a fixed $V$);
    \item $\mathcal{K}(r, V)$ is (jointly) continuous in $(r, V)$.
    \end{enumerate}
    \label{Kfunctionproperties}
\end{lemma}

\textit{Proof:} The proof is given in Appendix \ref{Kfuncpropertiesproof}.   

\begin{corollary}
    An equivalent definition of the function $\mathcal{K} : \mathbb{R} \times (0, \infty) \to (0, 1)$ is
\begin{align*}
    \mathcal{K}\left(r, V \right) &= \min_{\substack{\Pi:\\
    \mathbb{E}[\Pi] = r \\
    \text{Var}(\Pi) \leq V\\
    |\text{supp}(\Pi)| \leq 3
    }} \mathbb{E}\left [\Phi(\Pi) \right]. 
\end{align*}
    \label{corKfunc}
\end{corollary}
\textit{Proof:} The corollary follows from Lemma \ref{Kfunctionproperties}. 

The converse (Theorem \ref{Converse_Theorem}) and achievability (Theorem \ref{Achievability_Theorem}) results provide matching upper and lower bounds, characterizing the optimal second-order coding rate of the $(\Gamma, V)$ cost constraint in terms of the function $\mathcal{K}(r, V)$ as follows: 

\begin{align}
    \lim_{n \to \infty}\,\frac{\log M^*(n, \epsilon, \Gamma, V) - nC(\Gamma)}{\sqrt{n}} = \max \left \{r \in \mathbb{R} : \mathcal{K}\left(\frac{r}{\sqrt{V(\Gamma) }}, \frac{C'(\Gamma)^2 V}{V(\Gamma)} \right) \leq \epsilon \right \}. \label{opt_sec_order} 
\end{align}
The maximum on the right-hand side of $(\ref{opt_sec_order})$ is well-defined. In fact, we have the following general result. 
\begin{lemma}
    For any $V > 0$ and $0 < \epsilon  <1$, the supremum, 
\begin{align}
    \sup \left \{r' \in \mathbb{R} : \mathcal{K}(r', V) \leq \epsilon \right \},  \label{infr}
\end{align}
is achieved. Furthermore, the maximum, call it $r^*$, satisfies $\mathcal{K}(r^*, V) = \epsilon$.   
\label{fqq}
\end{lemma}
\textit{Proof:} The proof is given in Appendix \ref{fqq_proof}. 

We denote the optimal second-order coding rate in $(\ref{opt_sec_order})$ by
\begin{align}
    r^* := \max \left \{r \in \mathbb{R} : \mathcal{K}\left(\frac{r}{\sqrt{V(\Gamma) }}, \frac{C'(\Gamma)^2 V}{V(\Gamma)} \right) \leq \epsilon \right \}. \label{insight6312}
\end{align}
The converse (Theorem \ref{Converse_Theorem}) and achievability (Theorem \ref{Achievability_Theorem}) results also show that in the limit as $n \to \infty$,
\begin{align}
    \mathcal{K}\left(\frac{r}{\sqrt{V(\Gamma) }}, \frac{C'(\Gamma)^2 V}{V(\Gamma)} \right) \label{plotreview1}
\end{align}
is the minimum average error probability of $\left(n, R, \Gamma, V  \right)$ codes without feedback for $R = C(\Gamma) + \frac{r}{\sqrt{n}}$. It is insightful to compare $(\ref{insight6312})$ with the optimal SOCR of previous cost constraints, such as almost-sure constraint and expected cost constraint. 

\begin{itemize}
    \item \underline{Almost-sure constraint ($c(X^n) \leq \Gamma$ almost surely)}: 
    
    Recall that $r_{\text{a.s.}} := \sqrt{V(\Gamma)} \Phi^{-1}(\epsilon)$ is the optimal SOCR associated with an a.s. cost constraint\footnote{assuming a unique capacity-cost-achieving distribution.}. We note that $r^* \geq \sqrt{V(\Gamma)} \Phi^{-1}(\epsilon)$ with equality if a minimizing probability distribution in
$$\mathcal{K}\left(\frac{r^*}{\sqrt{V(\Gamma) }}, \frac{C'(\Gamma)^2 V}{V(\Gamma)} \right)$$
has only one point mass. One indeed has $r^* = r_{\text{a.s.}}$ for $V = 0$. While the $(\Gamma, V)$ cost constraint for $V = 0$ is stronger than the a.s. cost constraint, one nevertheless obtains $r^* = r_{\text{a.s.}}$ for $V = 0$ because in practice, optimal coding schemes for the a.s. cost constraint satisfy the $(\Gamma, V)$ cost constraint for $V = 0$. 

\item \underline{Expected cost constraint ($\mathbb{E}[c(X^n)] \le \Gamma$):}

This is a special case of $(\Gamma, V)$ cost constraint with $V = \infty$. As mentioned in the Introduction, the strong converse does not hold with an expected cost constraint, i.e., $\frac{1}{n} \log M^*(n, \epsilon, \Gamma, V) > C(\Gamma)$ for $V = \infty$ and sufficiently large $n$. Hence, the achievable SOCR in this case is infinite based on our definition of SOCR in $(\ref{oursocrhai})$.  One therefore expects $r^* \to \infty$ as $V \to \infty$. This is indeed the case and it suffices to show that $\mathcal{K}(r, \infty) = 0$ for all values of $r$. Since $\mathcal{K}(r, \infty)$ is non-decreasing in $r$, fix an arbitrarily large value of $r > 0$. Then for any $\epsilon > 0$, we can choose $m$ sufficiently large so that for $p=1-1/m$, $\pi_1 = - \sqrt{\log m}$ and $\pi_2 = \frac{r - p \pi_1}{1-p}$, we have $p \pi_1 + (1-p) \pi_2 = r$ and
\begin{align*}
    \mathcal{K}(r, \infty) &\leq p \Phi(\pi_1) + (1-p) \Phi(\pi_2)\\
    &\leq -\frac{p}{\pi_1} \phi(\pi_1) + 1 - p\\
    &\leq \epsilon. 
\end{align*}

\end{itemize}

\subsection{Non-Feedback Converse}

We now state the converse result for $(n, R, \Gamma, V)$ codes defined in Definition \ref{defwcostwofeedback}. 

\begin{theorem}

Fix an arbitrary $\epsilon \in (0, 1)$. Consider a channel $W$ with cost constraint $ (\Gamma,V) \in (\Gamma_0, \Gamma^*) \times (0, \infty)$ such that $P^*$ is unique and $V(\Gamma) > 0$. Then 
\begin{align*}
     \limsup_{n \to \infty}\,\frac{\log M^*(n, \epsilon, \Gamma, V) - nC(\Gamma)}{\sqrt{n}} \leq \max \left \{r \in \mathbb{R} : \mathcal{K}\left(\frac{r}{\sqrt{V(\Gamma) }}, \frac{C'(\Gamma)^2 V}{V(\Gamma)} \right) \leq \epsilon \right \}. 
\end{align*}
Alternatively, 
\begin{align*}
    \liminf_{n \to \infty} \bar{P}_{\text{e}}\left(n,C(\Gamma) + \frac{r}{\sqrt{n}},  \Gamma, V\right) &\geq \mathcal{K}\left (\frac{r}{\sqrt{V(\Gamma) }}, \frac{C'(\Gamma)^2 V}{V(\Gamma)}\right ). 
\end{align*}
    \label{Converse_Theorem}
\end{theorem}

\textit{Proof:} The proof is given in Section \ref{Converse_Theorem_proof}.

\textit{Proof Outline:} The starting point will be the result in Lemma \ref{mostgenconv}, where $\mathcal{P}_{\Gamma,V}(\mathcal{A}^n)$ is set equal to the set of distributions $\overline{P}$ such that the channel input $X^n \sim \overline{P}$ satisfies $\mathbb{E}\left [ \sum_{i=1}^n c(X_i) \right ] \leq n \Gamma$ and $\text{Var}\left(\sum_{i=1}^n c(X_i) \right) \leq n V$. 

Choosing $\rho = \exp \left(n C(\Gamma) + \sqrt{n} r \right)$ in Lemma \ref{mostgenconv} and a suitable choice of $q$ in $(\ref{qandp})$, we upper bound 
\begin{align}
     &\sup_{\overline{P}}\,(\overline{P} \circ W) \left( \frac{W(Y^n|X^n)}{q(Y^n)} > \rho \right) \notag \\
     &= \sup_{\overline{P}}\, \sum_{x^n \in \mathcal{A}^n} \overline{P}(x^n)  W \left( \log \frac{W(Y^n|x^n)}{q(Y^n)} > n C(\Gamma) + \sqrt{n} r \right) \notag \\
     &\lessapprox \sup_{\overline{P}}\, \mathbb{E}_{\overline{P}}\left [ 1-  \Phi\left(\frac{\sqrt{n} C'(\Gamma)}{\sqrt{V(\Gamma)}} \left(\Gamma - \frac{1}{n} \sum_{i=1}^n c(X_i)  \right)  + 
    \frac{ r}{\sqrt{V(\Gamma) }} \right) \right] \notag \\
    &= 1 - \inf_{\Pi} \mathbb{E}\left [ \Phi(\Pi) \right], \label{1and1together}
\end{align}
where the infimum is over random variables $\Pi$ satisfying $\mathbb{E}\left [ \Pi \right] \geq \frac{r}{\sqrt{V(\Gamma)}}$ and $\text{Var}(\Pi) \leq \frac{C'(\Gamma)^2 V}{V(\Gamma)}$. Lemma \ref{Kfunctionproperties} is then used in $(\ref{1and1together})$, giving rise to the $\mathcal{K}$-function. In view of $(\ref{qandp})$, the converse is given by the maximum value of $r$ in $(\ref{1and1together})$ for which the value of $\mathcal{K}$-function is at most $\epsilon$.

\subsection{Non-Feedback Achievability}

\begin{theorem}
    Fix an arbitrary $\epsilon \in (0, 1)$. Consider a channel $W$ with cost constraint $ (\Gamma,V) \in (\Gamma_0, \Gamma^*) \times (0, \infty)$ such that $P^*$ is unique and $V(\Gamma) > 0$. Then  
\begin{align*}
     \liminf_{n \to \infty}\,\frac{\log M^*(n, \epsilon, \Gamma, V) - nC(\Gamma)}{\sqrt{n}} \geq \max \left \{r : \mathcal{K}\left(\frac{r}{\sqrt{V(\Gamma) }}, \frac{C'(\Gamma)^2 V}{V(\Gamma)} \right) \leq \epsilon \right \}.  
\end{align*}
Alternatively, 
\begin{align*}
    \limsup_{n \to \infty} \bar{P}_{\text{e}}\left(n,C(\Gamma) + \frac{r}{\sqrt{n}},  \Gamma, V\right) &\leq \mathcal{K}\left (\frac{r}{\sqrt{V(\Gamma) }}, \frac{C'(\Gamma)^2 V}{V(\Gamma)}\right ). 
\end{align*}

    \label{Achievability_Theorem}
\end{theorem}
\textit{Proof:} The proof is given in Section \ref{Achievability_Theorem_proof}. 
\begin{remark}
We prove Theorem \ref{Achievability_Theorem} under a slightly stricter cost formulation given by  
\begin{align}
\begin{split}
    \max_{1 \leq i \leq n} \mathbb{E}\left [ c(X_i) \right ] &\leq \Gamma\\
    \text{Var}\left(\sum_{i=1}^n c(X_i) \right) &\leq nV,
\end{split}
     \label{strong_version}
\end{align}
which trivially implies the original cost formulation,
\begin{align}
\begin{split}
    \mathbb{E}\left[\sum_{i=1}^n c(X_i)\right] &\leq n\Gamma\\
    \text{Var}\left(\sum_{i=1}^n c(X_i) \right) &\leq nV. 
    \end{split}
    \label{weaker_version}
\end{align}
Despite the restriction, we obtain a matching lower bound in Theorem \ref{Achievability_Theorem} to the upper bound in Theorem \ref{Converse_Theorem}. This means that the distinction between $(\ref{strong_version})$ and $(\ref{weaker_version})$ is immaterial as far as the optimal SOCR is concerned. 

\textit{Proof Outline:} As a brief overview, the achievability scheme makes use of the solution (a three-point probability distribution) to the optimization problem in 
$$\mathcal{K}\left(\frac{r^*}{\sqrt{V(\Gamma) }}, \frac{C'(\Gamma)^2 V}{V(\Gamma)} \right),$$
where $r^*$ is defined in $(\ref{insight6312})$. 
Specifically, the minimizing probability distribution with three point masses, denoted by 
\begin{align}
    P_{\Pi}(\pi) = \begin{cases}
    p_1 & \pi = \pi_1\\
    p_2 & \pi = \pi_2\\
    p_3 & \pi = \pi_3,
    \end{cases} \label{PPigvq}
\end{align}
is mapped to three different cost values $\Gamma_1, \Gamma_2$ and $\Gamma_3$ as follows:
$$\Gamma_{j} = \Gamma - \frac{\sqrt{V(\Gamma) }}{C'(\Gamma)\sqrt{n}}\pi_j + \frac{r^*}{C'(\Gamma)\sqrt{n}}$$
for $j \in \{1,2,3 \}$. The cost values $(\Gamma_1,\Gamma_2,\Gamma_3)$ are in turn mapped to three types $T_1, T_2$ and $T_3$, each type $T_j$ being close to a capacity-cost-achieving distribution for cost $\Gamma_j$. Subsequently, we use a random coding scheme where the codewords are drawn randomly from one of the three type classes with probability weights $p_1, p_2$ and $p_3$.  

The analysis of constant composition codes, i.e., codes that generate random codewords uniformly from a fixed type class, is complicated by the fact that the induced output distribution is $Q^{cc}$ instead of $Q^*$, the former of which does not factorize into a product over the output sequence. In other words, the channel output is not i.i.d. The third part of Lemma \ref{combinedlemmas} bounding the ratio of $Q^{cc}$ and $Q^*$ is then helpful for such analysis. Specifically, it is helpful in effecting a change of measure from $Q^{cc}$ to $Q^*$, although it cannot be directly applied because the induced output distribution itself is a mixture of three "$Q^{cc}$'s".
\end{remark}

\subsection{Feedback Improves the SOCR}

From Theorems \ref{Converse_Theorem} and \ref{Achievability_Theorem}, we have that 
$$r^* = \max \left \{r : \mathcal{K}\left(\frac{r}{\sqrt{V(\Gamma) }}, \frac{C'(\Gamma)^2 V}{V(\Gamma)} \right) \leq \epsilon \right \}$$
is the optimal second-order coding rate for $(n, R, \Gamma, V)$ codes without feedback. The next theorem proves that feedback can improve the second-order coding rate as well as the average error probability. 

\begin{theorem}
    Fix an arbitrary $\epsilon \in (0, 1)$. Consider a channel $W$ with cost constraint $(\Gamma ,V) \in (\Gamma_0, \Gamma^*) \times (0, \infty)$ such that $P^*$ is unique and $V(\Gamma) > 0$. Then 
\begin{align*}
     \liminf_{n \to \infty}\,\frac{\log M_{\text{fb}}^*(n, \epsilon, \Gamma, V) - nC(\Gamma)}{\sqrt{n}} > \max \left \{r : \mathcal{K}\left(\frac{r}{\sqrt{V(\Gamma) }}, \frac{C'(\Gamma)^2 V}{V(\Gamma)} \right) \leq \epsilon \right \}.  
\end{align*}
Alternatively, 
\begin{align}
    \limsup_{n \to \infty} \bar{P}_{\text{e,fb}}\left(n,C(\Gamma) + \frac{r}{\sqrt{n}},  \Gamma, V\right) &< \mathcal{K}\left (\frac{r}{\sqrt{V(\Gamma) }}, \frac{C'(\Gamma)^2 V}{V(\Gamma)}\right ). \label{bestaaa}
\end{align}
    \label{Feedback_Achievability_Theorem}
\end{theorem}

\textit{Proof:} The proof is given in section \ref{Feedback_Achievability_Theorem_proof}.

\textit{Proof Outline:} We consider
a random feedback code in which feedback is only used once halfway through the transmission. As discussed in the Introduction section, the feedback scheme used in the proof is a variant of timid/bold coding. Recall from the discussion in the Introduction section, specifically from $(\ref{eq:compound}) - (\ref{blindarrow})$, that timid and bold channel input distributions are associated with low and high variance, respectively, of the information density.  During the first half, the distribution of the channel input is similar to the distribution used in the non-feedback achievability scheme (as described in the proof outline for Theorem \ref{Achievability_Theorem}). To reiterate, 
it is a mixture distribution of the three type classes emanating from a minimizing three-point probability distribution in
\begin{align}
\mathcal{K}\left(\frac{r}{\sqrt{V(\Gamma) }}, \frac{C'(\Gamma)^2 V}{V(\Gamma)} \right).  \label{vc5}  
\end{align}
Using feedback at $t = n/2$, the encoder determines whether the information density evaluated at $(x^{n/2}, y^{n/2})$ is above some suitable threshold. If it is, then the channel input for the second half is chosen to be constant-composition over only one type class with cost $\Gamma$; otherwise, the channel input is chosen to retain the mixture distribution of the three type classes over the whole blocklength. Since the mixture distribution of three type classes has a greater spread around the cost point $\Gamma$ than the constant-composition distribution over a single type class, the former can be considered a bold distribution and the latter a timid distribution. Hence, if the transmission has proceeded well, the encoder switches to a timid distribution. 

\begin{remark}
    Since the proof of Theorem \ref{Feedback_Achievability_Theorem} only uses feedback once over $n$ channel uses, the result in Theorem \ref{Feedback_Achievability_Theorem} is primarily meant to be a theoretical contribution and a proof of concept that feedback improvement is possible. This finding has significance in the context of previous results which showed that for a unique $P^*$, second-order feedback improvement is impossible for unconstrained channels \cite[Theorem 3]{9099482} as well as for channels with the a.s. cost constraint \cite[Theorem 2]{mahmood2024improvedchannelcodingperformance}.
\end{remark}  

\begin{remark}
An explicit expression for the improvement in error probability in $(\ref{bestaaa})$ can be obtained from $(\ref{digimon1})$ in terms of an
optimizable design parameter $\beta$. 
\end{remark}

\section{Supplementary Lemmas \label{suppsection}}

\begin{lemma}
    \begin{enumerate}
        \item $C(\Gamma)$ and $[C(\Gamma)]_n$: For sufficiently large $n$ so that $\text{supp}(P^*) = \text{supp}([P^*]_n)$, 
        \begin{align*}
             C(\Gamma) - \frac{2C'(\Gamma) J c_{\max}}{n} \leq  [C(\Gamma)]_n \leq   C(\Gamma). 
        \end{align*}
        \item $V(\Gamma)$ and $[V(\Gamma)]_n$:
        \begin{align*}
            V(\Gamma) - \frac{2J \nu_{\max}}{n} \leq [V(\Gamma)]_n \leq  V(\Gamma) + \frac{2J \nu_{\max}}{n}.
        \end{align*}
        \item $Q^{cc}$ and $Q^*$:
        There exist an integer $N > 0$ and some constant $\kappa$ such that for $n > N$, 
    \begin{align*}
        \log \frac{Q^*(y^{n})}{Q^{\text{cc}}(y^{n})} \geq - \frac{s(P^*) - 1}{2} \log (n) -\kappa
    \end{align*}
    for all $y^{n} \in \mathcal{B}^{n}$.
    \end{enumerate}
    \label{combinedlemmas}
\end{lemma}
\textit{Proof:} The proof is given in Appendix \ref{myrefinement}. 

\begin{remark}
    The third part of Lemma \ref{combinedlemmas} only requires that $[P^*]_n$ and $P^*$ are within $O(1/n)$ in $l^1$ distance. Furthermore, if we are only given that $[P^*]_n$ and $P^*$ are within $O(1/n)$ in $l^1$ distance, the first and second parts of Lemma \ref{combinedlemmas} can be written as 
    \begin{align*}
             C(\Gamma) - O\left( \frac{1}{n}\right) \leq  [C(\Gamma)]_n \leq   C(\Gamma) + O\left( \frac{1}{n}\right) 
        \end{align*}
        and 
        \begin{align*}
            V(\Gamma) - O\left( \frac{1}{n}\right) \leq [V(\Gamma)]_n \leq  V(\Gamma) + O\left(\frac{1}{n} \right),
        \end{align*}
        respectively. 
        \label{upeeupee}
\end{remark}

\begin{lemma}
    For any random variable $X$, we have 
    \begin{align*}
        \lim_{x \to \infty} \left [ \frac{1}{x} \log \mathbb{E}\left [ \Phi(X - x) \right] + \frac{x}{2} \right] = ||X||_\infty. 
    \end{align*}
    \label{aaronesssuplemma}
\end{lemma}
\begin{IEEEproof}
    Consider any $\bar{x} < ||X||_\infty$. Then we have, for $x > \bar{x}$, 
    \begin{align*}
        \mathbb{E}\left [ \Phi(X - x) \right] &\geq \mathbb{E}\left [ \Phi(X - x) \mathds{1}\left(X \geq \bar{x}\right) \right]\\
        &\geq \mathbb{P}\left(X \geq \bar{x} \right) \Phi(\bar{x} - x)\\
        &\geq \mathbb{P}\left(X \geq \bar{x} \right) \frac{x-\bar{x}}{1 + (x-\bar{x})^2} \phi(\bar{x} - x),
    \end{align*}
which implies 
\begin{align*}
    \lim_{x \to \infty} \left [ \frac{1}{x} \log \mathbb{E}\left [ \Phi(X - x) \right] + \frac{x}{2} \right] \geq ||X||_\infty
\end{align*}
since $\bar{x} < ||X||_\infty$ was arbitrary. If $||X||_\infty = \infty$ then the reverse inequality is trivial. Otherwise, write $\bar{x} = ||X||_\infty < \infty$. Then we have, for $x > \bar{x}$, 
\begin{align*}
    \mathbb{E}\left [ \Phi(X - x) \right] &\leq \Phi(\bar{x} - x)\\
    &\leq \frac{1}{x-\bar{x}} \phi(x-\bar{x}) 
\end{align*}
which yields the upper bound 
\begin{align*}
    \lim_{x \to \infty} \left [ \frac{1}{x} \log \mathbb{E}\left [ \Phi(X - x) \right] + \frac{x}{2} \right] \leq ||X||_\infty.
\end{align*}

\end{IEEEproof}

\begin{lemma}
Let $Y^n$ be i.i.d. according to $Q_n' \in \mathcal{P}(\mathcal{B})$. Let there exist $N > 0$ and $\gamma > 0$ such that $\min\{Q_n'(y): y \in \mathcal{B} \} \geq \gamma$ for all $n > N$. Let $Q_n \in \mathcal{P}(\mathcal{B})$ be such that $Q_n \neq Q_n'$ and $D(Q_n'||Q_n) \to 0$ as $n \to \infty$. Then 
    \begin{align}
    &\mathbb{P}\left(\log \frac{Q_n(Y^n)}{Q_n'(Y^n)} \geq c_n \right) \notag \\
    & \leq \exp \left( -\frac{n D(Q_n' || Q_n) \gamma^2}{9} \right) \exp \left(- \frac{ 2c_n \gamma^2}{9} \right) \exp \left( - \frac{c_n^2 \gamma^2}{9n D(Q_n' || Q_n)} \right)
    \label{showingoff2}
    \end{align}
    for sufficiently large $n$,
    where $c_n > 0$ is any sequence.
\label{proudlemma}
\end{lemma}

\begin{IEEEproof}
We have 
\begin{align}
    &\mathbb{P}\left( \log \frac{Q_n(Y^n)}{Q_n'(Y^n)} \geq c_n  \right) \notag \\
    &= \mathbb{P}\left(\sum_{m=1}^n \log \frac{Q_n(Y_m)}{Q_n'(Y_m)} \geq c_n  \right) \notag \\
    &= \mathbb{P}\left(\sum_{m=1}^n \left [ \log \frac{Q_n(Y_m)}{Q_n'(Y_m)}\right] - (-n D(Q_n'||Q_n))  \geq n D(Q_n' || Q_n) + c_n  \right). \label{hoeffee}
\end{align}
From Pinsker's inequality and the well-known relationships between $l^1$, $l^2$ and $l^\infty$ norms, we have \\
$||Q_n' - Q_n||_\infty \leq \sqrt{2 D(Q_n' || Q_n)}$. Hence, 
\begin{align*}
    \log \left( \frac{Q_n'(Y) - \sqrt{2 D(Q_n' || Q_n)}}{Q_n'(Y)}\right) &\leq \log \left( \frac{Q_n(Y)}{Q_n'(Y)}\right) \leq \log \left( \frac{Q_n'(Y) + \sqrt{2 D(Q_n' || Q_n)}}{Q_n'(Y)}\right) \\
    a = \log \left( \frac{\gamma - \sqrt{2 D(Q_n' || Q_n)}}{\gamma}\right) &\leq \log \left( \frac{Q_n(Y)}{Q_n'(Y)}\right) \leq \log \left( \frac{\gamma + \sqrt{2 D(Q_n' || Q_n)}}{\gamma}\right) = b
\end{align*}
almost surely, where the lower and upper bounds, called $a$ and $b$, respectively, are well-defined for sufficiently large $n$ so that $\gamma - \sqrt{2 D(Q_n' || Q_n)} > 0$. By a standard Taylor series expansion, we have 
\begin{align*}
    b-a &\leq \frac{3 \sqrt{2 D(Q_n' || Q_n)}}{\gamma} 
\end{align*}
for sufficiently large $n$ since $D(Q_n' || Q_n) \to 0$ as $n \to \infty$. Then by the application of Hoeffding's inequality \cite[Theorem 2]{omghoeffding} to $(\ref{hoeffee})$, we obtain  
\begin{align*}
    &\mathbb{P}\left(\log \frac{Q_n(Y^n)}{Q_n'(Y^n)} \geq c_n \right)\\
    &\leq  \exp \left( - \frac{2 \gamma^2\left(n D(Q_n' || Q_n) + c_n  \right )^2}{ 18n D(Q_n' || Q_n)} \right)\\
    &= \exp \left( -\frac{n D(Q_n' || Q_n) \gamma^2}{9} \right) \exp \left(- \frac{ 2c_n \gamma^2}{9} \right) \exp \left( - \frac{c_n^2 \gamma^2}{9n D(Q_n' || Q_n)} \right).
\end{align*}

\end{IEEEproof}

\begin{lemma}
Fix any $\Gamma \in (\Gamma_0, \Gamma^*)$, and let $P^*$ be the unique capacity-cost-achieving distribution for $C(\Gamma)$. Then
for any $\beta > 0$, there exists an $\epsilon > 0$ such that 
    $$\sup_{\substack{P \in \mathcal{P}(\mathcal{A}):\\
    ||P - P^*||_1 \geq \beta\\
    c(P) \leq \Gamma + \epsilon }} I(P, W) < C(\Gamma).$$
    \label{lemcont}
\end{lemma}
\begin{IEEEproof}
    We prove by contradiction. Assume there exists a sequence $P_n$ such that $||P_n - P^*||_1 \geq \beta$ for all $n$,\\ $\limsup_{n \to \infty} c(P_n) \leq \Gamma$ and $\liminf_{n \to \infty} I(P_n, W) \geq C(\Gamma)$. By considering a suitable convergent subsequence (Bolzano-Weierstrass Theorem), we may assume that $P_n \to P$ for some $P \in \mathcal{P}(\mathcal{A})$, where $c(P) \leq \Gamma$, $I(P, W) \geq C(\Gamma)$ and $||P - P^*||_1 \geq \beta$. But this contradicts the uniqueness of $P^*$.   
\end{IEEEproof}

\section{Proof of non-feedback converse \label{Converse_Theorem_proof}}

We start with Lemma \ref{mostgenconv} and first upper bound 
\begin{align}
\sup_{\overline{P} \in \mathcal{P}_{\Gamma,V}(\mathcal{A}^n)}\, \inf_{q \in \mathcal{P}(\mathcal{B}^n)} (\overline{P} \circ W) \left( \frac{W(Y^n|X^n)}{q(Y^n)} > \rho \right).  \label{vq0}   
\end{align}
Let $\rho = \exp\left(n C(\Gamma) + \sqrt{n} r \right)$ in $(\ref{vq0})$, where $r$ is a number to be specified later. 
Let $X^n \sim \overline{P}$ for any arbitrary $\overline{P}$ satisfying $\mathbb{E}\left [ \sum_{i=1}^n c(X_i) \right ] \leq n \Gamma$ and $\text{Var}\left(\sum_{i=1}^n c(X_i) \right) \leq n V$.

Choosing $q \in \mathcal{P}(\mathcal{B}^n)$ as 
\begin{align*}
    q(y^n) = \frac{1}{2} \prod_{i=1}^n Q^*(y_i) + \frac{1}{2} \frac{1}{|\mathcal{P}_n(\mathcal{A})|} \sum_{t \in \mathcal{P}_n(\mathcal{A})} \prod_{i=1}^n q_t(y_i),
\end{align*}
where 
\begin{align*}
    q_t(b) := \sum_{a \in \mathcal{A}} t(a) W(b|a),    
\end{align*}
we have 
\begin{align}
    &(\overline{P} \circ W) \left( \log \frac{W(Y^n|X^n)}{q(Y^n)} > n C(\Gamma) + \sqrt{n} r  \right) \notag \\
    &= \sum_{x^n \in \mathcal{A}^n} \overline{P}(x^n)  W \left( \log \frac{W(Y^n|x^n)}{q(Y^n)} > n C(\Gamma) + \sqrt{n} r \right), \label{fopv}
\end{align}
where in $(\ref{fopv})$, $Y_1, Y_2, \ldots, Y_n$ are independent random variables and each $Y_i \sim W(\cdot | x_i)$. 

Define 
\begin{align*}
    \nu_a^t :=  \text{var}\left( \log \frac{W(Y|a)}{q_t(Y)} \right),\quad  \text{ where } Y \sim W(\cdot|a).
\end{align*}

For any $\beta \geq 0$, define 
\begin{align*}
    \delta_\beta := \sup_{\substack{p \in \mathcal{P}(\mathcal{A}): \\||p-P^*||_1 \leq \beta}} \Bigg | \sum_{a \in \mathcal{A}} p(a) \nu_a^p  - V(\Gamma) \Bigg |
\end{align*}
and note that $\delta_\beta \to 0$ as $\beta \to 0$. Define the function  
\begin{align*}
    \psi_\beta\left(c \right ) := \begin{cases}
V(\Gamma) + \delta_\beta & \text{ if } c \leq \Gamma\\
V(\Gamma) - \delta_\beta & \text{ if } c > \Gamma.
\end{cases}    
\end{align*}
Fix any $0 < \Delta < 1/2$ and $\eta > 0$. Then select $\beta > 0$ such that $\delta_\beta < V(\Gamma)$, 
\begin{align}
    \sup_{c \in \mathbb{R}} \Bigg | \frac{1}{\sqrt{V(\Gamma)}} - \frac{1}{\sqrt{\psi_\beta(c)}} \Bigg | \leq \frac{\Delta}{C'(\Gamma) \sqrt{V}} \label{aaronerr1}
\end{align} 
and 
\begin{align}
    \sup_{c \in \mathbb{R}} \Bigg | \frac{1}{V(\Gamma)} - \frac{1}{\psi_\beta(c)} \Bigg | \leq \frac{\Delta/2}{C'(\Gamma)^2 V}. \label{aaronerr2}
\end{align} 
Also define 
\begin{align}
    \varphi(r) := \begin{cases}
V(\Gamma) - \delta_\beta & \text{ if } r \leq 0\\
V(\Gamma) + \delta_\beta & \text{ if } r > 0.
\end{cases} \label{varphidefinitionr}
\end{align}
With $\Delta$, $\eta$ and $\beta$ fixed as above\footnote{$\eta$ is used later.}, we divide the set of sequences $x^n \in \mathcal{A}^n$ into four subsets:
\begin{align*}
    \mathcal{P}_{n,1} &:= \left \{ x^n: c(x^n) \leq \Gamma - \beta \right \},\\
    \mathcal{P}_{n,2} &:= \left \{ x^n: \Gamma - \beta < c(x^n) \leq \Gamma + \sqrt{\frac{\ln n}{n}}, ||P^* - P_{x^n}||_1 \leq \beta \right \},\\
    \mathcal{P}_{n,3} &:= \left \{ x^n: \Gamma - \beta < c(x^n) \leq \Gamma + \sqrt{\frac{\ln n}{n}}, ||P^* - P_{x^n}||_1  > \beta \right \},\\
    \mathcal{P}_{n,4} &:= \left \{ x^n: c(x^n) > \Gamma + \sqrt{\frac{\ln n}{n}} \right \}.
\end{align*}

For $x^n \in \mathcal{P}_{n, 1}$ and denoting $t = t(x^n)$, we have 
\begin{align}
 &W \left( \log \frac{W(Y^n|x^n)}{q(Y^n)} > n C(\Gamma) + \sqrt{n} r  \right) \notag \\
 &\leq W \left( \log \frac{W(Y^n|x^n)}{Q^*(Y^n)} > n C(\Gamma) + \sqrt{n} r - \log 2 \right) \notag \\
    &= W \left( \sum_{i=1}^n  \log \frac{W(Y_i|x_i)}{Q^*(Y_i)} > n C(\Gamma) + \sqrt{n} r  - \log 2\right) \notag \\
    &\stackrel{(a)}{\leq} W \left( \sum_{i=1}^n \left [   \log \frac{W(Y_{i}|x_i)}{Q^*(Y_i)}  
 - \mathbb{E}_{W(\cdot|x_i)} \left [   \log \frac{W(Y|x_i)}{Q^*(Y)}  \right]\right] >     nC'(\Gamma)\left (\Gamma - \frac{1}{n}\sum_{i=1}^n c(x_i)\right )  + \sqrt{n} r  
 - \log 2\right) \notag \\
&\leq W \left( \sum_{i=1}^n  \left [   \log \frac{W(Y_{i}|x_i)}{Q^*(Y_i)}  
 - \mathbb{E}_{W(\cdot|x_i)} \left [   \log \frac{W(Y|x_i)}{Q^*(Y)}  \right]\right] >     nC'(\Gamma)\beta  + \sqrt{n} r  - \log 2 \right) \notag \\
 &\stackrel{(b)}{\leq} W \left( \sum_{i=1}^n \left [   \log \frac{W(Y_{i}|x_i)}{Q^*(Y_i)}  
 - \mathbb{E}_{W(\cdot|x_i)} \left [   \log \frac{W(Y|x_i)}{Q^*(Y)}  \right]\right] >     \frac{nC'(\Gamma)\beta}{2}    \right) \notag \\
 &\stackrel{(c)}{\leq} \frac{4  \sum_{a \in \mathcal{A}} t(a) \nu_a}{n (C'(\Gamma)^2 \beta^2)} \label{chebyguy}\\
 &\leq \frac{4  \nu_{\max}}{n (C'(\Gamma)^2 \beta^2)} \label{pn1}
\end{align}
for sufficiently large $n$. In inequality $(a)$ above, we used Proposition $\ref{getout}$. Inequality $(b)$ holds because $r$ is a constant. In inequality $(c)$, we used Chebyshev's inequality.   

For $x^n \in \mathcal{P}_{n, 2}$ and denoting $t = t(x^n)$, we have
\begin{align}
    &W \left( \log \frac{W(Y^n|x^n)}{q(Y^n)} > n C(\Gamma) + \sqrt{n} r \right) \notag \\
    &\leq W \left( \sum_{i=1}^n \log \frac{W(Y_i|x_i)}{q_t(Y_i)} > n C(\Gamma) + \sqrt{n} r  - \log 2(n+1)^J \right) \notag \\
    &\stackrel{(a)}{\leq} W \left( \sum_{i=1}^n \left [  \log \frac{W(Y_{i}|x_i)}{q_t(Y_{i})} - \mathbb{E}_W \left [\log \frac{W(Y|x_i)}{q_t(Y)}  \right ] \right ]  >  \right . \notag \\
    & \left . \quad \quad \quad \quad \quad \quad \quad   n C'(\Gamma) \left(\Gamma - \frac{1}{n} \sum_{i=1}^n c(x_i)  \right) +  n \left [  C\left (\frac{1}{n} \sum_{i=1}^n c(x_i)\right ) - I(t, W) \right ]  + \sqrt{n} r  - \log 2(n+1)^J \right) \notag \\
    &= W \left( \frac{1}{\sqrt{n \sum_a t(a) \nu^t_a}}\sum_{i=1}^n  \left [  \log \frac{W(Y_{i}|x_i)}{q_t(Y_{i})} - \mathbb{E}_W \left [\log \frac{W(Y|x_i)}{q_t(Y)}  \right ] \right ]  > - \frac{\log 2(n+1)^J}{\sqrt{n \sum_a t(a) \nu_a^t}} + \mbox{} \right . \notag \\
    & \left . \quad \quad \quad \quad \quad \quad \quad   \frac{\sqrt{n} C'(\Gamma)}{\sqrt{ \sum_a t(a) \nu^t_a}} \left(\Gamma - \frac{1}{n} \sum_{i=1}^n c(x_i)  \right) +  \frac{\sqrt{n}}{\sqrt{ \sum_a t(a) \nu^t_a}} \left [  C\left (\frac{1}{n} \sum_{i=1}^n c(x_i)\right ) - I(t, W) \right ]  + \frac{ r}{\sqrt{\sum_a t(a) \nu^t_a}}   \right) \notag\\
    &\stackrel{(b)}{\leq} 1-  \Phi\left(\frac{\sqrt{n} C'(\Gamma)}{\sqrt{ \sum_a t(a) \nu_a^t}} \left(\Gamma - \frac{1}{n} \sum_{i=1}^n c(x_i)  \right) +  \frac{\sqrt{n}}{\sqrt{ \sum_a t(a) \nu^t_a}} \left [  C\left (\frac{1}{n} \sum_{i=1}^n c(x_i)\right ) - I(t, W) \right ]   \right. \notag \\
    & \quad \quad \quad\quad\quad\quad \left .-  \frac{\log 2(n+1)^J}{\sqrt{n \sum_a t(a) \nu_a^t}} + 
    \frac{ r}{\sqrt{\sum_a t(a) \nu^t_a}} \right) + \frac{\kappa_t}{\sqrt{n}} \notag\\
    &\stackrel{(c)}{\leq} 1-  \Phi\left(\frac{\sqrt{n} C'(\Gamma)}{\sqrt{ \sum_a t(a) \nu_a^t}} \left(\Gamma - \frac{1}{n} \sum_{i=1}^n c(x_i)  \right) +  \frac{\sqrt{n}}{\sqrt{ \sum_a t(a) \nu^t_a}} \left [  C\left (\frac{1}{n} \sum_{i=1}^n c(x_i)\right ) - I(t, W) \right ]   \right. \notag\\
    & \quad \quad \quad\quad\quad\quad \left .-  \frac{\log 2(n+1)^J}{\sqrt{n \sum_a t(a) \nu_a^t}} + 
    \frac{ r}{\sqrt{\sum_a t(a) \nu^t_a}} \right) + \frac{\kappa }{\sqrt{n}}
    \notag\\
    &\leq 1-  \Phi\left(\frac{\sqrt{n} C'(\Gamma)}{\sqrt{ \sum_a t(a) \nu_a^t}} \left(\Gamma - \frac{1}{n} \sum_{i=1}^n c(x_i)  \right) -  \frac{\log 2(n+1)^J}{\sqrt{n \sum_a t(a) \nu_a^t}} + 
    \frac{ r}{\sqrt{\sum_a t(a) \nu^t_a}} \right) + \frac{\kappa}{\sqrt{n}} \notag\\
    &\leq
        1-  \Phi\left(\frac{\sqrt{n} C'(\Gamma)}{\sqrt{\psi_\beta\left(c(x^n)\right)}} \left(\Gamma - \frac{1}{n} \sum_{i=1}^n c(x_i)  \right) -  \frac{\log 2(n+1)^J}{\sqrt{n (V(\Gamma)- 
 \delta_\beta)}} + 
    \frac{ r}{\sqrt{\varphi(r)}} \right) + \frac{\kappa}{\sqrt{n}}  \label{pn2}
\end{align}
for sufficiently large $n$. In inequality $(a)$, we subtract $nI(t,W)$ on both sides and use the concavity of the capacity-cost function as follows: 
$$C(c(x^n)) \leq C(\Gamma) + C'(\Gamma)\left( c(x^n) - \Gamma \right).$$
In inequality $(b)$, we apply the Berry-Esseen Theorem \cite{esseen11}, where $\kappa_t$ is a constant depending on the second- and third-order moments of the following set of zero-mean random variables:
\begin{align*}
    \left \{ \log \frac{W(Y_i|x_i)}{q_t(Y_i)} - \mathbb{E}_{W(\cdot|x_i)} \left [\log  \frac{W(Y|x_i)}{q_t(Y_i)} \right ] \right \}_{i=1}^n.
\end{align*}
In inequality $(c)$, it is easy to see from the construction of $\mathcal{P}_{n,2}$ and the choice of $\beta$ that for every $x^n \in \mathcal{P}_{n,2}$, $\kappa_{t}$ can be upper bounded by a constant $\kappa$ depending only on $P^*$ and $\beta$ and not depending on $t = P_{x^n}$. 

For $x^n \in \mathcal{P}_{n,3}$ and denoting $t = t(x^n)$, it follows from Lemma \ref{lemcont} that there exists a $K > 0$ such that 
$$\sup_{\substack{t:\\
||t-P^*||_1 \geq \beta\\
c(t) \leq \Gamma + K}} I(t,W) < C(\Gamma) - K.$$
This implies that for sufficiently large $n$, 
$$\sup_{t(x^n): x^n \in \mathcal{P}_{n,3}  } I(t,W) < C(\Gamma) - K.$$
Hence, we have 
\begin{align}
    &W \left( \log \frac{W(Y^n|x^n)}{q(Y^n)} > n C(\Gamma) + \sqrt{n} r \right) \notag\\
    &\leq W \left( \sum_{i=1}^n \log \frac{W(Y_i|x_i)}{q_t(Y_i)} > n C(\Gamma) + \sqrt{n} r  - \log 2(n+1)^J \right)\notag\\
    &= W \left( \sum_{i=1}^n \left [  \log \frac{W(Y_{i}|x_i)}{q_t(Y_{i})} - \mathbb{E}_W \left [\log \frac{W(Y|x_i)}{q_t(Y)}  \right ] \right ]  >  \right . \notag\\
    &  \quad \quad \quad \quad \quad \quad \quad   n \left [ C(\Gamma) - I(t,W) \right]  + \sqrt{n} r - \log 2(n+1)^J  \Bigg )\notag\\
    &\leq W \left( \sum_{i=1}^n  \left [  \log \frac{W(Y_{i}|x_i)}{q_t(Y_{i})} - \mathbb{E}_W \left [\log \frac{W(Y|x_i)}{q_t(Y)}  \right ] \right ]  > n\frac{K}{2} \right) \label{maroise}
\end{align}
for sufficiently large $n$.
Let $i_{\max, t} := \max_{a, b: q_t(b)W(b|a) > 0} \big | \log \frac{ W(b|a)}{q_t(b)} \big |$. We now show that $i_{\max, t} \leq 2 \log n$ for all $t$. Let $W_{\min}:= \min_{a, b: W(b|a) > 0} W(b|a)$ and $q_{\min, t} := \min_{b:q_t(b) > 0} q_t(b)$. Then 
\begin{align*}
    q_{\min, t} &= \min_{b:q_t(b) > 0} \sum_{a \in \mathcal{A}} t(a) W(b|a)\\
    &\geq \min_{a, b: W(b|a) > 0} W(b|a) \min_{a:t(a) > 0} t(a)\\
    &= \frac{W_{\min}}{n}.
\end{align*}
Thus,
\begin{align*}
     i_{\max, t} &=  \max_{a, b: q_t(b)W(b|a) > 0} \big | \log \frac{ W(b|a)}{q_t(b)} \big |\\
    &\leq \max_{a, b: q_t(b)W(b|a) > 0} \big | \log W(b|a) \big | + \max_{b:q_t(b) > 0} \big | \log q_t(b) \big |\\
    &\leq \log \frac{n}{W_{\min}^2}\\
    &\leq 2 \log n
\end{align*}
for all sufficiently large $n$. 
Hence, we can use Azuma's inequality \cite[(33), p. 61]{azumaguy} to upper bound $(\ref{maroise})$, giving us  
\begin{align}
    &W \left( \log \frac{W(Y^n|x^n)}{q(Y^n)} > n C(\Gamma) + \sqrt{n} r \right) \notag\\
    &\leq \exp \left( -\frac{n K^2}{128  \log^2 n } \right). \label{pn4}
\end{align}

For $\mathcal{P}_{n, 4}$, the following lemma shows that the probability of this set under $\overline{P}$ is small.  
\begin{lemma}
    We have 
    \begin{align}
\overline{P}\left(\mathcal{P}_{n,4} \right) \leq \frac{V}{\ln n}. \label{pn5}    
    \end{align}
    \end{lemma}
\begin{IEEEproof}
Let $\mu_i = \mathbb{E}[c(X_i)]$. For any $\vartheta > 0$, 
\begin{align*}
    &\overline{P}\left( c(X^n) > \Gamma + \vartheta \right)\\
    &= \overline{P}\left( \sum_{i=1}^n \left (c(X_i) - \mu_i\right ) + \sum_{i=1}^n \mu_i \geq n\Gamma + n\vartheta \right)\\
    &\leq  \overline{P}\left( \sum_{i=1}^n \left (c(X_i) - \mu_i\right ) \geq  n\vartheta \right)\\
    &\leq \overline{P}\left( \left( \sum_{i=1}^n \left (c(X_i) - \mu_i\right )\right)^2 \geq  n^2\vartheta^2 \right)\\
    &\leq \frac{1}{n^2 \vartheta^2} \mathbb{E} \left [ \left( \sum_{i=1}^n \left (c(X_i) - \mu_i\right )\right)^2 \right ]\\
    &\leq \frac{V}{n \vartheta^2}.  
\end{align*}
Letting $\vartheta = \sqrt{\ln n / n}$ establishes the result. 
\end{IEEEproof}

Using the results in $(\ref{pn1})$, $(\ref{pn2})$, $(\ref{pn4})$ and $(\ref{pn5})$, we can upper bound $(\ref{fopv})$ as 
\begin{align}
    &(\overline{P} \circ W) \left( \log \frac{W(Y^n|X^n)}{q(Y^n)} > nC(\Gamma) + \sqrt{n} r \right) \label{y30x}\\
    &\leq \sum_{x^n \in \mathcal{P}_{n,1}} \overline{P}(x^n) \frac{4  \nu_{\max}}{n (C'(\Gamma)^2 \beta^2)} + \sum_{x^n \in \mathcal{P}_{n,3}} \overline{P}(x^n) \exp \left(- \frac{n K^2}{128 \log n} \right) + \frac{V}{\ln n} +  \mbox{}\notag\\
    & \sum_{x^n \in \mathcal{P}_{n,2}} \overline{P}(x^n)\left [1-  \Phi\left(\frac{\sqrt{n} C'(\Gamma)}{\sqrt{\psi_\beta(c(x^n))}} \left(\Gamma - \frac{1}{n} \sum_{i=1}^n c(x_i)  \right) -  \frac{\log 2(n+1)^J}{\sqrt{n (V(\Gamma)- 
 \delta_\beta)}} + 
    \frac{ r}{\sqrt{\varphi(r)}} \right) + \frac{\kappa}{\sqrt{n}} \right ] \notag\\
    &\leq  \frac{4  \nu_{\max}}{n (C'(\Gamma)^2 \beta^2)} +  \exp \left(- \frac{n K^2}{128 \log n} \right) + \frac{V}{\ln n} +  \mbox{}\notag\\
    & \mathbb{E}\left [1-  \Phi\left(\frac{\sqrt{n} C'(\Gamma)}{\sqrt{\psi_\beta(c(X^n))}} \left(\Gamma - \frac{1}{n} \sum_{i=1}^n c(X_i)  \right) -  \frac{\log 2(n+1)^J}{\sqrt{n (V(\Gamma)- 
 \delta_\beta)}} + 
    \frac{ r}{\sqrt{\varphi(r)}} \right) + \frac{\kappa}{\sqrt{n}} \right ] \label{iskocontkar}
\end{align}
To further upper bound the above expression, we need to obtain a lower bound to 
\begin{align*}
    \inf_{X^n} \mathbb{E} \left [\Phi\left( \frac{\sqrt{n} C'(\Gamma)}{\sqrt{\psi_\beta(c(X^n))}} \left(\Gamma - \frac{1}{n} \sum_{i=1}^n c(X_i)  \right) -  \frac{\log 2(n+1)^J}{\sqrt{n (V(\Gamma)- 
 \delta_\beta)}} + 
    \frac{ r}{\sqrt{\varphi(r)}} \right) \right ],
\end{align*}
where the infimum is over all random variables $X^n$ such that $\mathbb{E}\left[ \sum_{k=1}^n c(X_k) \right] \leq n\Gamma$ and $\text{Var}\left(\sum_{i=1}^n c(X_i) \right) \leq n V$. Without loss of generality, we can assume $\mathbb{E}\left[ \sum_{k=1}^n c(X_k) \right] = n\Gamma$ since $\Phi$ is monotonically increasing and the function 
$$c \mapsto \frac{\Gamma - c}{\sqrt{\psi_\beta(c)}}$$
is monotonically nonincreasing. 

Note that  
\begin{align}
    &\Bigg | \mathbb{E}\left [ \frac{\sqrt{n} C'(\Gamma)}{\sqrt{\psi_{\beta}(c(X^n))}} \left(\Gamma - c(X^n) \right) \right] - \mathbb{E}\left [ \frac{\sqrt{n} C'(\Gamma)}{\sqrt{V(\Gamma)}} \left(\Gamma - c(X^n) \right) \right] \Bigg | \notag \\
    &= \Bigg | \mathbb{E}\left [ \sqrt{n} C'(\Gamma)  \left(\Gamma - c(X^n) \right)\left [ \frac{1}{\sqrt{\psi_{\beta}(c(X^n))}} - \frac{1}{\sqrt{V(\Gamma)}} \right ] \right] \Bigg | \notag \\
    &\leq  \sqrt{n} C'(\Gamma) \mathbb{E} \left [ \big | \Gamma - c(X^n) \big | \cdot \Bigg |  \frac{1}{\sqrt{\psi_{\beta}(c(X^n))}} - \frac{1}{\sqrt{V(\Gamma)}} \Bigg | \right ] \notag \\
    &\stackrel{(a)}{\leq} \sqrt{n} C'(\Gamma) \frac{ \Delta}{C'(\Gamma) \sqrt{V}} \mathbb{E}\left [ \big | \Gamma - c(X^n) \big | \right] \notag \\
    &\stackrel{(b)}{\leq} \Delta. \label{bzuyv} 
\end{align}
In inequality $(a)$, we used $(\ref{aaronerr1})$. In inequality $(b)$, we used the fact that $\mathbb{E}\left [ \sqrt{ \left ( \Gamma - c(X^n) \right)^2 }\right] \leq \sqrt{\text{Var}(c(X^n))}$. Hence, from $(\ref{bzuyv})$, we have 
\begin{align}
    &\mathbb{E}\left [ \frac{\sqrt{n} C'(\Gamma)}{\sqrt{\psi_{\beta}(c(X^n))}} \left(\Gamma - c(X^n) \right) \right] \notag \\
    &\geq \mathbb{E}\left [ \frac{\sqrt{n} C'(\Gamma)}{\sqrt{V(\Gamma)}} \left(\Gamma - c(X^n) \right) \right] - \Delta \notag \\
    &= -\Delta. \label{Piprimemean}
\end{align}

Now define $X' = \Gamma - c(X^n)$ so that $\mathbb{E}[X'] = 0$ and $\text{Var}(X') \leq V/n$. Also define 
\begin{align*}
    \Psi_\beta( X' ) := \begin{cases}
V(\Gamma) + \delta_\beta & \text{ if } X' \geq 0\\
V(\Gamma) - \delta_\beta & \text{ if } X' < 0.
\end{cases}    
\end{align*}
Then note that  
\begin{align}
    & \Bigg |\text{Var}\left( \frac{\Gamma - c(X^n)}{\sqrt{\psi_\beta(c(X^n))}} \right) -
     \text{Var}\left( \frac{\Gamma - c(X^n)}{\sqrt{V(\Gamma)}} \right) \Bigg | \notag  \\
     &= \Bigg | \text{Var}\left( \frac{X'}{\sqrt{\Psi_\beta(X')}} \right) - \text{Var}\left( \frac{X'}{\sqrt{V(\Gamma)}} \right) \Bigg | \notag \\
     &= \Bigg | \mathbb{E}\left [ \frac{X'^2}{\Psi_\beta(X')}  - \frac{X'^2}{V(\Gamma)} \right ] +  \left(\mathbb{E}\left [  \frac{X'}{\sqrt{V(\Gamma)}} \right] \right)^2 - \left(\mathbb{E}\left [  \frac{X'}{\sqrt{\Psi_\beta(X')}} \right] \right)^2 \Bigg |\notag  \\
     &=  \Bigg | \mathbb{E}\left [ \left( \frac{X'}{\sqrt{\Psi_\beta(X')}} + \frac{X'}{\sqrt{V(\Gamma)}}\right) \left( \frac{X'}{\sqrt{\Psi_\beta(X')}}  -\frac{X'}{\sqrt{V(\Gamma)}}\right)   \right ] +  \mbox{} \notag \\
     & \quad \quad \quad \quad \quad \left(\mathbb{E}\left [  \frac{X'}{\sqrt{V(\Gamma)}} \right] + \mathbb{E}\left [  \frac{X'}{\sqrt{\Psi_\beta(X')}} \right] \right) \left (\mathbb{E}\left [  \frac{X'}{\sqrt{V(\Gamma)}} \right] - \mathbb{E}\left [  \frac{X'}{\sqrt{\Psi_\beta(X')}} \right] \right) \Bigg | \notag \\
     &\leq \Bigg | \mathbb{E}\left [ \left( \frac{X'}{\sqrt{\Psi_\beta(X')}} + \frac{X'}{\sqrt{V(\Gamma)}}\right) \left( \frac{X'}{\sqrt{\Psi_\beta(X')}}  -\frac{X'}{\sqrt{V(\Gamma)}}\right)   \right ] \Bigg | +  \mbox{} \notag  \\
     & \quad \quad \quad \quad \quad \Bigg |\left(\mathbb{E}\left [  \frac{X'}{\sqrt{V(\Gamma)}} \right] + \mathbb{E}\left [  \frac{X'}{\sqrt{\Psi_\beta(X')}} \right] \right) \left (\mathbb{E}\left [  \frac{X'}{\sqrt{V(\Gamma)}} \right] - \mathbb{E}\left [  \frac{X'}{\sqrt{\Psi_\beta(X')}} \right] \right) \Bigg | \notag \\
     &\stackrel{(a)}{\leq} \mathbb{E}[X'^2] \frac{ \Delta/2}{C'(\Gamma)^2 V} + \Bigg |\left(\mathbb{E}\left [  \frac{X'}{\sqrt{V(\Gamma)}} \right] + \mathbb{E}\left [  \frac{X'}{\sqrt{\Psi_\beta(X')}} \right] \right) \left (\mathbb{E}\left [  \frac{X'}{\sqrt{V(\Gamma)}} \right] - \mathbb{E}\left [  \frac{X'}{\sqrt{\Psi_\beta(X')}} \right] \right) \Bigg | \notag \\
     &\leq \mathbb{E}[X'^2] \frac{ \Delta/2}{C'(\Gamma)^2 V} + \Bigg |\mathbb{E}\left [  \frac{X'}{\sqrt{V(\Gamma)}} +  \frac{X'}{\sqrt{\Psi_\beta(X')}} \right] \Bigg | \cdot \Bigg |\mathbb{E}\left [  \frac{X'}{\sqrt{V(\Gamma)}}  -  \frac{X'}{\sqrt{\Psi_\beta(X')}} \right] \Bigg | \notag \\
     &\leq \mathbb{E}[X'^2] \frac{ \Delta/2}{C'(\Gamma)^2 V} + \Bigg | \mathbb{E}\left [   \frac{X'}{\sqrt{\Psi_\beta(X')}}   \right] \Bigg | \cdot \mathbb{E}\left [ |X'| \cdot \Bigg | \frac{1}{\sqrt{V(\Gamma)}} -  \frac{1}{\sqrt{\Psi_\beta(X')}} \Bigg | \right] \notag  \\
     &\stackrel{(b)}{\leq} \mathbb{E}[X'^2] \frac{ \Delta/2}{C'(\Gamma)^2 V} + \frac{\Delta}{\sqrt{n} C'(\Gamma)}  \cdot  \frac{ \Delta}{C'(\Gamma) \sqrt{V}}  \mathbb{E}\left [ |X'|   \right]  \notag \\
     &\stackrel{(c)}{\leq} \frac{V}{n} \frac{ \Delta/2}{C'(\Gamma)^2 V} + \frac{\Delta}{\sqrt{n} C'(\Gamma)}  \cdot  \frac{ \Delta}{C'(\Gamma) \sqrt{V}}  \sqrt{\frac{V}{n}}. \label{finbhikarle}  
\end{align}
In inequality $(a)$, we used $(\ref{aaronerr2})$. In inequality $(b)$, we used $(\ref{aaronerr1})$ and $(\ref{bzuyv})$, noting that $\mathbb{E}[X'] = 0$. In inequality $(c)$, we used the inequality $\mathbb{E}[|X'|] \leq \sqrt{\mathbb{E}[X'^2]} = \sqrt{\text{Var}(X')} \leq \sqrt{V/n}$. Therefore, from $(\ref{finbhikarle})$, we have 
\begin{align}
    &\text{Var}\left( \frac{\sqrt{n} C'(\Gamma)}{\sqrt{\psi_\beta(c(X^n))}} \left(\Gamma - c(X^n)  \right)\right) \notag \\
    &\leq \text{Var}\left( \frac{\sqrt{n} C'(\Gamma)}{\sqrt{V(\Gamma)}} \left(\Gamma - c(X^n) \right)\right) + \Delta \notag  \\
    &\leq \frac{C'(\Gamma)^2 V}{V(\Gamma)}  + \Delta. \label{Piprimevar}
\end{align}

Define the random variable $\Pi'$
as 
\begin{align*}
    \Pi' :=  \frac{\sqrt{n} C'(\Gamma)}{\sqrt{\psi_\beta(c(X^n))}} \left(\Gamma - \frac{1}{n} \sum_{i=1}^n c(X_i)  \right)
\end{align*}
so that, from $(\ref{Piprimemean})$ and $(\ref{Piprimevar})$, $\mathbb{E}[\Pi'] \geq -\Delta$ and $\text{Var}(\Pi') \leq \frac{C'(\Gamma)^2 V}{V(\Gamma)}  + \Delta$. Then, from $(\ref{iskocontkar})$, for sufficiently large $n$,
\begin{align}
    &(\overline{P} \circ W) \left( \log \frac{W(Y^n|X^n)}{q(Y^n)} > nC(\Gamma) + \sqrt{n} r \right) \notag \\
    &\leq 1 +  \frac{3V}{2\ln n} - \mathbb{E}\left [  \Phi\left(\Pi' -  \frac{\log 2(n+1)^J}{\sqrt{n (V(\Gamma)- 
 \delta_\beta)}} + \frac{ r}{\sqrt{\varphi(r)}} 
     \right)   \right ] \notag \\
    &\leq 1 + \frac{2V}{\ln n} - \mathbb{E}\left [  \Phi\left(\Pi' + 
    \frac{ r}{\sqrt{\varphi(r)}} \right)   \right ] \notag \\
    &\stackrel{(a)}{=}  1 +  \frac{2V}{\ln n} - \inf_\Pi \mathbb{E}\left [  \Phi\left(\Pi \right)   \right ] \notag \\
    &\stackrel{(b)}{=} 1 +  \frac{2V}{\ln n} - \mathcal{K}\left( -\Delta +  \frac{ r}{\sqrt{\varphi(r)}}, \frac{C'(\Gamma)^2 V}{V(\Gamma)}  + \Delta \right).  \label{iskosocrmain}
\end{align}
The infimum in equality $(a)$ above is over all random variables $\Pi$ such that $\mathbb{E}[\Pi] \geq -\Delta +  \frac{ r}{\sqrt{\varphi(r)}}$ and $\text{Var}(\Pi) \leq \frac{C'(\Gamma)^2 V}{V(\Gamma)}  + \Delta.$ Equality $(b)$ follows by the definition and properties of the $\mathcal{K}$ function given in Definition \ref{Kdef} and Lemma \ref{Kfunctionproperties}.

Using $(\ref{iskosocrmain})$ to upper bound $(\ref{vq0})$ and using that upper bound in the result of Lemma \ref{mostgenconv} (with $\rho = \exp\left(n C(\Gamma) + \sqrt{n} r \right)$), we obtain
\begin{align}
    &\log M^*(n, \epsilon, \Gamma, V) \leq n C(\Gamma) + \sqrt{n}r - \log \left [ \left(  - \epsilon   -  \frac{2V}{\ln n} + \mathcal{K}\left( -\Delta +  \frac{ r}{\sqrt{\varphi(r)}}, \frac{C'(\Gamma)^2 V}{V(\Gamma)}  + \Delta \right)  \right)^+\right] \notag \\
    &\frac{\log M^*(n, \epsilon, \Gamma, V) - nC(\Gamma)}{\sqrt{n}} \leq r - \frac{1}{\sqrt{n}} \log \left [ \left(  - \epsilon   -  \frac{2V}{\ln n} + \mathcal{K}\left( -\Delta +  \frac{ r}{\sqrt{\varphi(r)}}, \frac{C'(\Gamma)^2 V}{V(\Gamma)}  + \Delta \right)  \right)^+\right]. \label{rchoosekar}
\end{align}
For any given average error probability $\epsilon \in (0, 1)$, we choose $r$ in $(\ref{rchoosekar})$ as 
\begin{align}
    r &= \begin{cases}
       r_1^* + \eta & \text{ if } \epsilon \in \left(0, \epsilon^* \right ] \\
       r_2^* + \eta & \text{ if } \epsilon \in \left(\epsilon^* , 1 \right ),
    \end{cases}
    \label{rfuncepsilon}
\end{align}
where      
\begin{align}
    r^*_1 &:= \max \left \{r' : \mathcal{K}\left(-\Delta + \frac{r'}{\sqrt{V(\Gamma) - \delta_\beta }}, \frac{C'(\Gamma)^2 V}{V(\Gamma)}  + \Delta \right) \leq \epsilon \right \}, \label{r1star}\\ 
    r^*_2 &:= \max \left \{r' : \mathcal{K}\left(-\Delta + \frac{r'}{\sqrt{V(\Gamma) + \delta_\beta }}, \frac{C'(\Gamma)^2 V}{V(\Gamma)}  + \Delta \right) \leq \epsilon \right \}, \label{r2star} \\
    \epsilon^* &:= \mathcal{K}\left(-\Delta - \frac{\eta}{\sqrt{V(\Gamma) - \delta_\beta}}, \frac{C'(\Gamma)^2 V}{V(\Gamma)}  + \Delta \right). \notag  
\end{align}
Note that in $(\ref{rfuncepsilon})$, $r \leq 0$ for $\epsilon \leq \epsilon^*$ and $r > 0$ for $\epsilon > \epsilon^*$.

Therefore, for $\epsilon \leq \epsilon^*$, we have 
\begin{align}
    \frac{\log M^*(n, \epsilon, \Gamma, V) - nC(\Gamma)}{\sqrt{n}} \leq r_1^* + \eta - \frac{1}{\sqrt{n}} \log \left [ \left(  - \epsilon   -  \frac{2V}{\ln n} + \mathcal{K}\left( -\Delta +  \frac{ r_1^* + \eta}{\sqrt{V(\Gamma) - \delta_\beta}}, \frac{C'(\Gamma)^2 V}{V(\Gamma)}  + \Delta \right)  \right)^+\right].  \label{rchoosekarvv}
\end{align}
Since $\mathcal{K}(\cdot\,, \cdot)$ is strictly increasing in the first argument, we have 
\begin{align*}
    \mathcal{K}\left(-\Delta + \frac{r_1^* + \eta}{\sqrt{V(\Gamma) - \delta_\beta}}, \frac{C'(\Gamma)^2 V}{V(\Gamma)}  + \Delta \right) > \epsilon
\end{align*}
from the definition of $r_1^*$ in $(\ref{r1star})$ and the fact that $\eta > 0$. Hence, taking the limit supremum as $n \to \infty$ in $(\ref{rchoosekarvv})$, we obtain 
\begin{align}
    \limsup_{n \to \infty}\,\frac{\log M^*(n, \epsilon, \Gamma, V) - nC(\Gamma)}{\sqrt{n}} \leq r_1^* + \eta \label{pq2}
\end{align}
for $\epsilon \leq \epsilon^*$. For $\epsilon > \epsilon^*$, a similar derivation gives us 
\begin{align}
    \limsup_{n \to \infty}\,\frac{\log M^*(n, \epsilon, \Gamma, V) - nC(\Gamma)}{\sqrt{n}} \leq r_2^* + \eta. \label{pq3}
\end{align}
We now let $\Delta, \eta, \beta$ and $\delta_\beta$ go to zero in both $(\ref{pq2})$ and $(\ref{pq3})$. Then using the fact from Lemma \ref{Kfunctionproperties} that $\mathcal{K}(\cdot, \cdot)$ is continuous, we obtain 
\begin{align*}
    \limsup_{n \to \infty}\,\frac{\log M^*(n, \epsilon, \Gamma, V) - nC(\Gamma)}{\sqrt{n}} \leq \max \left \{r : \mathcal{K}\left(\frac{r}{\sqrt{V(\Gamma) }}, \frac{C'(\Gamma)^2 V}{V(\Gamma)}   \right) \leq \epsilon \right \}
\end{align*}
for all $\epsilon \in (0, 1)$.

The converse result can also be stated in terms of a lower bound on the minimum average probability of error of $(n, R, \Gamma, V)$ codes for a rate $R = C(\Gamma) + \frac{r}{\sqrt{n}}$.
From Lemma \ref{mostgenconv}, we have that for $(n, R, \Gamma, V)$ codes with minimum average error probability $\epsilon$, 
\begin{align}
    \log \lceil \exp(nR) \rceil &\leq \log \rho - \log \left [ \left( 1 - \epsilon - \sup_{\overline{P}}\, \inf_{q \in \mathcal{P}(\mathcal{B}^n)} (\overline{P} \circ W) \left( \frac{W(Y^n|X^n)}{q(Y^n)} > \rho \right)\right)^+\right], \label{qandp22}
\end{align}
where the supremum is over $\overline{P}$ such that $X^n \sim \overline{P}$ satisfies $\mathbb{E}\left [ \sum_{i=1}^n c(X_i) \right ] \leq n \Gamma$ and $\text{Var}\left(\sum_{i=1}^n c(X_i) \right) \leq n V$. In the first case, assume $r \leq 0$ and let $\rho = \exp\left(n C(\Gamma) + r' \sqrt{n} \right)$ for some arbitrary $r' < r$. It directly follows from  
$(\ref{iskosocrmain})$ and the definition of $\varphi(\cdot)$ in $(\ref{varphidefinitionr})$ that  
\begin{align}
    &\sup_{\overline{P}}\, \inf_{q \in \mathcal{P}(\mathcal{B}^n)} (\overline{P} \circ W) \left( \frac{W(Y^n|X^n)}{q(Y^n)} > \rho \right) \notag \\
    &\leq 1 +  \frac{2V}{\ln n} - \mathcal{K}\left(-\Delta +  \frac{r'}{\sqrt{V(\Gamma) - \delta_\beta}}, \frac{C'(\Gamma)^2 V}{V(\Gamma)}  + \Delta  \right). \label{nogames6}
\end{align}
From $(\ref{qandp22})$ and $(\ref{nogames6})$, we have 
\begin{align}
    \log \lceil \exp(nR) \rceil &\leq \log \rho - \log \left [ \left( - \epsilon -  \frac{2V}{\ln n} + \mathcal{K}\left(-\Delta +  \frac{r'}{\sqrt{V(\Gamma) - \delta_\beta}}, \frac{C'(\Gamma)^2 V}{V(\Gamma)}  + \Delta  \right)\right)^+\right] \label{z1s}
\end{align}
which evaluates to
\begin{align*}
     - \epsilon -  \frac{2V}{\ln n} + \mathcal{K}\left(-\Delta +  \frac{r'}{\sqrt{V(\Gamma) - \delta_\beta}}, \frac{C'(\Gamma)^2 V}{V(\Gamma)}  + \Delta  \right) \leq e^{-(r-r')\sqrt{n}}.
\end{align*}
Taking the limit as $n \to \infty$ and letting $r' \to r$, $\Delta \to 0$, $\beta \to 0$ and $\delta_\beta \to 0$, we obtain 
\begin{align}
    \epsilon \geq \mathcal{K}\left( \frac{r}{\sqrt{V(\Gamma) }}, \frac{C'(\Gamma)^2 V}{V(\Gamma)}   \right) \label{z1s2}
\end{align}
for any $r \leq 0$. For $r > 0$, let $\rho = \exp\left(n C(\Gamma) + r' \sqrt{n} \right)$ for some arbitrary $0 < r' < r$. Then from  
$(\ref{iskosocrmain})$ and the definition of $\varphi(\cdot)$ in $(\ref{varphidefinitionr})$, we have that  
\begin{align}
    &\sup_{\overline{P}}\, \inf_{q \in \mathcal{P}(\mathcal{B}^n)} (\overline{P} \circ W) \left( \frac{W(Y^n|X^n)}{q(Y^n)} > \rho \right) \notag \\
    &\leq 1 +  \frac{2V}{\ln n} - \mathcal{K}\left(-\Delta +  \frac{r'}{\sqrt{V(\Gamma) + \delta_\beta}}, \frac{C'(\Gamma)^2 V}{V(\Gamma)}  + \Delta  \right). \label{nogddames6}
\end{align} 
Then a similar derivation to that used from $(\ref{z1s})$ to $(\ref{z1s2})$ gives us 
\begin{align}
    \epsilon \geq \mathcal{K}\left( \frac{r}{\sqrt{V(\Gamma) }}, \frac{C'(\Gamma)^2 V}{V(\Gamma)}   \right) \notag 
\end{align}
for all $r > 0$.

\section{Proof of non-feedback Achievability \label{Achievability_Theorem_proof}}

Let $r < r^*$ be any real number where 
\begin{align}
    r^* = \max \left \{r : \mathcal{K}\left(\frac{r}{\sqrt{V(\Gamma) }}, \frac{C'(\Gamma)^2 V}{V(\Gamma)} \right) \leq \epsilon \right \}. \label{rstar9}
\end{align}
In view of Lemma \ref{Kfunctionproperties}, consider any distribution $P_\Pi$ which achieves the minimum in 
$$\mathcal{K}\left(\frac{r}{\sqrt{V(\Gamma) }}, \frac{C'(\Gamma)^2 V}{V(\Gamma)} \right).$$
Let $\Pi \sim P_\Pi$, where we write 
\begin{align*}
    P_{\Pi}(\pi) = \begin{cases}
    p_1 & \pi = \pi_1\\
    p_2 & \pi = \pi_2\\
    p_3 & \pi = \pi_3.
    \end{cases}
\end{align*}
Recall that $\mathbb{E}[\Pi] \geq \frac{r}{\sqrt{V(\Gamma) }}$ and $\text{Var}(\Pi) \leq \frac{C'(\Gamma)^2 V}{V(\Gamma)}$. For each $j \in \{1,2,3 \}$, let $$\Gamma_j = \Gamma - \frac{\sqrt{V(\Gamma) }}{C'(\Gamma)\sqrt{n}}\pi_j + \frac{r}{C'(\Gamma)\sqrt{n}}.$$
We assume sufficiently large $n$ so that $\Gamma_j \in (\Gamma_0, \Gamma^*)$ for all $j \in \{1,2,3 \}$.

Let $P^*_j$ be a capacity-cost-achieving input distribution for $C(\Gamma_j)$, $Q_j^*$ be the corresponding optimal output distribution,
\begin{align}
    [P_j^*]_n = \arg\min_{\substack{t \in \mathcal{P}_n(\mathcal{A}):\\
    c(t) \leq \Gamma_j}} ||t - P_j^*||_1
\end{align}
and $Q^{cc}_j$ be the induced output distribution when the channel input $X^n \sim \text{Unif}(T^n_{\mathcal{A}}([P_j^*]_n))$. 

Let the random channel input $X^n$ be such that with probability $p_j$, $X^n \sim \text{Unif}(T^n_{\mathcal{A}}([P_j^*]_n))$. For brevity, we will write $T_j = T^n_{\mathcal{A}}([P_j^*]_n)$ in the rest of the proof. Denoting the distribution of $X^n$ by $P$, we have 
\begin{align*}
    P(x^n) = \sum_{j=1}^3 p_j \mathds{1} \left(x^n \in T_j \right) \frac{1}{|T_j|}.
\end{align*}
It is easy to see that for each time instant $k$ ($1 \leq k \leq n$), $\mathbb{E}[c(X_k)] \leq \Gamma$ and $\text{Var}\left(\sum_{i=1}^n c(X_i) \right) \leq n V + O(1)$. If necessary, we can meet the variance constraint exactly by quantizing to types which are at most $O(1/n)$ away from $[P_j^*]_n$ in $l^1$ distance. The resulting $O(1/n)$ deviations will be negligible in the proof to follow. Hence, we assume that the types $[P_j^*]_n$ are chosen such that $\text{Var}\left(\sum_{i=1}^n c(X_i) \right) \leq n V$ without explicitly characterizing the $O(1/n)$ terms. 

The distribution of $Y^n$ induced by $P \circ W$ is 
\begin{align*}
    PW(y^n) &= \sum_{x^n} P(x^n) W(y^n|x^n)\\
    &= \sum_{j=1}^3 p_j \sum_{x^n \in T_j} \frac{1}{|T_j|} \prod_{i=1}^n W(y_i|x_i)\\
    &= \sum_{j=1}^3 p_j Q^{cc}_j(y^n).
\end{align*}
We now define 
\begin{align*}
    \mathcal{E}_n := (P \circ W)\left(\frac{1}{n} \log \frac{W(Y^n|X^n)}{PW(Y^n)} \leq C(\Gamma) + \frac{r}{\sqrt{n}} \right)
\end{align*}
and show that $\limsup_{n \to \infty} \mathcal{E}_n < \epsilon$ which, by Lemma \ref{aaron'slemma}, would show that the random coding scheme achieves a second-order coding rate of
$r$. We first write 
\begin{align}
    &\mathcal{E}_n \notag \\
    &\triangleq (P \circ W)\left(\frac{1}{n} \log \frac{W(Y^n|X^n)}{PW(Y^n)} \leq C(\Gamma) + \frac{r}{\sqrt{n}} \right) \notag \\
    &= \sum_{j=1}^3 p_j \mathbb{P}_{X^n \sim \text{Unif}(T_j)} \left(\frac{1}{n} \log \frac{W(Y^n|X^n)}{PW(Y^n)} \leq C(\Gamma) + \frac{r}{\sqrt{n}} \right),   \label{epsnu} 
\end{align}
where in the last equality above, the distribution of $X^n$ is specified in the subscript and the distribution of $Y^n \sim W(\cdot | X^n)$. To proceed further, we upper bound 
\begin{align}
    &\mathbb{P}_{X^n \sim \text{Unif}(T_j)} \left(\frac{1}{n} \log \frac{W(Y^n|X^n)}{PW(Y^n)} \leq C(\Gamma) + \frac{r}{\sqrt{n}}   \right) \label{referbackn}\\
    &= \sum_{y^n \in \mathcal{B}^n} Q_j^{cc}(y^n) \mathbb{P}_{X^n \sim \text{Unif}(T_j)} \left(\frac{1}{n} \log \frac{W(y^n|X^n)}{PW(y^n)} \leq C(\Gamma) + \frac{r}{\sqrt{n}} \Big | Y^n = y^n \right) \notag \\
    &\leq \sum_{y^n \in \mathcal{B}^n } Q_j^{cc}(y^n) \mathbb{P}_{X^n \sim \text{Unif}(T_j)} \left(\frac{1}{n} \log \frac{W(y^n|X^n)}{Q_i^{cc}(y^n)} \leq C(\Gamma) + \frac{r}{\sqrt{n}} \Big |  Y^n = y^n \right) \notag 
\end{align}
where $i \in \{1,2,3 \}$ depends on $y^n$ and is such that $Q_i^{cc}(y^n)$ assigns the highest probability to $y^n$. Continuing,

\begin{align}
&\mathbb{P}_{X^n \sim \text{Unif}(T_j)} \left(\frac{1}{n} \log \frac{W(Y^n|X^n)}{PW(Y^n)} \leq C(\Gamma) + \frac{r}{\sqrt{n}}  \right) \notag \\
    &\leq \sum_{y^n \in \mathcal{B}^n} Q_j^{cc}(y^n) \mathbb{P}_{X^n \sim \text{Unif}(T_j)} \left( \log \frac{W(y^n|X^n)}{Q_i^{cc}(y^n)} \leq nC(\Gamma) + r \sqrt{n} \Big |  Y^n = y^n \right) \notag \\
    &= \sum_{y^n \in \mathcal{B}^n} Q_j^{cc}(y^n) \mathbb{P}_{X^n \sim \text{Unif}(T_j)} \left( \log \frac{W(y^n|X^n)}{Q_i^{*}(y^n)}  \leq nC(\Gamma) + r \sqrt{n} - \log \frac{Q_i^{*}(y^n)}{Q_i^{cc}(y^n)} \Big |  Y^n = y^n \right) \notag \\
    &\stackrel{(a)}{\leq} \sum_{y^n \in \mathcal{B}^n} Q_j^{cc}(y^n) \mathbb{P}_{X^n \sim \text{Unif}(T_j)} \left( \log \frac{W(y^n|X^n)}{Q_i^{*}(y^n)}  \leq nC(\Gamma) + r \sqrt{n} + \frac{s(P^*)}{2} \log n \Big |  Y^n = y^n \right) \notag \\
    &= \sum_{y^n \in \mathcal{B}^n} Q_j^{cc}(y^n) \mathbb{P}_{X^n \sim \text{Unif}(T_j)} \left( \log \frac{W(y^n|X^n)}{Q_j^{*}(y^n)} + \log \frac{Q_j^{*}(y^n)}{Q_i^{*}(y^n)} \leq nC(\Gamma) + r \sqrt{n} + \frac{s(P^*)}{2} \log n \Big | Y^n = y^n \right) \notag \\
    &\stackrel{(b)}{\leq} \sum_{\substack{y^n \in \mathcal{B}^n: \\
    \log \frac{Q_i^{*}(y^n)}{Q_j^{*}(y^n)} < c_n }} Q_j^{cc}(y^n) \mathbb{P}_{X^n \sim \text{Unif}(T_j)} \left( \log \frac{W(y^n|X^n)}{Q_j^{*}(y^n)} \leq nC(\Gamma) + r \sqrt{n} + \frac{s(P^*)}{2} \log n +  c_n  \Big | Y^n = y^n \right) + \notag \\
    & \quad \quad \quad \quad \quad \quad \quad \sum_{y^n \in \mathcal{B}^n} Q_j^{cc}(y^n) \mathds{1}\left(\log \frac{Q_i^{*}(y^n)}{Q_j^{*}(y^n)} \geq c_n  \right)\notag \\
    &\leq \mathbb{P}_{X^n \sim \text{Unif}(T_j)} \left( \log \frac{W(Y^n|X^n)}{Q_j^{*}(Y^n)} \leq nC(\Gamma) + r \sqrt{n} + \frac{s(P^*)}{2} \log n  + c_n  \right) + \notag  \\
    & \quad \quad \quad \quad \quad \quad \quad \sum_{y^n \in \mathcal{B}^n} Q_j^{cc}(y^n) \mathds{1}\left(\log \frac{Q_i^{*}(y^n)}{Q_j^{*}(y^n)} \geq c_n  \right) \notag  \\
    &= \mathbb{P}_{X^n \sim \text{Unif}(T_j)} \left( \log \frac{W(Y^n|X^n)}{Q_j^{*}(Y^n)} \leq nC(\Gamma) + r \sqrt{n} + \frac{s(P^*)}{2} \log n  + c_n   \right) + \notag  \\
    & \quad \quad \quad \quad \quad \quad \quad \sum_{y^n \in \mathcal{B}^n} \frac{Q_j^{cc}(y^n)}{Q_j^*(y^n)} Q_j^*(y^n) \mathds{1}\left(\log \frac{Q_i^{*}(y^n)}{Q_j^{*}(y^n)} \geq c_n  \right)  \notag \\
    &\stackrel{(c)}{\leq} \mathbb{P}_{X^n \sim \text{Unif}(T_j)} \left( \log \frac{W(Y^n|X^n)}{Q_j^{*}(Y^n)} \leq nC(\Gamma) + r \sqrt{n} + \frac{s(P^*)}{2} \log n  + c_n   \right) + \notag  \\
    & \quad \quad \quad \quad \quad \quad \quad \kappa \,n^{s(P^*)/2} \sum_{y^n \in \mathcal{B}^n}  Q_j^*(y^n) \mathds{1}\left(\log \frac{Q_i^{*}(y^n)}{Q_j^{*}(y^n)} \geq c_n  \right)  \notag\\
    &\leq \mathbb{P}_{X^n \sim \text{Unif}(T_j)} \left( \log \frac{W(Y^n|X^n)}{Q_j^{*}(Y^n)} \leq nC(\Gamma) + r \sqrt{n} + \frac{s(P^*)}{2} \log n  + c_n  \right) + \notag  \\
    & \quad \quad \quad \quad \quad \quad \quad \sum_{i'=1}^3 \kappa n^{s(P^*)/2}\,\mathbb{P}_{X^n \text{ i.i.d. } P_j^*}\left(\log \frac{Q_{i'}^{*}(Y^n)}{Q_j^{*}(Y^n)} \geq c_n  \right) \label{choicematters}\\
    &\stackrel{(d)}{\leq} \mathbb{P}_{X^n \sim \text{Unif}(T_j)} \left( \log \frac{W(Y^n|X^n)}{Q_j^{*}(Y^n)} \leq nC(\Gamma) + r \sqrt{n} + \frac{s(P^*)}{2} \log n  + \sqrt{\frac{n}{\log n}}  \right) + \delta_n. \label{subkaravg}
\end{align}
In inequality $(a)$, we use the third part of Lemma \ref{combinedlemmas} and assume sufficiently large $n$. In inequality $(b)$, $c_n > 0$ is an $o(\sqrt{n})$ sequence to be specified later. In inequality $(c)$, we again use the third part of Lemma $\ref{combinedlemmas}$, assume sufficiently large $n$ and choose an appropriate constant $\kappa > 0$ depending only on $P^*$ since $s(P_j^*) = s(P^*)$ for large enough $n$. In inequality $(d)$,
\begin{align*}
  \delta_n = 2 \kappa n^{s(P^*)/2} \exp \left(-\frac{2\gamma^2}{9} \sqrt{\frac{n}{\log n}} \right),  
\end{align*}
where $\gamma = 1/2\min\{Q^*(y) : y \in \mathcal{B}  \} > 0$, and $\delta_n \to 0$ as $n \to \infty$. To obtain inequality $(d)$, we first choose $c_n = \sqrt{n/\log n}$ in $(\ref{choicematters})$. The term in the sum in $(\ref{choicematters})$ with $i' = j$ is equal to zero. For $i' \neq j$, we have $Q_{i'}^*, Q_j^* \to Q^*$ as $n \to \infty$, so we can use Lemma \ref{proudlemma} to upper bound 
$$\mathbb{P}_{X^n \text{ i.i.d. } P_j^*}\left(\log \frac{Q_{i'}^{*}(Y^n)}{Q_j^{*}(Y^n)} \geq c_n  \right).$$
Specifically, in Lemma \ref{proudlemma}, choose $Q=Q_{i'}^*$, $Q' = Q_j^*$ and $c_n = \sqrt{n/\log n}$.

Substituting $(\ref{subkaravg})$ in $(\ref{epsnu})$, we have 
\begin{align}
    &\mathcal{E}_n \notag \\
    &\leq \sum_{j=1}^3 p_j \mathbb{P}_{X^n \sim \text{Unif}(T_j)} \left( \log \frac{W(Y^n|X^n)}{Q_j^{*}(Y^n)} \leq nC(\Gamma) + r \sqrt{n} + \frac{s(P^*)}{2} \log n  + \sqrt{\frac{n}{\log n}} \right) + \delta_n \notag \\
    &= \sum_{j=1}^3 p_j \frac{1}{|T_j|} \sum_{x^n \in T_j} \mathbb{P} \left( \log \frac{W(Y^n|x^n)}{Q_j^{*}(Y^n)} \leq nC(\Gamma) + r \sqrt{n} + \frac{s(P^*)}{2} \log n  + \sqrt{\frac{n}{\log n}} \Big |X^n= x^n \right) + \delta_n \notag \\
    &= \sum_{j=1}^3 p_j \frac{1}{|T_j|} \sum_{x^n \in T_j} \mathbb{P} \left( \sum_{m=1}^n \log \frac{W(Y_m|x_m)}{Q^*_j(Y_m)} \leq nC\left(\Gamma_j \right) +  nC(\Gamma) - \mbox{} \right . \notag  \\
    & \left . \quad \quad \quad \quad  nC\left(\Gamma_j \right) + r \sqrt{n} + \frac{s(P^*)}{2} \log n  + \sqrt{\frac{n}{\log n}} \Big | X^n = x^n \right) + \delta_n \notag  \\
    &\stackrel{(a)}{\leq} \sum_{j=1}^3 p_j \frac{1}{|T_j|} \sum_{x^n \in T_j} \mathbb{P} \left( \sum_{m=1}^n \left [ \log \frac{W(Y_m|x_m)}{Q^*_j(Y_m)} - \mathbb{E}\left [ \log \frac{W(Y_m|x_m)}{Q_j^*(Y_m)} \right ] \right ]  \leq   nC(\Gamma) - nC\left(\Gamma_j\right) + \mbox{} \right . \notag \\
    & \left . \quad \quad\quad\quad\quad \quad r \sqrt{n} + \frac{s(P^*)}{2} \log n  + \sqrt{\frac{n}{\log n}} + \lambda C'(\Gamma_j)  \Big | X^n = x^n \right) + \delta_n \notag \\
    &\stackrel{(b)}{=} \sum_{j=1}^3 p_j \frac{1}{|T_j|} \sum_{x^n \in T_j} \mathbb{P} \left(\frac{1}{\sqrt{n V_j}} \sum_{m=1}^n \left [ \log \frac{W(Y_m|x_m)}{Q^*_j(Y_m)} - \mathbb{E}\left [ \log \frac{W(Y_m|x_m)}{Q_j^*(Y_m)} \right ] \right ]  \leq   \frac{1}{\sqrt{n V_j}}\left ( nC(\Gamma) - nC\left( \Gamma_j \right)   \right)   + \mbox{} \right . \notag  \\
    & \left . \quad \quad\quad \quad \quad \frac{r}{\sqrt{V_j}}  +   \frac{s(P^*)}{2} \frac{\log n}{\sqrt{n V_j}}  + \sqrt{\frac{1}{V_j\,\log n}} + \frac{\lambda C'(\Gamma_j) }{\sqrt{n V_j}} \Big | X^n = x^n\right) + \delta_n \notag \\
    &\stackrel{(c)}{\leq} \sum_{j=1}^3 p_j \frac{1}{|T_j|} \sum_{x^n \in T_j} \left [ \Phi\left( \frac{1}{\sqrt{n V_j}}\left ( nC(\Gamma) - nC\left( \Gamma_j \right)   \right) + \frac{r}{\sqrt{V_j}}  +   \frac{s(P^*)}{2} \frac{\log n}{\sqrt{n V_j}}  + \mbox{} \right . \right. \notag\\
    & \left. \left . \quad \quad \quad \quad \quad \quad \quad \sqrt{\frac{1}{V_j\,\log n}} + \frac{\lambda C'(\Gamma_j) }{\sqrt{n V_j}} \right) + \frac{\beta_j}{\sqrt{n}} \right] + \delta_n \notag \\
    &\stackrel{(d)}{\leq} \sum_{j=1}^3 p_j   \Phi\left( \frac{1}{\sqrt{n V_j}}\left ( nC(\Gamma) - nC\left( \Gamma_j \right)   \right) + \frac{r}{\sqrt{V_j}}  +   \frac{s(P^*)}{2} \frac{\log n}{\sqrt{n V_j}}  + \mbox{} \right .  \notag\\
    &  \left . \quad \quad \quad \quad \quad \quad \quad \sqrt{\frac{1}{V_j\,\log n}} + \frac{\lambda C'(\Gamma_j) }{\sqrt{n V_j}} \right)  + \frac{\beta}{\sqrt{n}} +  \delta_n \notag \\
    &\leq \sum_{j=1}^3 p_j   \Phi\left( \frac{C'(\Gamma_j)}{C'(\Gamma)}\frac{\sqrt{V(\Gamma)}}{\sqrt{V_j}} \pi_j - \frac{r}{\sqrt{V_j}} \frac{C'(\Gamma_j)}{C'(\Gamma)} + \frac{r}{\sqrt{V_j}}  +   \frac{s(P^*)}{2} \frac{\log n}{\sqrt{n V_j}}  + \sqrt{\frac{1}{V_j\,\log n}} + \frac{\lambda C'(\Gamma_j) }{\sqrt{n V_j}} \right)  + \frac{\beta}{\sqrt{n}} +  \delta_n \notag \\
    &\stackrel{(e)}{\leq} \sum_{j=1}^3 p_j   \Phi\left(\pi_j \right)  + \frac{\beta}{\sqrt{n}} +  \delta_n + \delta_n' \notag \\
    &= \mathcal{K}\left (\frac{r}{\sqrt{V(\Gamma) }}, \frac{C'(\Gamma)^2 V}{V(\Gamma)}\right ) + \frac{\beta}{\sqrt{n}} +  \delta_n + \delta_n' \label{extendtoerror} \\
    &\stackrel{(f)}{<} \epsilon + \frac{\beta}{\sqrt{n}} +  \delta_n + \delta_n' \label{endingemma} 
\end{align}
In inequality $(a)$, we use Proposition \ref{getout} for $C(\Gamma_j)$, assuming sufficiently large $n$ so that $\text{supp}([P^*_j]_n) = \text{supp}(P^*_j)$ and use the lower bound $$\min_{j=1,2,3} \left [ c([P^*_j]_n) - \Gamma_j\right] \geq - \lambda/n$$ for some suitable constant $\lambda > 0$. This holds because $c(P_j^*) = \Gamma_j$ and $[P_j^*]_n$ and $P_j^*$ are at most $O(1/n)$ away from each other in $l^1$ distance. In equality $(b)$, we define 
\begin{align*}
    V_j &= \frac{1}{n}\text{Var}\left( \sum_{m=1}^n \log \frac{W(Y_m|x_m)}{Q_j^*(Y_m)} \right)\\
    &= \sum_{a \in \mathcal{A}} [P^*_j]_n(a) \text{Var} \left(\log \frac{W(Y_a|a)}{Q_j^*(Y_a)} \right),
\end{align*}
where $Y_a \sim W(\cdot|a)$. In inequality $(c)$, we apply the Berry-Esseen Theorem \cite{esseen11}, where $\beta_j$ is a constant depending on the second- and third-order moments of the following set of zero-mean random variables:
\begin{align*}
    \left \{ \log \frac{W(Y_m|x_m)}{Q^*_j(Y_m)} - \mathbb{E}_{W(\cdot|x_m)} \left [\log  \frac{W(Y|x_m)}{Q^*_j(Y_m)} \right ] \right \}_{m=1}^n,
\end{align*}
where the dependence on $x^n$ is only through its type $[P^*_j]_n$. Furthermore, since $Q_j^* \to Q^*$ and $[P_j^*]_n \to P^*$ as $n \to \infty$, in inequality $(d)$, we have $\max_j \beta_j \leq \beta$ for sufficiently large $n$ for some constant $\beta$ depending only on $P^*$. Inequality $(e)$ follows from a basic continuity argument since $V_j \to V(\Gamma)$ and $C'(\Gamma_j) \to C'(\Gamma)$ as $n \to \infty$, where $\delta_n'$ is a suitable sequence satisfying $\delta_n' \to 0$ as $n \to \infty$. Inequality $(f)$ follows because $r < r^*$, where $r^*$ is defined in $(\ref{rstar9})$, and because
$\mathcal{K}(\cdot,\cdot)$ is a strictly increasing function in the first argument from Lemma \ref{Kfunctionproperties}. Taking the limit as $n \to \infty$ in $(\ref{endingemma})$ establishes that $\limsup_{n \to \infty} \mathcal{E}_n < \epsilon$ for $r < r^*$, thus establishing that any $r < r^*$ is an achievable second-order coding rate. Finally, letting $r \to r^*$ establishes an achievable second-order coding rate of $r^*$, matching the converse in Theorem \ref{Converse_Theorem}. 

The achievability result can also be stated in terms of an upper bound on the minimum average probability of error of $(n, R, \Gamma, V)$ codes for a rate $R = C(\Gamma) + \frac{r}{\sqrt{n}}$. From Lemma \ref{aaron'slemma}, we have for $\theta = 1/n^{\vartheta}$ for $1 > \vartheta > 1/2$,
\begin{align}
    \bar{P}_{\text{e}}\left(n,C(\Gamma) + \frac{r}{\sqrt{n}},  \Gamma, V\right) \leq (P \circ W) \left(\frac{1}{n} \log \frac{W(Y^n|X^n)}{PW(Y^n)} \leq C(\Gamma) + \frac{r}{\sqrt{n}} + \frac{1}{n^\vartheta} \right) + e^{-n^{1-\vartheta}}.  \label{notplaying}
\end{align}
From $(\ref{extendtoerror})$, we have 
\begin{align}
    &(P \circ W)\left(\frac{1}{n} \log \frac{W(Y^n|X^n)}{PW(Y^n)} \leq C(\Gamma) + \frac{r}{\sqrt{n}} \right) \notag \\
    &\leq \mathcal{K}\left (\frac{r}{\sqrt{V(\Gamma) }}, \frac{C'(\Gamma)^2 V}{V(\Gamma)}\right ) + \frac{\beta}{\sqrt{n}} +  \delta_n + \delta_n'. \label{notplaying2}
\end{align}
From $(\ref{notplaying})$ and $(\ref{notplaying2})$, we obtain 
\begin{align*}
    \limsup_{n \to \infty} \bar{P}_{\text{e}}\left(n,C(\Gamma) + \frac{r}{\sqrt{n}},  \Gamma, V\right) &\leq \mathcal{K}\left (\frac{r}{\sqrt{V(\Gamma) }}, \frac{C'(\Gamma)^2 V}{V(\Gamma)}\right ). 
\end{align*}

\section{Proof of Feedback Achievability \label{Feedback_Achievability_Theorem_proof}}

We consider
a random feedback code in which feedback is only 
used once halfway through the transmission,
following~\cite[Theorem 1]{9099482}.
In principle, the
approach can be applied to codes that adjust 
their signaling scheme many times over the course of
transmission, as in~\cite[Theorem 2]{9099482}, at the expense of more technical 
arguments. Recall from Theorems \ref{Converse_Theorem} and \ref{Achievability_Theorem} that
$$r^* = \max \left \{r : \mathcal{K}\left(\frac{r}{\sqrt{V(\Gamma) }}, \frac{C'(\Gamma)^2 V}{V(\Gamma)} \right) \leq \epsilon \right \}$$
is equal to the optimal second-order coding rate of $(n, R, \Gamma,V)$ codes without feedback. From \cite[Theorem 1]{mahmood2024improvedchannelcodingperformance}, we have $r^* > \sqrt{V(\Gamma)} \Phi^{-1}(\epsilon)$.

We propose the following controller-based feedback code, where feedback is only 
used once halfway through the transmission. The controller $F$ divides the channel input $X^n$ into two $n/2$-length sequences, $X^{n/2}$ and $X^{n/2+1:n}$.

Let $r$ be a real number to be specified later. From Corollary \ref{corKfunc}, we have 
\begin{align}
    \mathcal{K}\left(\frac{r}{\sqrt{V(\Gamma) }}, \frac{C'(\Gamma)^2 V}{V(\Gamma)} \right) &= \min_{\substack{\Pi:\\
    \mathbb{E}[\Pi] = \frac{r}{\sqrt{V(\Gamma)}} \\
    \text{Var}(\Pi) \leq \frac{C'(\Gamma)^2 V}{V(\Gamma)}\\
    |\text{supp}(\Pi)| \leq 3
    }} \mathbb{E}\left [\Phi(\Pi) \right]. \label{FKr}
\end{align}
Let $\mathcal{F}_{\mathcal{K}(r)}$ denote the feasible set in $(\ref{FKr})$.
Consider any arbitrary distribution $P_{\Pi} \in \mathcal{F}_{\mathcal{K}(r)}$. We write $P_{\Pi}$ as 
\begin{align}
    P_{\Pi}(\pi) = \begin{cases}
    p_1 & \pi = \pi_1\\
    p_2 & \pi = \pi_2\\
    p_3 & \pi = \pi_3,
    \end{cases} \label{PPi}
\end{align}
and we let the random variable $\Pi \sim P_{\Pi}$. 

\textbf{Distribution of $X^{n/2}$}

For each $j \in \{1,2,3 \}$, let $$\Gamma_{j} = \Gamma - \frac{\sqrt{V(\Gamma) }}{C'(\Gamma)\sqrt{n}}\pi_j + \frac{r}{C'(\Gamma)\sqrt{n}}.$$
We assume sufficiently large $n$ so that $\Gamma_j \in (\Gamma_0, \Gamma^*)$ for all $j \in \{1,2,3 \}$.

Let $P^*_j$ be a capacity-cost-achieving input distribution for $C(\Gamma_j)$, $Q_j^*$ be the corresponding optimal output distribution,
\begin{align*}
    [P_j^*]_{n/2} = \arg\min_{\substack{t \in \mathcal{P}_{n/2}(\mathcal{A}):\\
    c(t) \leq \Gamma_j}} ||t - P_j^*||_1
\end{align*}
and $Q_j^{cc} \in \mathcal{P}(\mathcal{B}^{n/2})$ be the induced output distribution when the channel input $X^{n/2} \sim \text{Unif}(T^{n/2}_{\mathcal{A}}([P_j^*]_{n/2}))$.

Let the random channel input $X^{n/2}$ be such that with probability $p_j$, $X^{n/2} \sim \text{Unif}(T^{n/2}_{\mathcal{A}}([P_j^*]_{n/2}))$. For brevity, we will write $T = T_{\mathcal{A}}^{n/2}([P^*]_{n/2})$ and $T_j = T^{n/2}_{\mathcal{A}}([P_j^*]_{n/2})$ in the rest of the proof. Denoting the distribution of $X^{n/2}$ by $P$, we have 
\begin{align}
    P(x^{n/2}) = \sum_{j=1}^3 p_j \mathds{1} \left(x^{n/2} \in T_j \right) \frac{1}{|T_j|}. \label{XPdistn}
\end{align}
It is easy to see that for each time instant $k$ ($1 \leq k \leq n/2$), $\Gamma - O(1/n) \leq \mathbb{E}[c(X_k)] \leq \Gamma$. For the variance constraint, we have 
\begin{align}
    &\text{Var}\left( \sum_{i=1}^{n/2} c(X_i)\right) \notag \\
    &= \mathbb{E} \left [ \left(\sum_{i=1}^{n/2} c(X_i) - \mathbb{E}\left [ \sum_{i=1}^{n/2} c(X_i) \right ] \right)^2 \right ] \notag \\
    &= \sum_{j=1}^3 p_j \left(\frac{n}{2}\Gamma_j - \mathbb{E}\left [ \sum_{i=1}^{n/2} c(X_i) \right ] \right)^2  \notag \\
    &\leq \sum_{j:\Gamma_j \leq \Gamma} p_j \left(\frac{n}{2}\Gamma_j - \frac{n}{2}\Gamma \right)^2 + \sum_{j:\Gamma_j > \Gamma} p_j \left(\frac{n}{2}\Gamma_j - \frac{n}{2} \Gamma + O\left(1 \right) \right)^2  \notag \\
    &\leq \sum_{j=1}^3 p_j \left( \frac{n}{2}\Gamma_j - \frac{n}{2}\Gamma \right)^2 + O(\sqrt{n}) \notag \\
    &= \frac{n^2}{4} \sum_{j=1}^3 p_j \left(\Gamma_j - \Gamma  \right)^2 + O(\sqrt{n}) \notag  \\
    &\leq \frac{n}{4}V + O(\sqrt{n}). \label{varparta}
\end{align}

\textbf{Distribution of $X^{n/2+1:n}$}

Based on feedback, the encoder selects the distribution of $X^{n/2+1:n}$ as follows. Let $\beta_1, \beta_2$ and $\beta_3$ be positive constants to be specified later. Then given that $X^{n/2} \in T_j$ for $j \in \{1,2,3 \}$, if the event $\mathcal{V}_j$ given by  
\begin{align}
     \frac{1}{n/2} \log \frac{W(Y^{n/2}|X^{n/2})}{Q_j^*(Y^{n/2})} > C(\Gamma_j) + \frac{\beta_j}{\sqrt{n/2}}, \label{ovzbf1}
\end{align}
happens, then choose the distribution of $X^{n/2+1:n} \sim \text{Unif}(T^{n/2}_\mathcal{A}([P^*]_{n/2}))$. If the event $\mathcal{V}_j$ given $X^{n/2} \in T_j$ does not happen, then choose the distribution of $X^{n/2+1:n} \sim \text{Unif}(T_j)$.

We write the probability of the event $\mathcal{V}_j$ given $X^{n/2} \in T_j$ as
\begin{align*}
    \mathcal{U}_j &:= \mathbb{P}_{X^{n/2} \sim \text{Unif}(T_j)}\left( \frac{1}{n/2} \sum_{m=1}^{n/2} \log \frac{W(Y_{m}|X_{m})}{Q_j^*(Y_{m})} > C(\Gamma_j) + \frac{\beta_j}{\sqrt{n/2}}   \right)\\
    &= 1 - \mathbb{P}_{X^{n/2} \sim \text{Unif}(T_j)}\left( \frac{1}{n/2} \sum_{m=1}^{n/2} \log \frac{W(Y_{m}|X_{m})}{Q_j^*(Y_{m})} \leq C(\Gamma_j) + \frac{\beta_j}{\sqrt{n/2}}  \right)\\
    &= 1 - \underline{G}_{n/2}^j\left(\frac{\beta_j}{\sqrt{ V_j}} + \sqrt{\frac{n}{2 V_j}}\left(C(\Gamma_j) - [C(\Gamma_j)]_{n/2} \right) \right),
\end{align*}
where $V_j = [V(\Gamma_j)]_{n/2}$ and $\underline{G}_{n/2}^j$ is the CDF of the random variable 
defined in $(\ref{Gn2junder1})$.

\textbf{Meeting the Cost Constraint}

For every $k = n/2 + 1, \ldots, n$, we have 
\begin{align}
    &\mathbb{E}[c(X_{k})] \notag \\
    &\leq \sum_{j=1}^3 p_j \left [ \mathcal{U}_j \Gamma + (1  - \mathcal{U}_j) \Gamma_j \right ]\notag \\
    &= \Gamma + \sum_{j=1}^3 p_j \mathcal{U}_j(\Gamma - \Gamma_j)\notag \\
    &= \Gamma + \sum_{j: \Gamma_j \leq \Gamma} p_j \mathcal{U}_j(\Gamma - \Gamma_j)  + \sum_{j: \Gamma_j > \Gamma} p_j \mathcal{U}_j(\Gamma - \Gamma_j). \label{mzgf1}
\end{align}
From $(\ref{lemma7kaalternative})$, we have 
\begin{align}
    &\mathcal{U}_j \in  \notag \\
    &\left [ 1 - \Phi\left(\frac{\beta_j}{\sqrt{ V_j}} + \sqrt{\frac{n}{2 V_j}}\left(C(\Gamma_j) - [C(\Gamma_j)]_{n/2} \right)  \right) - \frac{\underline{\kappa}_{n/2}^j}{\sqrt{n/2}}, 1 - \Phi\left(\frac{\beta_j}{\sqrt{ V_j}} + \sqrt{\frac{n}{2 V_j}}\left(C(\Gamma_j) - [C(\Gamma_j)]_{n/2} \right) \right) + \frac{\underline{\kappa}_{n/2}^j}{\sqrt{n/2}}  \right ]. \label{3386}
\end{align} 
From the first part of Lemma \ref{combinedlemmas}, we have $C(\Gamma_j) - \frac{\psi}{n} \leq  [C(\Gamma_j)]_{n/2} \leq C(\Gamma_j)$ for some constant $\psi$ which can be chosen independent of $j$ since $\Gamma_j \to \Gamma$ for all $j \in \{ 1,2,3\}$. By a similar argument, since $\underline{\kappa}_{n/2}^j$, defined in $(\ref{kappanjdef})$, is continuous as a function of $\Gamma_j$, $\underline{\kappa}_{n/2}^j \in [\underline{\kappa}_{n/2}^{-}, \underline{\kappa}_{n/2}^{+}]$, where $\underline{\kappa}_{n/2}^{-}$ and $ \underline{\kappa}_{n/2}^{+}$ are independent  of $j$ and $|\underline{\kappa}_{n/2}^{-} - \underline{\kappa}_{n/2}^{+}| \to 0$ as $n \to \infty$. Hence, we can write $(\ref{3386})$ as 
\begin{align}
    &\mathcal{U}_j \in  \notag \\
    &\left [ 1 - \Phi\left(\frac{\beta_j}{\sqrt{ V(\Gamma))}}  \right) - \mu_n,  1 - \Phi\left(\frac{\beta_j}{\sqrt{ V(\Gamma)}}  \right) + \mu_n  \right ], \label{3e386}
\end{align} 
for some sequence $\mu_n$ independent of $j$ such that $\mu_n \to 0$ as $n \to \infty$. Using $(\ref{3e386})$, we can upper bound $(\ref{mzgf1})$ as 
\begin{align}
    &\mathbb{E}\left [c(X_{k}) \right] \notag \\
    &\leq  \Gamma + \sum_{j: \Gamma_j \leq \Gamma} p_j \left(   1 - \Phi\left(\frac{\beta_j}{\sqrt{ V(\Gamma)}}  \right) + \mu_n \right)(\Gamma - \Gamma_j)  + \sum_{j: \Gamma_j > \Gamma} p_j \left(   1 - \Phi\left(\frac{\beta_j}{\sqrt{ V(\Gamma)}}  \right) - \mu_n\right)(\Gamma - \Gamma_j) \notag \\
    &= \Gamma + \sum_{j: \Gamma_j \leq \Gamma} p_j \left(   1 - \Phi\left(\frac{\beta + \delta_n^{(1)}}{\sqrt{ V(\Gamma)}}  \right) + \mu_n \right)(\Gamma - \Gamma_j)  + \sum_{j: \Gamma_j > \Gamma} p_j \left(   1 - \Phi\left(\frac{\beta}{\sqrt{ V(\Gamma)}}  \right) - \mu_n\right)(\Gamma - \Gamma_j) \notag \\
    &= \Gamma. \label{annoyingthe}
\end{align}
The last two equalities above can be obtained by choosing $\beta_1, \beta_2$ and $\beta_3$ as follows. For all $j \in \{1,2,3 \}$ such that $\Gamma_j > \Gamma$, we choose $\beta_j = \beta$, where $\beta \geq 0$ is a constant to be specified later. For all $j \in \{1,2,3 \}$ such that $\Gamma_j \leq \Gamma$, we choose $\beta_j = \beta + \delta_n^{(1)}$, where $\delta_n^{(1)} \geq 0$ is a sequence that satisfies 
\begin{align*}
    1 - \Phi\left( \frac{\beta + \delta_n^{(1)}}{\sqrt{V(\Gamma)}} \right) + \mu_n = 1 - \Phi\left( \frac{\beta }{\sqrt{V(\Gamma)}} \right) - \mu_n.  
\end{align*}
Clearly, $\delta_n^{(1)} \to 0$ as $n \to \infty$. With these choices of $\beta_1, \beta_2$ and $\beta_3$, we have $\mathbb{E}\left [ c(X_k) \right] \leq \Gamma$. Hence, the mean constraint is satisfied. Furthermore, we have for each time instant $k = n/2 + 1,\ldots, n$,  
\begin{align}
    &\mathbb{E}\left [ c(X_k) \right] \notag \\
    &\geq \sum_{j=1}^3 p_j \left [ \mathcal{U}_j \Gamma + (1  - \mathcal{U}_j) \Gamma_j \right ] - O\left(\frac{1}{n} \right) \notag \\
    &= \Gamma + \sum_{j: \Gamma_j \leq \Gamma} p_j \mathcal{U}_j(\Gamma - \Gamma_j)  + \sum_{j: \Gamma_j > \Gamma} p_j \mathcal{U}_j(\Gamma - \Gamma_j) - O\left(\frac{1}{n} \right) \notag \\
    &\stackrel{(a)}{\geq} \Gamma + \sum_{j: \Gamma_j \leq \Gamma} p_j \left( 1 - \Phi\left(\frac{\beta + \delta_n^{(1)}}{\sqrt{ V(\Gamma))}}  \right) - \mu_n \right)(\Gamma - \Gamma_j)  + \sum_{j: \Gamma_j > \Gamma} p_j \left( 1 - \Phi\left(\frac{\beta}{\sqrt{ V(\Gamma))}}  \right) + \mu_n \right)(\Gamma - \Gamma_j) - O\left(\frac{1}{n} \right) \notag \\
    &\stackrel{(b)}{=} \Gamma + \sum_{j: \Gamma_j \leq \Gamma} p_j \left(  - 2\mu_n \right)(\Gamma - \Gamma_j)  + \sum_{j: \Gamma_j > \Gamma} p_j \left( 2 \mu_n \right)(\Gamma - \Gamma_j) - O\left(\frac{1}{n} \right) \notag \\
    &\stackrel{(c)}{\geq} \Gamma - o\left(\frac{1}{\sqrt{n}} \right). \label{khatambhhihoja}
\end{align}
Inequality $(a)$ above follows by using the bounds given in $(\ref{3e386})$ and the aforementioned choices of $\beta_1, \beta_2$ and $\beta_3$. Equality $(b)$ follows from the equality in $(\ref{annoyingthe})$. Inequality $(c)$ follows because $\max_j |\Gamma_j - \Gamma| \leq O(1/\sqrt{n})$ and $\mu_n \to 0$ as $n \to \infty$.  

From $(\ref{annoyingthe})$ and $(\ref{khatambhhihoja})$, we have 
$\Gamma - o(1/\sqrt{n}) \leq \mathbb{E}[c(X_k)] \leq \Gamma$ for all time instants $k$ ($n/2 + 1 \leq k \leq n$). For the variance constraint, we have  
\begin{align}
    &\text{Var}\left(\sum_{i=n/2+1}^n c(X_i) \right) \notag \\
    &= \mathbb{E}\left [ \left(\sum_{i=n/2+1}^n c(X_i) - \mathbb{E}\left [ \sum_{i=n/2+1}^n c(X_i) \right]  \right)^2 \right ] \notag \\
    &= \sum_{j=1}^3 p_j \left [ \mathcal{U}_j \left( \frac{n}{2}\Gamma + O(1) - \mathbb{E}\left [ \sum_{i=n/2+1}^n c(X_i) \right]   \right)^2 + (1-\mathcal{U}_j)\left( \frac{n}{2}\Gamma_j + O(1) - \mathbb{E}\left [ \sum_{i=n/2+1}^n c(X_i) \right] \right)^2 \right] \notag \\
    &\stackrel{(a)}{\leq} \sum_{j=1}^3 p_j  \mathcal{U}_j \left( \frac{n}{2}\Gamma + O(1) - \frac{n}{2}\Gamma + o\left( \sqrt{n}\right)   \right)^2 + \sum_{j:\Gamma_j \leq \Gamma} p_j (1-\mathcal{U}_j)\left( \frac{n}{2}\Gamma_j + O(1) - \frac{n}{2}\Gamma \right)^2 + \mbox{} \notag \\
    & \quad \quad \quad \quad \quad \quad \sum_{j:\Gamma_j > \Gamma} p_j (1-\mathcal{U}_j)\left( \frac{n}{2}\Gamma_j + O(1) - \frac{n}{2}\Gamma + o(\sqrt{n}) \right)^2 \notag \\
    &\leq \frac{n^2}{4} \sum_{j:\Gamma_j \leq \Gamma} p_j (1-\mathcal{U}_j)\left( \Gamma_j - \Gamma + O\left(\frac{1}{n} \right)  \right)^2 + \frac{n^2}{4} \sum_{j:\Gamma_j > \Gamma} p_j (1-\mathcal{U}_j)\left( \Gamma_j - \Gamma  + o\left(\frac{1}{\sqrt{n}}\right) \right)^2 + o(n) \notag \\
    &\leq \frac{n^2}{4} \sum_{j:\Gamma_j \leq \Gamma} p_j (1-\mathcal{U}_j)\left( \Gamma_j - \Gamma   \right)^2 + \frac{n^2}{4} \sum_{j:\Gamma_j > \Gamma} p_j (1-\mathcal{U}_j)\left( \Gamma_j - \Gamma   \right)^2 + o(n) \label{huen} \\
    &\stackrel{(b)}{\leq} \frac{n^2}{4}\left ( \Phi\left( \frac{\beta}{\sqrt{V(\Gamma)}}  \right) + \theta_n \right) \sum_{j=1}^3 p_j \left( \Gamma_j - \Gamma   \right)^2 + o(n) \notag \\
    &\leq \Phi\left(  \frac{\beta}{\sqrt{V(\Gamma)}} \right) \frac{n}{4} V  + o(n). \label{combinedvar}
\end{align}
Inequality $(a)$ above follows by using the bounds $\Gamma - o(1/\sqrt{n}) \leq \mathbb{E}[c(X_k)] \leq \Gamma$, for $n/2 + 1 \leq k \leq n$. Inequality $(b)$ follows by using the bounds given in $(\ref{3e386})$ and the aforementioned choices of $\beta_1, \beta_2$ and $\beta_3$, where $\theta_n$ is a suitable sequence satisfying $\theta_n \to 0$ as $n \to \infty$. 

We also have from $(\ref{varparta})$ and  $(\ref{combinedvar})$ that  
\begin{align}
    &\text{Cov}\left( \sum_{i=1}^{n/2} c(X_i),  \sum_{i=n/2+1}^{n} c(X_i) \right) \notag \\
    &\leq \sqrt{\text{Var}\left( \sum_{i=1}^{n/2} c(X_i)\right) \cdot \text{Var}\left( \sum_{i=n/2+1}^{n} c(X_i) \right) } \notag \\
    &\leq \sqrt{\frac{n^2 V^2}{16} \Phi\left(  \frac{\beta}{\sqrt{V(\Gamma)}} \right) + o(n^2) }\notag \\
    &\leq \frac{n}{4} V \sqrt{\Phi\left(  \frac{\beta}{\sqrt{V(\Gamma)}} \right)} \sqrt{1 + o(1)}   \label{covvvv}
 \end{align}

In view of $(\ref{varparta})$, $(\ref{combinedvar})$ and $(\ref{covvvv})$, we have  
\begin{align*}
    &\text{Var}\left(\sum_{i=1}^n c(X_i) \right)\\
    &= \text{Var}\left(\sum_{i=1}^{n/2} c(X_i) \right) + \text{Var}\left(\sum_{i=n/2+1}^{n} c(X_i) \right) + 2\, \text{Cov}\left( \sum_{i=1}^{n/2} c(X_i),  \sum_{i=n/2+1}^{n} c(X_i) \right) \\
    &\leq \frac{n}{4} V + \Phi\left(  \frac{\beta}{\sqrt{V(\Gamma)}} \right) \frac{n}{4} V + \frac{n}{2} V \sqrt{\Phi\left(  \frac{\beta}{\sqrt{V(\Gamma)}} \right)} \sqrt{1 + o(1)} + o(n)\\
    &\leq nV
\end{align*}
for sufficiently large $n$, since $\beta < \infty$ is a constant independent of $n$.

\textbf{Analysis of the Feedback Scheme}

With the aforementioned feedback scheme based on controller $F$ where feedback is used only once halfway through transmission, we define 
\begin{align*}
    \mathcal{E}_n \triangleq (F \circ W)\left(\frac{1}{n} \log \frac{W(Y^n|X^n)}{FW(Y^n)} \leq C(\Gamma) + \frac{r}{\sqrt{n}} \right).
\end{align*}
We will now show that $\limsup_{n \to \infty} \mathcal{E}_n < \epsilon$ which, by Lemma \ref{aaron'slemma}, would show that our feedback scheme achieves a second-order coding rate of $r$. We will only require that the limit superior is taken along the even integers. We first write
\begin{align}
    &\mathcal{E}_n = \notag \\
    &\,\,(F \circ W)\left(\frac{1}{n} \log \frac{W(Y^{n/2}|X^{n/2})\cdot W(Y^{n/2+1:n}|X^{n/2+1:n})}{PW(Y^{n/2})\cdot FW(Y^{n/2+1:n}|Y^{n/2})} \leq C(\Gamma) + \frac{r}{\sqrt{n}} \right) \notag \\
    &= \sum_{j=1}^3 p_j \mathbb{P}\left( \log \frac{W(Y^{n/2}|X^{n/2})}{PW(Y^{n/2})} + \log \frac{W(Y^{n/2+1:n}|X^{n/2+1:n})}{FW(Y^{n/2+1:n}|Y^{n/2})} \leq nC(\Gamma) + r \sqrt{n} \right). \label{bppee1} 
\end{align}
In $(\ref{bppee1})$,   
\begin{align}
    PW(y^{n/2}) &= \sum_{j=1}^3 p_j Q_j^{cc}(y^{n/2}) \notag \\
    &= \sum_{j=1}^3 p_j \sum_{x^{n/2} \in T_j} \frac{1}{|T_j|} W(y^{n/2} | x^{n/2}) \notag 
\end{align}
and
\begin{align*}
   FW(y^{n/2+1:n}|y^{n/2}) &= P^0_{n/2}(y^{n/2})\,  Q^{\text{cc}}(y^{n/2+1:n}) + \sum_{j=1}^3 P^j_{n/2}(y^{n/2})\,  Q_j^{\text{cc}}(y^{n/2+1:n}),  
\end{align*}
where $P_{n/2}^i(y^{n/2})$, for $i=1,2,3,4$, are positive weights which sum to one.

In $(\ref{bppee1})$, we use a similar derivation to the one used in going from $(\ref{referbackn})$ to $(\ref{subkaravg})$ to obtain 
\begin{align}
    &\mathcal{E}_n \notag \\
    &\leq \sum_{j=1}^3 p_j \mathbb{P}\left( \log \frac{W(Y^{n/2}|X^{n/2})}{Q_j^*(Y^{n/2})} + \log \frac{W(Y^{n/2+1:n}|X^{n/2+1:n})}{FW(Y^{n/2+1:n}|Y^{n/2})} \leq nC(\Gamma) + r \sqrt{n} \right. \notag \\
    & \left . \quad \quad \quad + \frac{s(P^*)}{2} \log (n/2)  + \sqrt{\frac{n/2}{\log (n/2)}}  \right) + \delta_{n/2}, \label{3335} 
\end{align}
where $\delta_{n/2} \to 0$ as $n \to \infty$. Continuing from $(\ref{3335})$,
\begin{align}
&\mathcal{E}_n \notag \\   
    &\leq \sum_{j=1}^3 p_j \mathbb{P}\left( \left [ \log \frac{W(Y^{n/2}|X^{n/2})}{Q_j^*(Y^{n/2})} - \frac{n}{2} C(\Gamma_j)\right] + \log \frac{W(Y^{n/2+1:n}|X^{n/2+1:n})}{FW(Y^{n/2+1:n}|Y^{n/2})} \leq \frac{n}{2}( C(\Gamma) - C(\Gamma_j) )  + \frac{n}{2}C(\Gamma) + \mbox{} \right. \notag \\
    & \left . \quad \quad \quad  r \sqrt{n} + \frac{s(P^*)}{2} \log (n/2)  + \sqrt{\frac{n/2}{\log (n/2)}}  \right) + \delta_{n/2} \notag \\
    &\leq \sum_{j=1}^3 p_j \mathbb{P}\left( \left [ \log \frac{W(Y^{n/2}|X^{n/2})}{Q_j^*(Y^{n/2})} - \frac{n}{2} [C(\Gamma_j)]_{n/2}\right] + \left [ \log \frac{W(Y^{n/2+1:n}|X^{n/2+1:n})}{FW(Y^{n/2+1:n}|Y^{n/2})} - \frac{n}{2} C(\Gamma) \right ]  \leq \frac{n}{2}( C(\Gamma) - C(\Gamma_j) )   + \mbox{}  \right. \notag \\
    & \left . \quad \quad \quad  r \sqrt{n} + \frac{s(P^*)}{2} \log (n/2)  + \sqrt{\frac{n/2}{\log (n/2)}} + 2 C'(\Gamma_j) J c_{\max}  \right) + \delta_{n/2}\notag \\
    &= \sum_{j=1}^3 p_j \mathbb{P}\left( \frac{1}{\sqrt{\frac{n}{2}V_j}}\left [ \log \frac{W(Y^{n/2}|X^{n/2})}{Q_j^*(Y^{n/2})} - \frac{n}{2} [C(\Gamma_j)]_{n/2}\right] + \frac{1}{\sqrt{\frac{n}{2}V_j}}\left [ \log \frac{W(Y^{n/2+1:n}|X^{n/2+1:n})}{FW(Y^{n/2+1:n}|Y^{n/2})} - \frac{n}{2} C(\Gamma) \right ]  \leq \mbox{}\right . \notag  \\
    & \left . \quad \quad \quad \quad \frac{\sqrt{n}}{\sqrt{2V_j}}( C(\Gamma) - C(\Gamma_j) )   +   \frac{r\sqrt{2}}{\sqrt{V_j}}  + s(P^*) \frac{\log (n/2)}{\sqrt{2n V_j}}  + \sqrt{\frac{n}{n V_j\log (n/2)}} + \frac{2 C'(\Gamma_j) J c_{\max}}{\sqrt{\frac{n}{2} V_j}}  \right) + \delta_{n/2}, \label{partitionv1}
\end{align}
where we used the following: 
\begin{align}
    \frac{n}{2} [C(\Gamma_j)]_{n/2} &:= \mathbb{E} \left [ \sum_{i=1}^{n/2} \log \frac{W(Y_i|X_i)}{Q_j^*(Y_i)} \right ] \notag  \\
    &\in  \left [ \frac{n}{2} C(\Gamma_j) - 2C'(\Gamma_j) J c_{\max}, \frac{n}{2} C(\Gamma_j)   \right ] \label{Cquantf1}
\end{align}
and
\begin{align}
\begin{split}
    V_j = [V(\Gamma_j)]_{n/2} &= \text{Var}\left( \log \frac{W(Y^{n/2}|X^{n/2})}{Q_j^*(Y^{n/2})} \right)\\
    &= \frac{n}{2} \sum_{a \in \mathcal{A}} [P_j^*]_{n/2}(a) \text{Var}\left( \log \frac{W(Y_a|a)}{Q_j^*(Y_a)} \right),
    \end{split}
    \label{Vj1and21}
\end{align}
where $X^{n/2} \sim \text{Unif}(T_j)$ and $Y_a \sim W(\cdot | a)$. 
We now upper bound the following term from $(\ref{partitionv1})$: 
\begin{align}
    &\mathbb{P}\left( \frac{1}{\sqrt{\frac{n}{2}V_j}}\left [ \log \frac{W(Y^{n/2}|X^{n/2})}{Q_j^*(Y^{n/2})} - \frac{n}{2} [C(\Gamma_j)]_{n/2}\right] + \mbox{} \right . \notag \\
    & \quad \quad \quad \left .  \frac{1}{\sqrt{\frac{n}{2}V_j}}\left [ \log \frac{W(Y^{n/2+1:n}|X^{n/2+1:n})}{FW(Y^{n/2+1:n}|Y^{n/2})} - \frac{n}{2} C(\Gamma) \right ]  \leq d_{n,j}  \right), \label{iopvey1}
\end{align}
where we have defined
\begin{align*}
    d_{n, j} := \frac{\sqrt{n}}{\sqrt{2V_j}}( C(\Gamma) - C(\Gamma_j) )   +   \frac{r\sqrt{2}}{\sqrt{V_j}}  + s(P^*) \frac{\log (n/2)}{\sqrt{2n V_j}}  + \sqrt{\frac{1}{ V_j\log (n/2)}} + \frac{2 C'(\Gamma_j) J c_{\max}}{\sqrt{\frac{n}{2} V_j}}  
\end{align*}
for convenience. 

Define a new random variable $T^j_{n/2}$ as 
\begin{align}
    T^j_{n/2} \triangleq \frac{1}{\sqrt{\frac{n}{2}V_j}}\left [ \sum_{m=1}^{n/2} \log \frac{W(Y_m|X_m)}{Q_j^*(Y_m)} - \frac{n}{2} [C(\Gamma_j)]_{n/2}\right], \label{Gn2junder1}
\end{align}
where $X^{n/2} \sim \text{Unif}(T_j)$, $Y^{n/2} \sim W(\cdot|X^{n/2})$,
and let $\underline{G}^j_{n/2}$ denote its CDF. Note that the distribution of 
$$\sum_{i=1}^{n/2} \log \frac{W(Y_i|X_i)}{Q_j^*(Y_i)}$$
depends on $\{X_i\}_{i=1}^{n/2}$ only through their $n/2$-type which is fixed as $[P^*_j]_{n/2}$. Accordingly, by defining a sequence of independent random variables 
\begin{align}
    \left \{Y_{i, a}: i \in \mathbb{N}, 1 \leq i \leq \frac{n}{2}[P_j^*]_{n/2}(a), a \in \mathcal{A} \right \}, \label{ixp11166}
\end{align}   
where $Y_{i,a} \sim W(\cdot | a)$, we have 
\begin{align}
    \sum_{i=1}^{n/2} \log \frac{W(Y_i|X_i)}{Q_j^*(Y_i)} \stackrel{d}{=} \sum_{a \in \mathcal{A}} \sum_{i=1}^{\frac{n}{2}[P_j^*]_{n/2}(a)} \log \frac{W(Y_{i,a}|a)}{Q_j^*(Y_{i, a})}. \label{ixp211}
\end{align}
Since $\underline{G}_{n/2}^j$ is the CDF of a normalized sum of independent random variables, Berry-Esseen Theorem \cite{esseen11} for non-identically distributed summands gives us 
\begin{align}
    \sup_{x \in \mathbb{R}} \big | \underline{G}_{n/2}^j(x) - \Phi(x)  \big | \leq \frac{\underline{\kappa}_{n/2}^j}{\sqrt{n/2}}, \label{lemma7kaalternative}
\end{align}
where $\underline{\kappa}_{n/2}^j$ depends on the second- and third-order moments as follows: 
\begin{align}
    \underline{\kappa}^j_{n/2} := \frac{\max_{a \in \mathcal{A}} \mathbb{E} \left [ 
 \Big | \log \frac{W(Y|a)}{Q_j^*(Y)} - \mathbb{E}\left[ \log \frac{W(Y|a)}{Q_j^*(Y)} \right]  \Big |^3\right ]}{[V(\Gamma_j)]_{n/2}^{3/2}}.  \label{kappanjdef}   
\end{align}
We have $\underline{\kappa}_{n/2}^j < \infty$ since $i_{\max}$ in $(\ref{basicdefs})$ is assumed to be finite, and $\Gamma_j \to \Gamma$ as $n \to \infty$.

Let $\overline{G}_{n/2}^j$ denote the CDF of  
$$\frac{1}{\sqrt{\frac{n}{2}V_j}}\left [ \sum_{m=1}^{n/2} \log \frac{W(Y_m|X_m)}{Q^*(Y_m)} - \frac{n}{2} [C(\Gamma)]_{n/2}\right],$$
where $X^{n/2} \sim \text{Unif}(T)$ and $Y^{n/2} \sim W(\cdot | X^{n/2})$. By a similar argument as above, we have 
\begin{align*}
    \sup_{x \in \mathbb{R}} \big | \overline{G}_{n/2}^j(x) - \Phi(R_{n/2}^j x)  \big | \leq \frac{\overline{\kappa}_{n/2}}{\sqrt{n/2}}, 
\end{align*}
where 
\begin{align*}
    R_{n/2}^j &= \sqrt{\frac{[V(\Gamma_j)]_{n/2}}{[V(\Gamma)]_{n/2}}},\\
\overline{\kappa}_{n/2} &:= \frac{\max_{a \in \mathcal{A}} \mathbb{E} \left [ 
 \Big | \log \frac{W(Y|a)}{Q^*(Y)} - \mathbb{E}\left [ \log \frac{W(Y|a)}{Q^*(Y)} \right ]  \Big |^3\right ]}{[V(\Gamma)]_{n/2}^{3/2}}.    
\end{align*}

Define 
\begin{align*}
    d'_{n, j} := \frac{\sqrt{n}}{\sqrt{2V_j}} \left( C(\Gamma_j) - [C(\Gamma_j)]_{n/2} \right) + \frac{\beta_j}{\sqrt{V_j}}.
\end{align*}
Then $(\ref{iopvey1})$ is equal to 
\begin{align}
    &\int_{-\infty}^{\infty} d\underline{G}^j_{n/2}(x) \mathbb{P}\left( \frac{1}{\sqrt{\frac{n}{2}V_j}}\left [ \log \frac{W(Y^{n/2+1:n}|X^{n/2+1:n})}{FW(Y^{n/2+1:n}|Y^{n/2})} - \frac{n}{2} C(\Gamma) \right ]  \leq d_{n, j} - x \Big | T^j_{n/2} = x \right)  \notag \\
    &= \int_{d'_{n,j}}^\infty d\underline{G}^j_{n/2}(x) \mathbb{P}\left( \frac{1}{\sqrt{\frac{n}{2}V_j}}\left [ \log \frac{W(Y^{n/2+1:n}|X^{n/2+1:n})}{FW(Y^{n/2+1:n}|Y^{n/2})} - \frac{n}{2} C(\Gamma) \right ]  \leq d_{n, j} - x \Big | T^j_{n/2} = x \right)   + \mbox{} \notag \\
    & \int_{-\infty}^{d'_{n,j}} d\underline{G}^j_{n/2}(x) \mathbb{P}\left( \frac{1}{\sqrt{\frac{n}{2}V_j}}\left [ \log \frac{W(Y^{n/2+1:n}|X^{n/2+1:n})}{FW(Y^{n/2+1:n}|Y^{n/2})} - \frac{n}{2} C(\Gamma) \right ]  \leq d_{n, j} - x \Big |  T^j_{n/2} = x \right). \label{uxxx1}
\end{align}
Consider the first integral in $(\ref{uxxx1})$ which corresponds to the case when the information density at halftime exceeds the threshold in $(\ref{ovzbf1})$. In this case, we have $X^{n/2+1:n} \sim \text{Unif}(T^{n/2}_\mathcal{A}([P^*]_{n/2}))$, or $X^{n/2+1:n} \sim \text{Unif}(T)$ for brevity. Since we have the Markov Chain $Y^{n/2} - X^{n/2+1:n} - Y^{n/2+1:n}$, 
\begin{align}
    &\int_{d'_{n,j}}^\infty d\underline{G}^j_{n/2}(x) \mathbb{P}\left( \frac{1}{\sqrt{\frac{n}{2}V_j}}\left [ \log \frac{W(Y^{n/2+1:n}|X^{n/2+1:n})}{FW(Y^{n/2+1:n}|Y^{n/2})} - \frac{n}{2} C(\Gamma) \right ]  \leq d_{n, j} - x \Big | T^j_{n/2} = x \right)  \notag \\
    &= \int_{d'_{n,j}}^\infty d\underline{G}^j_{n/2}(x) \mathbb{P}_{X^{n/2+1:n} \sim \text{Unif}(T)}\left( \frac{1}{\sqrt{\frac{n}{2}V_j}}\left [ \log \frac{W(Y^{n/2+1:n}|X^{n/2+1:n})}{FW(Y^{n/2+1:n}|Y^{n/2})} - \frac{n}{2} C(\Gamma) \right ]  \leq d_{n, j} - x  \right)  \notag \\
    &\stackrel{(a)}{\leq} \int_{d'_{n,j}}^\infty d\underline{G}^j_{n/2}(x) \mathbb{P}_{X^{n/2+1:n} \sim \text{Unif}(T)}\left( \frac{1}{\sqrt{\frac{n}{2}V_j}}\left [ \log \frac{W(Y^{n/2+1:n}|X^{n/2+1:n})}{Q^*(Y^{n/2+1:n})} - \frac{n}{2} C(\Gamma) \right ]  \leq d_{n, j} - x  + \mbox{}\right . \notag   \\
    & \quad \quad \quad \quad \left .   \frac{s(P^*)}{2} \frac{\log(n/2)}{\sqrt{\frac{n}{2}V_j}} + \sqrt{\frac{1}{V_j \log(n/2)}}  \right)  + \delta_{n/2} \notag \\ 
    &= \int_{d'_{n,j}}^\infty d\underline{G}^j_{n/2}(x) \mathbb{P}_{X^{n/2+1:n} \sim \text{Unif}(T)}\left( \frac{1}{\sqrt{\frac{n}{2}V_j}}\left [ \log \frac{W(Y^{n/2+1:n}|X^{n/2+1:n})}{Q^*(Y^{n/2+1:n})} - \frac{n}{2} [C(\Gamma)]_{n/2} \right ]  \leq d_{n, j} - x \right . \notag   \\
    & \quad \quad \quad \quad \left . + \frac{n}{2} \frac{C(\Gamma) - [C(\Gamma)]_{n/2}}{\sqrt{\frac{n}{2}V_j}} + \frac{s(P^*)}{2} \frac{\log(n/2)}{\sqrt{\frac{n}{2}V_j}} + \sqrt{\frac{1}{V_j \log(n/2)}}  \right)  + \delta_{n/2} \label{nepaln1}
\end{align}
for sufficiently large $n$, where $Y^{n/2 + 1:n} \sim W(\cdot | X^{n/2 + 1:n})$, inequality $(a)$ follows from a similar derivation to that used in going from $(\ref{referbackn})$ to $(\ref{subkaravg})$, and $\delta_{n/2} \to 0$ as $n \to \infty$. $(\ref{nepaln1})$ can be written as 
\begin{align*}
    \int_{d'_{n,j}}^\infty d\underline{G}^j_{n/2}(x) \overline{G}_{n/2}^j\left(  d_{n, j} - x + \frac{n}{2} \frac{C(\Gamma) - [C(\Gamma)]_{n/2}}{\sqrt{\frac{n}{2}V_j}} + \frac{s(P^*)}{2} \frac{\log(n/2)}{\sqrt{\frac{n}{2}V_j}} + \sqrt{\frac{1}{V_j \log(n/2)}}  \right)  + \delta_{n/2}.
\end{align*}

Consider the second integral in $(\ref{uxxx1})$. We have 
\begin{align*}
    &\int_{-\infty}^{d'_{n,j}} d\underline{G}^j_{n/2}(x) \mathbb{P}\left( \frac{1}{\sqrt{\frac{n}{2}V_j}}\left [ \log \frac{W(Y^{n/2+1:n}|X^{n/2+1:n})}{FW(Y^{n/2+1:n}|Y^{n/2})} - \frac{n}{2} C(\Gamma) \right ]  \leq d_{n, j} - x \Big | T^j_{n/2} = x \right) \\
    &= \int_{-\infty}^{d'_{n,j}} d\underline{G}^j_{n/2}(x) \mathbb{P}_{X^{n/2+1:n} \sim \text{Unif}(T_j)}\left( \frac{1}{\sqrt{\frac{n}{2}V_j}}\left [ \log \frac{W(Y^{n/2+1:n}|X^{n/2+1:n})}{FW(Y^{n/2+1:n}|Y^{n/2})} - \frac{n}{2} C(\Gamma) \right ]  \leq d_{n, j} - x \right) \\
    &\leq \int_{-\infty}^{d'_{n,j}} d\underline{G}^j_{n/2}(x)  \mathbb{P}_{X^{n/2+1:n} \sim \text{Unif}(T_j)}\left( \frac{1}{\sqrt{\frac{n}{2}V_{j}}} \left [  \log \frac{W(Y^{n/2+1:n}|X^{n/2+1:n})}{Q^*_{j}(Y^{n/2+1:n})} - \frac{n}{2}[C(\Gamma_{j})]_{n/2} \right]  \leq d_{n, j} - x   + \mbox{} \right. \\
    & \quad \quad \quad \quad  \left . \frac{n}{2} \frac{C(\Gamma) - [C(\Gamma_{j})]_{n/2}}{\sqrt{\frac{n}{2}V_{j}}} + \frac{s(P^*)}{2} \frac{\log(n/2)}{\sqrt{\frac{n}{2}V_{j}}} + \sqrt{\frac{1}{V_{j}\log(n/2)}} \right)  + \delta_{n/2}\\
    &= \int_{-\infty}^{d'_{n,j}} d\underline{G}^j_{n/2}(x)  \underline{G}_{n/2}^{j}\left(  d_{n, j} - x   +  \frac{n}{2} \frac{C(\Gamma) - [C(\Gamma_{j})]_{n/2}}{\sqrt{\frac{n}{2}V_{j}}} + \frac{s(P^*)}{2} \frac{\log(n/2)}{\sqrt{\frac{n}{2}V_{j}}} + \sqrt{\frac{1}{V_{j}\log(n/2)}}  \right)  + \delta_{n/2}\\
    &= \int_{-\infty}^{\infty} d\underline{G}^j_{n/2}(x)  \underline{G}_{n/2}^{j}\left(  d_{n, j} - x   +  \frac{n}{2} \frac{C(\Gamma) - [C(\Gamma_{j})]_{n/2}}{\sqrt{\frac{n}{2}V_{j}}} + \frac{s(P^*)}{2} \frac{\log(n/2)}{\sqrt{\frac{n}{2}V_{j}}} + \sqrt{\frac{1}{V_{j}\log(n/2)}}  \right)   - \mbox{}\\
    & \int_{d'_{n, j}}^{\infty} d\underline{G}^j_{n/2}(x)  \underline{G}_{n/2}^{j}\left(  d_{n, j} - x   +  \frac{n}{2} \frac{C(\Gamma) - [C(\Gamma_{j})]_{n/2}}{\sqrt{\frac{n}{2}V_{j}}} + \frac{s(P^*)}{2} \frac{\log(n/2)}{\sqrt{\frac{n}{2}V_{j}}} + \sqrt{\frac{1}{V_{j}\log(n/2)}}  \right)  + \delta_{n/2}.
\end{align*}
Note that 
\begin{align}
    &\int_{-\infty}^{\infty} d\underline{G}^j_{n/2}(x)  \underline{G}_{n/2}^{j}\left(  d_{n, j} - x   +  \frac{n}{2} \frac{C(\Gamma) - [C(\Gamma_{j})]_{n/2}}{\sqrt{\frac{n}{2}V_{j}}} + \frac{s(P^*)}{2} \frac{\log(n/2)}{\sqrt{\frac{n}{2}V_{j}}} + \sqrt{\frac{1}{V_{j}\log(n/2)}}  \right)  \notag \\
    &= \mathbb{P}\left( \frac{1}{\sqrt{\frac{n}{2}V_j}}\left [ \sum_{m=1}^{n/2} \log \frac{W(Y_m|X_m)}{Q_j^*(Y_m)} - \frac{n}{2} [C(\Gamma_j)]_{n/2}\right] +  \frac{1}{\sqrt{\frac{n}{2}V_j}}\left [ \sum_{m=1}^{n/2} \log \frac{W(Y_m'|X_m')}{Q_j^*(Y_m')} - \frac{n}{2} [C(\Gamma_j)]_{n/2}\right] \right. \notag  \\
    & \left . \quad \quad \quad \quad  \leq    d_{n, j}   +  \frac{n}{2} \frac{C(\Gamma) - [C(\Gamma_{j})]_{n/2}}{\sqrt{\frac{n}{2}V_{j}}} + \frac{s(P^*)}{2} \frac{\log(n/2)}{\sqrt{\frac{n}{2}V_{j}}} + \sqrt{\frac{1}{V_{j}\log(n/2)}} \right), \label{tiring1}
\end{align}
where $\{(X_m',Y_m') \}_{m=1}^{n/2}$ are independent copies of $\{(X_m,Y_m) \}_{m=1}^{n/2}$, $X^{n/2} \sim \text{Unif}(T_j)$ and $Y^{n/2}$ is the output of the DMC $W$ when $X^{n/2}$ is the input. For convenience, define 
$$d_{n, j}^{(2)} = d_{n, j}   +  \frac{n}{2} \frac{C(\Gamma) - [C(\Gamma_{j})]_{n/2}}{\sqrt{\frac{n}{2}V_{j}}} + \frac{s(P^*)}{2} \frac{\log(n/2)}{\sqrt{\frac{n}{2}V_{j}}} + \sqrt{\frac{1}{V_{j}\log(n/2)}}. $$
For any $\epsilon^{(1)} > 0$,
we can rewrite $(\ref{tiring1})$ as 
\begin{align}
    &\mathbb{P}\left( \sum_{m=1}^n \log \frac{W(Y_m|X_m)}{Q_j^*(Y_m)} - n[C(\Gamma_j)]_{n/2} \leq \sqrt{\frac{n}{2}V_j} \,d_{n, j}^{(2)} \right)\notag \\
    &= \mathbb{P}\left( \frac{1}{\sqrt{n V_j}} \left [ \sum_{m=1}^n \log \frac{W(Y_m|X_m)}{Q_j^*(Y_m)} - n[C(\Gamma_j)]_{n/2}\right] \leq \frac{1}{\sqrt{2}} \,d_{n, j}^{(2)} \right) \notag  \\
    &\leq \Phi\left( \frac{1}{\sqrt{2}} d_{n,j}^{(2)}\right) + \frac{\underline{\kappa}^j_{n/2}}{\sqrt{n}}\notag \\
    &\leq \Phi\left(   \frac{\sqrt{n}(C(\Gamma) - C(\Gamma_j))}{\sqrt{V_j}} + \frac{r}{\sqrt{V_j}} \right) + \frac{\underline{\kappa}^j_{n/2}}{\sqrt{n}} + \epsilon^{(1)} \notag \\
    &\leq \Phi\left( \pi_j  \right) + \frac{\underline{\kappa}^j_{n/2}}{\sqrt{n}} + 2\epsilon^{(1)} \notag 
\end{align}
for sufficiently large $n$, where $X^n \sim \text{Unif}(T^n_{\mathcal{A}}([P^*_j]_{n/2}))$. 

Combining the derivation carried out thus far, we have that $(\ref{iopvey1})$ is upper bounded as   
\begin{align}
    &\mathbb{P}\left( \frac{1}{\sqrt{\frac{n}{2}V_j}}\left [ \log \frac{W(Y^{n/2}|X^{n/2})}{Q_j^*(Y^{n/2})} - \frac{n}{2} [C(\Gamma_j)]_{n/2}\right] + \frac{1}{\sqrt{\frac{n}{2}V_j}}\left [ \log \frac{W(Y^{n/2+1:n}|X^{n/2+1:n})}{FW(Y^{n/2+1:n}|Y^{n/2})} - \frac{n}{2} C(\Gamma) \right ]  \leq d_{n,j}  \right) \notag \\
    &\leq \Phi\left( \pi_j \right) + \frac{\underline{\kappa}_{n/2}^j}{\sqrt{n}} + 2\epsilon^{(1)} + 2\delta_{n/2} 
 - \mbox{} \notag  \\  
    & \quad \quad \quad \quad \int_{d'_{n, j}}^{\infty} d\underline{G}^j_{n/2}(x)  \left [ \underline{G}_{n/2}^{j}\left(  d_{n, j} - x   +  \frac{n}{2} \frac{C(\Gamma) - [C(\Gamma_{j})]_{n/2}}{\sqrt{\frac{n}{2}V_{j}}} + \frac{s(P^*)}{2} \frac{\log(n/2)}{\sqrt{\frac{n}{2}V_{j}}} + \sqrt{\frac{1}{V_{j}\log(n/2)}}  \right) - \right . \notag  \\
    & \left . \quad \quad \quad \quad \quad \quad \quad \quad \quad \quad \quad \quad \quad \quad \quad  \overline{G}_{n/2}^j\left(  d_{n, j} - x + \frac{n}{2} \frac{C(\Gamma) - [C(\Gamma)]_{n/2}}{\sqrt{\frac{n}{2}V_j}} + \frac{s(P^*)}{2} \frac{\log(n/2)}{\sqrt{\frac{n}{2}V_j}} + \sqrt{\frac{1}{V_j \log(n/2)}}  \right) \right ]  
    \label{lwbndint1}
\end{align}
We now lower bound the integral in $(\ref{lwbndint1})$. Before doing so, we first write 
\begin{align*}
    &d_{n, j} - x   +  \frac{n}{2} \frac{C(\Gamma) - [C(\Gamma_{j})]_{n/2}}{\sqrt{\frac{n}{2}V_{j}}} + \frac{s(P^*)}{2} \frac{\log(n/2)}{\sqrt{\frac{n}{2}V_{j}}} + \sqrt{\frac{1}{V_{j}\log(n/2)}}\\
    &\quad \quad = \frac{\sqrt{2n}}{\sqrt{V_j}}(C(\Gamma) - C(\Gamma_j)) + \frac{r \sqrt{2}}{\sqrt{V_j}} - x + \cdots 
\end{align*}
and 
\begin{align*}
      &d_{n, j} - x + \frac{n}{2} \frac{C(\Gamma) - [C(\Gamma)]_{n/2}}{\sqrt{\frac{n}{2}V_j}} + \frac{s(P^*)}{2} \frac{\log(n/2)}{\sqrt{\frac{n}{2}V_j}} + \sqrt{\frac{1}{V_j \log(n/2)}}\\
      &\quad \quad = \frac{\sqrt{n}}{\sqrt{2V_j}}(C(\Gamma) - C(\Gamma_j)) + \frac{r \sqrt{2}}{\sqrt{V_j}} - x + \cdots, 
\end{align*}
where "$\cdots$" denotes all the $o(1)$ terms, i.e., terms which go to zero as $n \to \infty$.    

For any $\epsilon^{(2)} > 0$, we use multiple applications of Berry-Esseen Theorem and integration by parts to obtain 
\begin{align*}
    &\int_{d'_{n, j}}^{\infty} d\underline{G}^j_{n/2}(x)  \left [ \underline{G}_{n/2}^{j}\left(  \frac{\sqrt{2n}}{\sqrt{V_j}}(C(\Gamma) - C(\Gamma_j)) + \frac{r \sqrt{2}}{\sqrt{V_j}} - x + \cdots  \right) - \mbox{} \right . \\
    & \left . \quad \quad \quad \quad \quad \quad \quad   \overline{G}_{n/2}^j\left(  \frac{\sqrt{n}}{\sqrt{2V_j}}(C(\Gamma) - C(\Gamma_j)) + \frac{r \sqrt{2}}{\sqrt{V_j}} - x + \cdots  \right) \right ]  \\
    &\geq \int_{d'_{n, j}}^{\infty} d\underline{G}^j_{n/2}(x) \left[\Phi\left(  \frac{\sqrt{2n}}{\sqrt{V_j}}(C(\Gamma) - C(\Gamma_j)) + \frac{r \sqrt{2}}{\sqrt{V_j}} - x + \cdots \right) - \mbox{} \right . \\
    & \left . \quad \quad \quad \quad \Phi\left(R_{n/2}^j\left(  \frac{\sqrt{n}}{\sqrt{2V_j}}(C(\Gamma) - C(\Gamma_j)) + \frac{r \sqrt{2}}{\sqrt{V_j}} - x + \cdots\right) \right)   \right] - \frac{\underline{\kappa}^j_{n/2} + \overline{\kappa}_{n/2}}{\sqrt{n/2}}\\
    &= -\underline{G}_{n/2}^j(d'_{n,j}) \left [\Phi\left(  \frac{\sqrt{2n}}{\sqrt{V_j}}(C(\Gamma) - C(\Gamma_j)) + \frac{r \sqrt{2}}{\sqrt{V_j}} - d'_{n,j} + \cdots  \right) - \mbox{} \right . \\
    & \left . \quad \quad \quad \quad \Phi\left(R_{n/2}^j\left(  \frac{\sqrt{n}}{\sqrt{2V_j}}(C(\Gamma) - C(\Gamma_j)) + \frac{r \sqrt{2}}{\sqrt{V_j}} - d'_{n,j} + \cdots \right) \right)  \right ] + \mbox{}\\
    & \int_{d'_{n, j}}^{\infty} \underline{G}^j_{n/2}(x) \left[\phi\left(  \frac{\sqrt{2n}}{\sqrt{V_j}}(C(\Gamma) - C(\Gamma_j)) + \frac{r \sqrt{2}}{\sqrt{V_j}} - x  + \cdots  \right) - \mbox{} \right . \\
    & \left . \quad \quad \quad \quad R_{n/2}^j  \phi\left(R_{n/2}^j\left(  \frac{\sqrt{n}}{\sqrt{2V_j}}(C(\Gamma) - C(\Gamma_j)) + \frac{r \sqrt{2}}{\sqrt{V_j}} - x + \cdots \right) \right)   \right]dx - \frac{\underline{\kappa}^j_{n/2} + \overline{\kappa}_{n/2}}{\sqrt{n/2}}\\
    &\geq -\underline{G}_{n/2}^j\left ( d'_{n,j} \right) \left [\Phi\left(  \frac{\sqrt{2n}}{\sqrt{V_j}}(C(\Gamma) - C(\Gamma_j)) + \frac{r \sqrt{2}}{\sqrt{V_j}} -d'_{n,j} + \cdots   \right) - \mbox{} \right. \\
    & \left . \quad \quad  \Phi\left(R_{n/2}^j\left(  \frac{\sqrt{n}}{\sqrt{2V_j}}(C(\Gamma) - C(\Gamma_j)) + \frac{r \sqrt{2}}{\sqrt{V_j}} - d'_{n,j} + \cdots \right) \right)  \right ] + \mbox{}\\
    & \int_{d'_{n, j}}^{\infty} \Phi(x) \left[\phi\left(  \frac{\sqrt{2n}}{\sqrt{V_j}}(C(\Gamma) - C(\Gamma_j)) + \frac{r \sqrt{2}}{\sqrt{V_j}} - x  + \cdots  \right) - R_{n/2}^j  \phi\left(R_{n/2}^j\left(  \frac{\sqrt{n}}{\sqrt{2V_j}}(C(\Gamma) - C(\Gamma_j)) + \frac{r \sqrt{2}}{\sqrt{V_j}} - x + \cdots\right) \right)   \right]dx \\
    & \quad \quad \quad \quad \quad \quad \quad \quad \quad \quad \quad \quad -\frac{(2 + R^j_{n/2})\underline{\kappa}^j_{n/2} + \overline{\kappa}_{n/2}}{\sqrt{n/2}}\\
    &= -\underline{G}_{n/2}^j\left ( d'_{n,j} \right) \left [\Phi\left(  \frac{\sqrt{2n}}{\sqrt{V_j}}(C(\Gamma) - C(\Gamma_j)) + \frac{r \sqrt{2}}{\sqrt{V_j}} - d'_{n,j} + \cdots   \right) - \mbox{} \right. \\
    & \left . \quad \quad \quad \quad \Phi\left(R_{n/2}^j\left(  \frac{\sqrt{n}}{\sqrt{2V_j}}(C(\Gamma) - C(\Gamma_j)) + \frac{r \sqrt{2} }{\sqrt{V_j}} - d'_{n,j} + \cdots \right) \right)  \right ] + \mbox{}\\
    & \Phi\left (d'_{n,j} \right ) \left[\Phi\left(  \frac{\sqrt{2n}}{\sqrt{V_j}}(C(\Gamma) - C(\Gamma_j)) + \frac{r \sqrt{2}}{\sqrt{V_j}} -d'_{n,j} + \cdots  \right) -   \Phi\left(R_{n/2}^j\left(  \frac{\sqrt{n}}{\sqrt{2V_j}}(C(\Gamma) - C(\Gamma_j)) + \frac{r \sqrt{2} }{\sqrt{V_j}} - d'_{n,j} + \cdots\right) \right)   \right] + \mbox{} \\
    & \int_{d'_{n, j}}^{\infty} \phi(x) \left[\Phi\left(  \frac{\sqrt{2n}}{\sqrt{V_j}}(C(\Gamma) - C(\Gamma_j)) + \frac{r \sqrt{2}}{\sqrt{V_j}} - x + \cdots \right) -   \Phi\left(R_{n/2}^j\left(  \frac{\sqrt{n}}{\sqrt{2V_j}}(C(\Gamma) - C(\Gamma_j)) + \frac{r \sqrt{2}}{\sqrt{V_j}} - x + \cdots\right) \right)   \right]dx  \\
    & \quad \quad \quad \quad \quad \quad \quad \quad \quad \quad \quad \quad -\frac{(2 + R^j_{n/2})\underline{\kappa}^j_{n/2} + \overline{\kappa}_{n/2}}{\sqrt{n/2}}\\
    &\geq \int_{d'_{n, j}}^{\infty} \phi(x) \left[\Phi\left(  \frac{\sqrt{2n}}{\sqrt{V_j}}(C(\Gamma) - C(\Gamma_j)) + \frac{r \sqrt{2}}{\sqrt{V_j}} - x + \cdots \right) -   \Phi\left(R_{n/2}^j\left(  \frac{\sqrt{n}}{\sqrt{2V_j}}(C(\Gamma) - C(\Gamma_j)) + \frac{r \sqrt{2}}{\sqrt{V_j}} - x + \cdots\right) \right)   \right]dx  \\
    & \quad \quad \quad \quad \quad \quad \quad \quad \quad \quad \quad \quad - \frac{(3 + R^j_{n/2})\underline{\kappa}^j_{n/2} + \overline{\kappa}_{n/2}}{\sqrt{n/2}}
    \end{align*}
    \begin{align}
    &\geq \int_{d'_{n, j}}^{\infty} \phi(x) \left[\Phi\left(  \frac{\sqrt{2n}}{\sqrt{V_j}}(C(\Gamma) - C(\Gamma_j)) + \frac{r \sqrt{2}}{\sqrt{V_j}} - x  \right) -   \Phi\left(R_{n/2}^j\left(  \frac{\sqrt{n}}{\sqrt{2V_j}}(C(\Gamma) - C(\Gamma_j)) + \frac{r \sqrt{2}}{\sqrt{V_j}} - x \right) \right)   \right]dx - \mbox{} \notag \\
    & \quad \quad \quad \quad \quad \quad \quad \quad \quad \quad \quad \quad \frac{(3 + R^j_{n/2})\underline{\kappa}^j_{n/2} + \overline{\kappa}_{n/2}}{\sqrt{n/2}} - \epsilon^{(2)} \label{gbxx}
\end{align}
for sufficiently large $n$. $(\ref{gbxx})$ can be further lower bounded by 
\begin{align}
    &\int_{d'_{n, j}}^{\infty} \phi(x) \left[\Phi\left(\sqrt{2} \pi_j - \frac{r \sqrt{2}}{\sqrt{V(\Gamma)}} + \frac{r \sqrt{2}}{\sqrt{V(\Gamma)}} -x \right) -   \Phi\left(\frac{\pi_j}{\sqrt{2}} - \frac{r}{\sqrt{2V(\Gamma)}} + \frac{r \sqrt{2}}{\sqrt{V(\Gamma)}} - R_{n/2}^j x   \right)   \right]dx - \mbox{} \notag \\
    & \quad \quad \quad \quad \quad \quad \quad \quad \quad \quad \quad \quad \frac{(3 + R^j_{n/2})\underline{\kappa}^j_{n/2} + \overline{\kappa}_{n/2}}{\sqrt{n/2}} - \epsilon^{(2)} - \epsilon^{(3)} \label{faraz}
\end{align}
for sufficiently large $n$ for some suitable $\epsilon^{(3)} > 0$ depending on $\epsilon^{(1)}$ such that $\epsilon^{(3)} \to 0$ as $\epsilon^{(1)} \to 0$ in view of the fact that  
\begin{align*}
    &\frac{\sqrt{2n}}{\sqrt{V_j}}(C(\Gamma) - C(\Gamma_j)) + \frac{r \sqrt{2}}{\sqrt{V_j}} - x\\
    &\geq \frac{\sqrt{2n}}{\sqrt{V_j}}C'(\Gamma)(\Gamma - \Gamma_j) + \frac{r \sqrt{2}}{\sqrt{V_j}} - x\\
    &= \frac{\sqrt{2n}}{\sqrt{V_j}}C'(\Gamma)\left( \frac{\sqrt{ V(\Gamma)}}{C'(\Gamma) \sqrt{n}} \pi_j - \frac{r}{C'(\Gamma) \sqrt{n}}\right) + \frac{r \sqrt{2}}{\sqrt{V_j}} - x\\
    &\geq \sqrt{2} \pi_j - \frac{r \sqrt{2}}{\sqrt{V(\Gamma)}} + \frac{r \sqrt{2}}{\sqrt{V(\Gamma)}} - x - \epsilon^{(1)}
\end{align*}
and 
\begin{align*}
    & R_{n/2}^j\left(  \frac{\sqrt{n}}{\sqrt{2V_j}}(C(\Gamma) - C(\Gamma_j)) + \frac{r \sqrt{2}}{\sqrt{V_j}} - x\right)\\
    &\leq R_{n/2}^j\left(  \frac{\sqrt{n}C'(\Gamma_j)}{\sqrt{2V_j}}\left (\Gamma - \Gamma_j\right ) + \frac{r \sqrt{2}}{\sqrt{V_j}} - x\right)\\
    &= R_{n/2}^j\left(  \frac{\sqrt{n}C'(\Gamma_j)}{\sqrt{2V_j}}\left (\frac{\sqrt{V(\Gamma)}}{C'(\Gamma) \sqrt{n}} \pi_j - \frac{r}{C'(\Gamma) \sqrt{n}}\right ) + \frac{r \sqrt{2}}{\sqrt{V_j}} - x\right)\\
    &\leq   \frac{\pi_j}{\sqrt{2}} - \frac{r}{\sqrt{2V(\Gamma)}} + \frac{r \sqrt{2}}{\sqrt{V(\Gamma)}} - R_{n/2}^j x  + \epsilon^{(1)}
\end{align*}
for sufficiently large $n$, since $R_{n/2}^j \to 1$ as $n \to \infty$.

Using the lower bound obtained in $(\ref{faraz})$ in $(\ref{lwbndint1})$, we have 
\begin{align}
    &\mathbb{P}\left( \frac{1}{\sqrt{\frac{n}{2}V_j}}\left [ \log \frac{W(Y^{n/2}|X^{n/2})}{Q_j^*(Y^{n/2})} - \frac{n}{2} [C(\Gamma_j)]_{n/2}\right] + \frac{1}{\sqrt{\frac{n}{2}V_j}}\left [ \log \frac{W(Y^{n/2+1:n}|X^{n/2+1:n})}{FW(Y^{n/2+1:n}|Y^{n/2})} - \frac{n}{2} C(\Gamma) \right ]  \leq d_{n,j}  \right) \notag \\
    &\leq \Phi\left( \pi_j \right) - \int_{d'_{n, j}}^{\infty} \phi(x) \left[\Phi\left( \frac{\pi_j}{\sqrt{2}} + \frac{\pi_j}{\sqrt{2}}  -x \right) -   \Phi\left(  \frac{\pi_j}{\sqrt{2}} + \frac{r}{\sqrt{2V(\Gamma)}}  - R_{n/2}^jx    \right)   \right]dx  + \mbox{} \notag \\
    & \quad \quad \quad \quad \quad \quad \quad  \frac{\underline{\kappa}_{n/2}^j}{\sqrt{n}} + 2\epsilon^{(1)} + 2\delta_{n/2} 
 +  \frac{(3 + R^j_{n/2})\underline{\kappa}^j_{n/2} + \overline{\kappa}_{n/2}}{\sqrt{n/2}} + \epsilon^{(2)} + \epsilon^{(3)} \label{stillnotdone1}
\end{align}
which gives us 
\begin{align}
    &\mathcal{E}_n \notag \\
    &\leq \sum_{j=1}^3 p_j \Phi(\pi_j) - \sum_{j=1}^3 p_j \int_{d'_{n,j}}^{\infty} \phi(x) \left[\Phi\left(\frac{\pi_j}{\sqrt{2}} + \frac{\pi_j}{\sqrt{2}} -x \right) -   \Phi\left(  \frac{\pi_j}{\sqrt{2}} + \frac{r}{\sqrt{2V(\Gamma)}}  - R_{n/2}^jx     \right)   \right]dx + \mbox{} \notag \\
    & \quad \frac{\underline{\kappa}_{n/2}^j}{\sqrt{n}} + 2\epsilon^{(1)} + 3\delta_{n/2} 
 +  \frac{(3 + R^j_{n/2})\underline{\kappa}^j_{n/2} + \overline{\kappa}_{n/2}}{\sqrt{n/2}} + \epsilon^{(2)} + \epsilon^{(3)} \notag \\
 &\stackrel{(a)}{\leq} \sum_{j=1}^3 p_j \Phi(\pi_j) - \sum_{j=1}^3 p_j \int_{\frac{\beta}{\sqrt{V(\Gamma)}}}^{\infty} \phi(x) \left[\Phi\left(\frac{\pi_j}{\sqrt{2}} + \frac{\pi_j}{\sqrt{2}} -x \right) -   \Phi\left(  \frac{\pi_j}{\sqrt{2}} + \frac{r}{\sqrt{2V(\Gamma)}}  - R_{n/2}^jx     \right)   \right]dx + \mbox{} \notag  \\
    & \quad \frac{\underline{\kappa}_{n/2}^j}{\sqrt{n}} + 3\epsilon^{(1)} + 3\delta_{n/2} 
 +  \frac{(3 + R^j_{n/2})\underline{\kappa}^j_{n/2} + \overline{\kappa}_{n/2}}{\sqrt{n/2}} + \epsilon^{(2)} + \epsilon^{(3)}  \notag \\
 &\stackrel{(b)}{\leq } \sum_{j=1}^3 p_j \Phi(\pi_j) - \sum_{j=1}^3 p_j \int_{\frac{\beta}{\sqrt{V(\Gamma)}}}^{\infty} \phi(x) \left[\Phi\left(\frac{\pi_j}{\sqrt{2}} + \frac{\pi_j}{\sqrt{2}} -x \right) -   \Phi\left(  \frac{\pi_j}{\sqrt{2}} + \frac{r}{\sqrt{2V(\Gamma)}}  - x     \right)   \right]dx + \mbox{} \notag  \\
    & \quad \frac{\underline{\kappa}_{n/2}^j}{\sqrt{n}} + 3\epsilon^{(1)} + 3\delta_{n/2} 
 +  \frac{(3 + R^j_{n/2})\underline{\kappa}^j_{n/2} + \overline{\kappa}_{n/2}}{\sqrt{n/2}} + \epsilon^{(2)} + \epsilon^{(3)}  + \frac{1}{2\pi} \max_{j=1,2,3} |1 - R_{n/2}^j| \notag \\
 &= \mathbb{E}[\Phi(\Pi)] - \int_{\frac{\beta}{\sqrt{V(\Gamma)}}}^{\infty} \phi(x) \left[ \mathbb{E}\left [ \Phi\left(\sqrt{2}\Pi -x \right) \right] -   \mathbb{E}\left[ \Phi\left(  \frac{\Pi}{\sqrt{2}} + \frac{r}{\sqrt{2V(\Gamma)}}  - x     \right)\right]   \right]dx + \mbox{} \notag  \\
    & \quad \frac{\underline{\kappa}_{n/2}^j}{\sqrt{n}} + 3\epsilon^{(1)} + 3\delta_{n/2} 
 +  \frac{(3 + R^j_{n/2})\underline{\kappa}^j_{n/2} + \overline{\kappa}_{n/2}}{\sqrt{n/2}} + \epsilon^{(2)} + \epsilon^{(3)}  + \frac{1}{2\pi} \max_{j=1,2,3} |1 - R_{n/2}^j| \notag \\
 &\stackrel{(c)}{\leq}  \inf_{P_{\Pi} \in \mathcal{F}_{\mathcal{K}(r)}} \left [ \mathbb{E}[\Phi(\Pi)] - \int_{\frac{\beta}{\sqrt{V(\Gamma)}}}^{\infty} \phi(x) \left[ \mathbb{E}\left [ \Phi\left(\sqrt{2}\Pi -x \right) \right] -   \mathbb{E}\left[ \Phi\left(  \frac{\Pi}{\sqrt{2}} + \frac{r}{\sqrt{2V(\Gamma)}}  - x     \right)\right]   \right]dx \right] + \mbox{} \notag  \\
    & \quad \frac{\underline{\kappa}_{n/2}^j}{\sqrt{n}} + 4\epsilon^{(1)} + 3\delta_{n/2} 
 +  \frac{(3 + R^j_{n/2})\underline{\kappa}^j_{n/2} + \overline{\kappa}_{n/2}}{\sqrt{n/2}} + \epsilon^{(2)} + \epsilon^{(3)}  + \frac{1}{2\pi} \max_{j=1,2,3} |1 - R_{n/2}^j| \label{gokuvsjiren}
\end{align}
for sufficiently large $n$.
Inequality $(a)$ holds since $d'_{n,j} \to \frac{\beta}{\sqrt{V(\Gamma)}}$ as $n \to \infty$ for all $j \in \{1,2,3 \}$. Inequality $(b)$ follows from the following derivation (recall that $\beta \geq 0$): 
\begin{align*}
    &\Bigg | \int_{\frac{\beta}{\sqrt{V(\Gamma)}}}^\infty \phi(x) \Phi\left(  \frac{\pi_j}{\sqrt{2}} + \frac{r}{\sqrt{2V(\Gamma)}}  - R_{n/2}^jx \right) dx - \int_{\frac{\beta}{\sqrt{V(\Gamma)}}}^\infty \phi(x) \Phi\left(  \frac{\pi_j}{\sqrt{2}} + \frac{r}{\sqrt{2V(\Gamma)}}  - x \right) dx \Bigg | \\
    &\leq  \int_{0}^\infty \phi(x) \Bigg |  \Phi\left(  \frac{\pi_j}{\sqrt{2}} + \frac{r}{\sqrt{2V(\Gamma)}}  - R_{n/2}^jx \right) - \Phi\left(  \frac{\pi_j}{\sqrt{2}} + \frac{r}{\sqrt{2V(\Gamma)}}  - x \right)\Bigg | dx\\
    &\leq \frac{1}{2\pi}|1-R_{n/2}^j|,
\end{align*}
where the last inequality above follows from a simple application of Taylor's Theorem. Inequality $(c)$ follows because $P_{\Pi}$ was an arbitrary element of $\mathcal{F}_{\mathcal{K}(r)}$ (recall $(\ref{FKr})$ and $(\ref{PPi})$).    

To continue from $(\ref{gokuvsjiren})$, we define a function
\begin{align}
    L_2(r, \beta) := \inf_{P_{\Pi} \in \mathcal{F}_{\mathcal{K}(r)}} \left [ \mathbb{E}[\Phi(\Pi)] - \int_{\frac{\beta}{\sqrt{V(\Gamma)}}}^{\infty} \phi(x) \left[ \mathbb{E}\left [ \Phi\left(\sqrt{2}\Pi -x \right) \right] -   \mathbb{E}\left[ \Phi\left(  \frac{\Pi}{\sqrt{2}} + \frac{r}{\sqrt{2V(\Gamma)}}  - x     \right)\right]   \right]dx \right]. \label{Ldef6}
\end{align}

Define $\mathcal{S}_{\mathcal{K}(r)} \subset \mathcal{F}_{\mathcal{K}(r)}$ to be the set of minimizers in $(\ref{FKr})$. 

\begin{proposition}
For any real number $r$, if there exists $P_{\Pi} \in \mathcal{S}_{\mathcal{K}(r)}$ such that $\text{Var}(\Pi) > 0$ for $\Pi \sim P_{\Pi}$, then for sufficiently large $\beta$, we have $$L_2(r, \beta) < \mathcal{K}\left(\frac{r}{\sqrt{V(\Gamma) }}, \frac{C'(\Gamma)^2 V}{V(\Gamma)} \right).$$
    \label{g21nc}
\end{proposition}
\begin{IEEEproof}
Let $P_{\Pi} \in \mathcal{S}_{\mathcal{K}(r)}$ be such that $\text{Var}(\Pi) > 0$. Then
\begin{align}
    L_2(r, \beta) &\leq \mathcal{K}\left(\frac{r}{\sqrt{V(\Gamma) }}, \frac{C'(\Gamma)^2 V}{V(\Gamma)} \right) -  \int_{\frac{\beta}{\sqrt{V(\Gamma)}}}^{\infty} \phi(x) \left[ \mathbb{E}\left [ \Phi\left(\sqrt{2}\Pi -x \right) \right] -   \mathbb{E}\left[ \Phi\left(  \frac{\Pi}{\sqrt{2}} + \frac{r}{\sqrt{2V(\Gamma)}}  - x     \right)\right]   \right]dx.  \label{ewvm}
\end{align}
It then suffices to show that for sufficiently large $\beta$, the integral in $(\ref{ewvm})$ is strictly positive. But this holds by Lemma \ref{aaronesssuplemma} since 
     $$\big | \big |\sqrt{2} \Pi \big | \big |_\infty > \Bigg| \Bigg |\frac{\Pi}{\sqrt{2}} + \frac{r}{\sqrt{2 V(\Gamma)}} \Bigg | \Bigg |_\infty.$$     
\end{IEEEproof}

\begin{proposition}
For any real number $r$, suppose there exists $P_{\Pi} \in \mathcal{S}_{\mathcal{K}(r)}$ such that $\text{Var}(\Pi) > 0$ for $\Pi \sim P_{\Pi}$. Then there exist $\beta$ and $r' > r$ such that 
\begin{align}
    L_2(r', \beta) < \mathcal{K}\left(\frac{r}{\sqrt{V(\Gamma) }}, \frac{C'(\Gamma)^2 V}{V(\Gamma)} \right).
\end{align}
    \label{fwn}
\end{proposition}

\begin{IEEEproof}
     From Proposition \ref{g21nc},  fix $\beta > 0$ such that $L_2(r, \beta) < \mathcal{K}\left(\frac{r}{\sqrt{V(\Gamma) }}, \frac{C'(\Gamma)^2 V}{V(\Gamma)} \right)$. Let $P_{\Pi} \in \mathcal{F}_{\mathcal{K}(r)}$ be a distribution that is within 
     $$\frac{1}{2}\left(\mathcal{K}\left(\frac{r}{\sqrt{V(\Gamma) }}, \frac{C'(\Gamma)^2 V}{V(\Gamma)} \right) - L_2(r, \beta) \right)$$
    of the infimum in $(\ref{Ldef6})$ so that for $\Pi \sim P_{\Pi}$,
    \begin{align}
        &\,\,\mathbb{E}[\Phi(\Pi)] - \int_{\frac{\beta}{\sqrt{V(\Gamma)}}}^{\infty} \phi(x) \left[ \mathbb{E}\left [ \Phi\left(\sqrt{2}\Pi -x \right) \right] -   \mathbb{E}\left[ \Phi\left(  \frac{\Pi}{\sqrt{2}} + \frac{r}{\sqrt{2V(\Gamma)}}  - x     \right)\right]   \right]dx \notag \\
        &\leq L_2(r, \beta) + \frac{1}{2}\left(\mathcal{K}\left(\frac{r}{\sqrt{V(\Gamma) }}, \frac{C'(\Gamma)^2 V}{V(\Gamma)} \right) - L_2(r, \beta) \right) \notag \\
        &< \mathcal{K}\left(\frac{r}{\sqrt{V(\Gamma) }}, \frac{C'(\Gamma)^2 V}{V(\Gamma)} \right).
    \label{unbelievablegenius}
    \end{align}
    For $r' > r$, consider the new random variable $\Pi' \sim P_{\Pi'}$ such that  
$$\Pi' = \Pi - \frac{r}{\sqrt{V(\Gamma)}} + \frac{r'}{\sqrt{V(\Gamma)}}.$$
Clearly, $P_{\Pi'} \in \mathcal{F}_{\mathcal{K}(r')}$ so that 
\begin{align*}
    L_2(r', \beta) &\leq \mathbb{E}[\Phi(\Pi')] - \int_{\frac{\beta}{\sqrt{V(\Gamma)}}}^{\infty} \phi(x) \left[ \mathbb{E}\left [ \Phi\left(\sqrt{2}\Pi' -x \right) \right] -   \mathbb{E}\left[ \Phi\left(  \frac{\Pi'}{\sqrt{2}} + \frac{r'}{\sqrt{2V(\Gamma)}}  - x     \right)\right]   \right]dx.
\end{align*}
The expression on the right-hand side of the above inequality is evidently continuous in $r'$ at $r' = r$; thus, from $(\ref{unbelievablegenius})$, the proposition follows.

\end{IEEEproof}

Recall that $r^* > \sqrt{V(\Gamma)} \Phi^{-1}(\epsilon)$ which implies that the hypothesis of Propositions \ref{g21nc} and \ref{fwn} is satisfied for $r = r^*$. Using Proposition \ref{fwn} in $(\ref{gokuvsjiren})$, we obtain 
\begin{align*}
    \mathcal{E}_n &< \epsilon + \frac{\underline{\kappa}_{n/2}^j}{\sqrt{n}} + 4\epsilon^{(1)} + 2\delta_{n/2} 
 +  \frac{(3 + R^j_{n/2})\underline{\kappa}^j_{n/2} + \overline{\kappa}_{n/2}}{\sqrt{n/2}} + \epsilon^{(2)} + \epsilon^{(3)}  + \frac{1}{2\pi} \max_{j=1,2,3} |1 - R_{n/2}^j|
\end{align*}
for some $r' > r^*$. Choosing $\epsilon^{(1)}, \epsilon^{(2)}$ and $\epsilon^{(3)}$ small enough, we obtain $\limsup_{n \to \infty} \mathcal{E}_n < \epsilon$, thus establishing an achievable SOCR of $r' > r^*$ via timid/bold coding. 

Feedback also improves the average error probability of an $(n, R, \Gamma, V)$ code. Fix $R = C(\Gamma) + \frac{r}{\sqrt{n}}$. From Lemma \ref{aaron'slemma}, we have for $\theta = 1/n^{\vartheta}$ for $1 > \vartheta > 1/2$,
\begin{align}
\bar{P}_{\text{e,fb}}\left(n,C(\Gamma) + \frac{r}{\sqrt{n}},  \Gamma, V\right) \leq (F \circ W) \left(\frac{1}{n} \log \frac{W(Y^n|X^n)}{PW(Y^n)} \leq C(\Gamma) + \frac{r}{\sqrt{n}} + \frac{1}{n^\vartheta} \right) + e^{-n^{1-\vartheta}}.  \label{not55playing}
\end{align}
From $(\ref{gokuvsjiren})$, we have for any $\tilde{r}$ and some $\tilde{\epsilon} > 0$ which can be made arbitrarily small for sufficiently large $n$, 
\begin{align}
    &(F \circ W)\left(\frac{1}{n} \log \frac{W(Y^n|X^n)}{PW(Y^n)} \leq C(\Gamma) + \frac{\tilde{r}}{\sqrt{n}} \right) \notag \\
    &\leq L_2(\tilde{r}, \beta) + \tilde{\epsilon}, \label{tilderkitang}
\end{align}
where the function $L_2$ is defined in $(\ref{Ldef6})$. Let $\tilde{r} > r$ be arbitrary. Then from $(\ref{not55playing})$ and $(\ref{tilderkitang})$, we obtain 
\begin{align}
&\bar{P}_{\text{e,fb}}\left(n,C(\Gamma) + \frac{r}{\sqrt{n}},  \Gamma, V\right) \notag \\
&\leq L_2(\tilde{r}, \beta) + 2\tilde{\epsilon} \notag 
\end{align}
for sufficiently large $n$. Then taking the limit supremum as $n \to \infty$ followed by letting $\tilde{\epsilon} \to 0$ and $\tilde{r} \to r$, we obtain 
\begin{align}
&\limsup_{n \to \infty} \bar{P}_{\text{e,fb}}\left(n,C(\Gamma) + \frac{r}{\sqrt{n}},  \Gamma, V\right) \notag \\
&\leq L_2(r, \beta), \label{digimon1}
\end{align}
where the last inequality above follows from the fact that $L_2(r, \beta)$ is upper semicontinuous in $r$ for any fixed $\beta$.  

Recall that $\beta$ is a free parameter, and it is possible to choose $\beta$ to obtain a strict improvement over the optimal error probability in the non-feedback case. Specifically, since
\begin{align*}
    \mathcal{K}\left(\frac{r}{\sqrt{V(\Gamma) }}, \frac{C'(\Gamma)^2 V}{V(\Gamma)} \right) < \Phi\left(\frac{r}{\sqrt{V(\Gamma)}} \right)
\end{align*}
for $V > 0$ and $V(\Gamma) > 0$ (a result which follows from \cite[Theorem 1]{mahmood2024improvedchannelcodingperformance}), there exists $P_{\Pi} \in \mathcal{S}_{\mathcal{K}(r)}$ such that $\text{Var}(\Pi) > 0$ for $\Pi \sim P_{\Pi}$. Hence, from Proposition \ref{g21nc},  
we have 
\begin{align}
    &L_2(r, \beta) <\mathcal{K}\left(\frac{r}{\sqrt{V(\Gamma) }}, \frac{C'(\Gamma)^2 V}{V(\Gamma)} \right) \label{digimon3}
\end{align}
for some choice of $\beta$. From $(\ref{digimon1})$ and $(\ref{digimon3})$, we obtain 
\begin{align*}
    \limsup_{n \to \infty} \bar{P}_{\text{e,fb}}\left(n,C(\Gamma) + \frac{r}{\sqrt{n}},  \Gamma, V\right) &< \mathcal{K}\left(\frac{r}{\sqrt{V(\Gamma) }}, \frac{C'(\Gamma)^2 V}{V(\Gamma)} \right). 
\end{align*}

\appendices

\section{Statement and Proof of Lemma \ref{lemcon3t} \label{lemcon3t_proof}}

\begin{lemma}
    Fix any $\Gamma \in (\Gamma_0, \Gamma^*)$. Let $P^* = P^*[\Gamma]$ be a capacity-cost-achieving distribution for $C(\Gamma)$. If $P^*$ is unique, then for every $\epsilon > 0$, there exists a $\delta > 0$ such that $|\Gamma' - \Gamma| < \delta$ implies $$\sup_{P \in \mathcal{S}_{C(\Gamma')}}||P - P^*[\Gamma]||_1 < \epsilon,$$ where $\mathcal{S}_{C(\Gamma')}$ is the set of all capacity-cost-achieving distributions for $C(\Gamma')$.  
    \label{lemcon3t}
\end{lemma}
\textit{Proof:} 

Let $\Gamma_n$ be a sequence converging to $\Gamma \in (\Gamma_0, \Gamma^*)$. Let $\mathcal{S}_{C(\Gamma_n)}$ be the set of all capacity-cost-achieving distributions for $C(\Gamma_n)$. For any $\eta > 0$, let $P^*[\Gamma_n] \in \mathcal{S}_{C(\Gamma_n)} $ be such that 
    \begin{align}
        ||P^*[\Gamma_n] - P^*[\Gamma]||_1 \geq \sup_{P \in \mathcal{S}_{C(\Gamma_n)}} ||P - P^*[\Gamma]||_1 - \eta. \label{stopalready}
    \end{align}
    Let $P^*_\infty$ be any subsequential limit of the sequence $P^*[\Gamma_n]$ w.r.t. the $||\cdot||_1$ metric. Mathematically, this implies that there exists a subsequence $n_l$ such that 
    \begin{align*}
        \lim_{l \to \infty} ||P^*_\infty - P^*[\Gamma_{n_l}]||_2 = 0. 
    \end{align*}
    It suffices to show that $P^*_\infty = P^*[\Gamma]$. Clearly, we have 
    \begin{align*}
        \sum_{a \in \mathcal{A}} P^*[\Gamma_{n_l}](a) c(a) \leq \Gamma_{n_l}.
    \end{align*}
    Taking the limit as $l$ goes to infinity gives 
    \begin{align*}
        \sum_{a \in \mathcal{A}} P^*_\infty(a) c(a) \leq \Gamma. 
    \end{align*}
    We already know from the optimality of $P^*[\Gamma]$ that 
\begin{align}
    I(P^*[\Gamma], W) \geq I(P^*_\infty, W).  \label{patanahi1}
\end{align}
Furthermore, we have 
\begin{align*}
    I(P^*[\Gamma], W) &= C(\Gamma)\\
    &\leq C(\Gamma_{n_l}) + C'(\Gamma_{n_l})(\Gamma - \Gamma_{n_l})\\
    &= I(P^*[\Gamma_{n_l}], W) + C'(\Gamma_{n_l})(\Gamma - \Gamma_{n_l}). 
\end{align*}
Taking the limit as $l$ goes to infinity gives us 
\begin{align}
    I(P^*[\Gamma], W) \leq \lim_{l \to \infty} I(P^*[\Gamma_{n_l}], W) \notag \\
    I(P^*[\Gamma], W) \leq I(P^*_\infty, W) \label{patanahi2} 
\end{align}
since mutual information $I(p,w)$ is continuous in $p$. Combining $(\ref{patanahi1})$ and $(\ref{patanahi2})$, we have $P^*_\infty = P^*[\Gamma]$ since $P^*[\Gamma]$ is unique. Since $P^*_\infty$ was an arbitrary subsequential limit of $P^*[\Gamma_n]$, we have that every subsequential limit of $P^*[\Gamma_n]$ is equal to $P^*[\Gamma]$. Furthermore, since $P^*[\Gamma_n]$ is a bounded sequence with respect to $l^1$ norm, it follows that $$\lim_{n \to \infty} ||P^*[\Gamma_n] - P^*[\Gamma]||_1 = 0.$$ 
Using this result in $(\ref{stopalready})$, we obtain 
\begin{align*}
    \lim_{n \to \infty} \sup_{P \in \mathcal{S}_{C(\Gamma_n)}} ||P - P^*[\Gamma]||_1 \leq \eta.
\end{align*}
Since $\eta > 0$ was arbitrary, the result of the lemma follows.

\section{Proof of Lemma \ref{Kfunctionproperties} \label{Kfuncpropertiesproof}}

To prove the first part of the lemma, we assume that in the infimum in $(\ref{jzz})$, the random variable $\Pi$ satisfies the inequality constraint $\mathbb{E}[\Pi] \geq r$ with equality. This is without loss of generality since the mean of any candidate probability distribution in the infimum in $(\ref{jzz})$ can be shifted to the left while keeping the variance unchanged and causing no increase in the value of the objective since $\Phi(\cdot)$ is an increasing function. Consider the mapping $\zeta : \mathbb{R} \to \mathcal{S} \subset  \mathbb{R}^3$ given by 
\begin{align}
    \zeta(\pi) = \left( \Phi(\pi), \pi, (\pi - r)^2 \right). \label{zetamapping}
\end{align}
For any $P_{\Pi}$ in the feasible set of the infimum in $(\ref{jzz})$, 
\begin{align*}
    \mathbb{E}\left [ \zeta(\Pi) \right ] = \left( \mathbb{E}\left [ \Phi(\Pi)\right] , \mathbb{E} \left [ \Pi\right], \mathbb{E}\left [ (\Pi - r)^2\right]  \right)
\end{align*}
is in the convex closure of $\mathcal{S} \subset \mathbb{R}^3$; hence, by Caratheodory's Theorem \cite[Ch. 2.1]{car1}, $\mathbb{E}\left [ \zeta(\Pi) \right ] $ can be expressed as a convex combination of at most $4$ points in $\mathcal{S}$, i.e., 
\begin{align*}
    \mathbb{E}\left [ \zeta(\Pi) \right ] = \begin{bmatrix}
    p_1\Phi(x_1) +p_2\Phi(x_2) +p_3\Phi(x_3) +p_4\Phi(x_4) \\
    p_1 \,x_1 +p_2 \,x_2 +p_3 \,x_3 +p_4 \,x_4 \\
    p_1(x_1 - r)^2 +p_2(x_2 - r)^2 +p_3(x_3 - r)^2 +p_4(x_4 - r)^2  
    \end{bmatrix}.  
\end{align*}
Therefore, $\mathcal{K}(r, V)$ can equivalently be written as the
infimum of 
\begin{align*}
    p_1\Phi(x_1) +p_2\Phi(x_2) +p_3\Phi(x_3) +p_4\Phi(x_4)
\end{align*}
w.r.t. $x_1, x_2, x_3, x_4, p_1, p_2, p_3, p_4$
subject to 
\begin{align}
    &p_1 \geq 0, p_2 \geq 0, p_3 \geq 0, p_4 \geq 0 \label{prob1},\\
    &p_1 + p_2 + p_3 + p_4 = 1 \label{prob2},\\
    &p_1 (x_1 - r) +p_2 (x_2 - r ) +p_3 (x_3 - r) +p_4 (x_4-r) = 0, \label{meancons}\\
    &p_1(x_1 - r)^2 +p_2(x_2 - r)^2 +p_3(x_3 - r)^2 +p_4(x_4 - r)^2 \leq V. \label{varcons}
\end{align}
Now we show that the infimum above is attained. Fix any $\epsilon > 0$. Let $\mathcal{K}^*(r, V)$ denote the infimum. Let \\
$(x_1^n,x_2^n,x_3^n,x_4^n,p_1^n,p_2^n,p_3^n,p_4^n)$ be a sequence such that 
\begin{align*}
     p_1^n\Phi(x_1^n) +p_2^n\Phi(x_2^n) +p_3^n\Phi(x_3^n) +p_4^n\Phi(x_4^n)  \to \mathcal{K}^*(r,V).
\end{align*}
Consider a subsequence (indexed by $m \in \mathbb{N}$ for simplicity) such that for $j \in \{1,2,3,4 \}$, $x_j^m \to x_j^* \in [-\infty, \infty]$ and $p_j^m \to p_j^* \in [0, 1]$. From $(\ref{varcons})$, we have 
\begin{align}
    p_j^m(x_j^m - r)^2 \leq V \label{chatgpte}
\end{align}
for all $j \in \{1,2,3,4 \}$. Let $\mathcal{J} \subset \{1,2,3,4 \}$ be such that $x_j^* \in \mathbb{R}$ for $j \in \mathcal{J}$, and let $\mathcal{J}^c = \{ 1,2,3,4\} \setminus \mathcal{J}$. From $(\ref{chatgpte})$, it follows that for $j \in \mathcal{J}^c$, $p_j^m \to p_j^* = 0$ and $p_j^m (x_j^m - r) \to 0$ since  
\begin{align*}
    |p_j^m (x_j^m - r)| = \frac{p_j^m (x_j^m - r)^2}{|x_j^m - r|} \leq \frac{V}{|x_j^m - r|} \to 0
\end{align*}
as $x_j^m \to \pm \infty$. Define 
\begin{align*}
    \tilde{x}_j = \begin{cases}
        x_j^*  & \text{ if } j \in \mathcal{J},\\
        r & \text{ if } j \in \mathcal{J}^c.
    \end{cases}
\end{align*}
We now show that the point $(\tilde{x}_1,\tilde{x}_2,\tilde{x}_3,\tilde{x}_4, p_1^*,p_2^*,p_3^*,p_4^*)$ attains the infimum $\mathcal{K}^*(r, V)$ and satisfies the constraints $(\ref{prob1}) - (\ref{varcons})$. It is easy to see that $p_1^*,p_2^*,p_3^*$ and $p_4^*$ satisfy $(\ref{prob1})$ and $(\ref{prob2})$. For $(\ref{meancons})$, we have 
\begin{align*}
    &\sum_{j=1}^4  p_j^* (\tilde{x}_j - r)\\ 
    &= \sum_{j \in \mathcal{J}}  p_j^* (\tilde{x}_j - r)\\
    &= \sum_{j \in \mathcal{J}}  p_j^* (x_j^* - r)\\
    &= \lim_{m \to \infty} \sum_{j \in \mathcal{J}} p_j^m(x_j^m - r)\\
    &= \lim_{m \to \infty} \sum_{j=1}^4 p_j^m(x_j^m - r) = 0.
\end{align*}
For $(\ref{varcons})$, we have 
\begin{align*}
    &\sum_{j=1}^4  p_j^* (\tilde{x}_j - r)^2\\
    &= \sum_{j \in \mathcal{J}}  p_j^* (\tilde{x}_j - r)^2\\
    &= \sum_{j \in \mathcal{J}}  p_j^* (x_j^* - r)^2\\
    &= \lim_{m \to \infty} \sum_{j \in \mathcal{J}}  p_j^m (x_j^m - r)^2\\
    &\leq \limsup_{m \to \infty} \sum_{j=1}^4  p_j^m (x_j^m - r)^2\\
    &\leq V.
\end{align*}
Finally, 
\begin{align*}
    &\sum_{j=1}^4 p_j^* \Phi(\tilde{x}_j)\\
    &= \sum_{j \in \mathcal{J}} p_j^* \Phi(\tilde{x}_j)\\
    &= \sum_{j \in \mathcal{J}} p_j^* \Phi(x_j^*)\\
    &= \lim_{m \to \infty} \sum_{j \in \mathcal{J}} p_j^m \Phi(x_j^m)\\
    &= \lim_{m \to \infty} \sum_{j=1}^4 p_j^m \Phi(x_j^m)\\
    &= \mathcal{K}^*(r, V).
\end{align*}
This concludes the proof that the infimum is attained by the point $(\tilde{x}_1,\tilde{x}_2,\tilde{x}_3,\tilde{x}_4, p_1^*,p_2^*,p_3^*,p_4^*)$. So far, we have shown that the minimizer is a probability distribution with at most 4 point masses. We denote this probability distribution by $\tilde{p}$, i.e., $\tilde{p}(\tilde{x}_i) = p_i^*$ for $j \in \{1,2,3,4 \}$. For the closed interval $$I^* = \left [ \min_j \tilde{x}_j, \max_j \tilde{x}_j \right ], $$ we now redefine the mapping $\zeta(\cdot)$ in $(\ref{zetamapping})$ to have $I^*$ as the new domain, i.e., $\zeta : I^* \to \mathcal{S}^* \subset \mathbb{R}^3$ and 
\begin{align}
    \zeta(\pi) = \left( \Phi(\pi), \pi, (\pi - r)^2 \right). \label{ze2tamapping}
\end{align}
Using $\mathbb{E}_{\tilde{p}}$ to denote expectation w.r.t. $\tilde{p}$, we have that 
\begin{align*}
    \mathbb{E}_{\tilde{p}}\left [ \zeta(\Pi) \right ] = \left(\mathbb{E}_{\tilde{p}} \left [  \Phi(\Pi)\right ] , \mathbb{E}_{\tilde{p}}\left [ \Pi\right] , \mathbb{E}_{\tilde{p}}\left [ (\Pi - r)^2\right] \right)
\end{align*}
is in the convex closure of $\mathcal{S}^*$. Since $\mathcal{S}^*$ is a connected, compact subset of $\mathbb{R}^3$, by Carathéodory–Fenchel's theorem \cite[Appendix A]{el_gamal_kim_2011}, $\mathbb{E}_{\tilde{p}}\left [ \zeta(\Pi) \right ]$ can be written as the convex combination of at most $3$ points in $\mathcal{S}^*$. This proves that the minimizer is a probability distribution with at most 3 point masses, thus concluding the proof of the first part of the lemma.

The proof of the second part of the lemma is straightforward. Let $r_2 < r_1$ and $\delta = r_1 - r_2$. Let   
\begin{align*}
    \mathcal{K}(r_1, V) = p_1^* \Phi(x_1^*) + p_2^* \Phi(x_2^*) + p_3^* \Phi(x_3^*), 
\end{align*}
where the probability distribution $(p_1^*, p_2^*, p_3^*)$ with point masses at $x_1^*, x_2^*$ and $x_3^*$, respectively, achieves the minimum, and $\sum_j p_j^* x_j^* = r_1$. Shifting each of the point masses to the left by $\delta$ while keeping the probabilities unchanged strictly decreases the objective since $\Phi(\cdot)$ is a strictly increasing function. Hence, 
\begin{align*}
    \mathcal{K}(r_1, V) &= p_1^* \Phi(x_1^*) + p_2^* \Phi(x_2^*) + p_3^* \Phi(x_3^*)\\
    &> p_1^* \Phi(x_1^* - \delta) + p_2^* \Phi(x_2^* - \delta) + p_3^* \Phi(x_3^* - \delta)\\
    &\geq \mathcal{K}(r_2, V). 
\end{align*}
The last inequality above follows because the shifted distribution has mean $r_2$. 

We now prove the third part of the lemma. Fix any $\epsilon > 0$ and a point $(r, V)$ with $V > 0$. Let $\delta > 0$ be such that $V - \delta > 0$. For all $(r',V')$ satisfying 
\begin{align}
    ||(r, V) - (r', V')||_2 &= \sqrt{(r - r')^2 + (V - V')^2} \leq \delta, \label{lilicop}
\end{align}
we have $\mathcal{K}(r', V') \leq \mathcal{K}(r + \delta, V - \delta).$ Let $\tilde{p}$ denote a probability distribution which is a minimizer in $\mathcal{K}(r, V)$ so that the random variable $X \sim \tilde{p}$ takes at most 3 possible values, has mean equal to $r$ and has variance upper bounded by $V$. Consider a random variable 
\begin{align*}
    X_\delta^{(u)} = \sqrt{\frac{V - \delta}{V}} X  + r + \delta - r \sqrt{\frac{V - \delta}{V}} 
\end{align*}
so that $\mathbb{E}[X_\delta^{(u)}] = r + \delta$ and $\text{Var}(X_\delta^{(u)}) \leq V - \delta$. Then we have 
\begin{align}
    \mathcal{K}(r', V') &\leq \mathcal{K}(r + \delta, V - \delta) \notag \\
    &\leq \mathbb{E}[\Phi(X_\delta^{(u)})]\notag \\
    &\leq \mathcal{K}(r, V) + \epsilon \label{cont1}
\end{align}
for small enough $\delta$ because $\mathbb{E}[\Phi(X_\delta^{(u)})]$, viewed as a function of $\delta$, is continuous in $\delta$; hence, $\mathbb{E}[\Phi(X_\delta^{(u)})] \to \mathbb{E}[\Phi(X)]$ as $\delta \to 0$. 

On the other hand, we have $\mathcal{K}(r', V') \geq \mathcal{K}(r - \delta, V + \delta)$ for all $(r',V')$ satisfying $(\ref{lilicop})$. Let  $X_\delta$ be a minimizer of $\mathcal{K}(r - \delta, V + \delta)$ so that $\mathbb{E}[X_\delta] = r - \delta$, $\text{Var}(X_\delta) \leq V + \delta$ and $X_\delta$ takes at most 3 possible values. Define 
\begin{align*}
    Y_\delta = \sqrt{\frac{V}{V  +\delta}} X_\delta + r - r \sqrt{\frac{V}{V + \delta}} + \delta \sqrt{\frac{V}{V + \delta}}
\end{align*}
so that $\mathbb{E}[Y_\delta] = r$ and $\text{Var}(Y_\delta) \leq V$. We then have 
\begin{align}
    \mathcal{K}(r, V) &\leq \mathbb{E}[\Phi(Y_\delta)] \notag \\
    &= \mathbb{E} \left [ \Phi\left(\sqrt{\frac{V}{V  +\delta}} X_\delta + r - r \sqrt{\frac{V}{V + \delta}} + \delta \sqrt{\frac{V}{V + \delta}} \right) \right ]\notag \\
    &= \mathbb{E} \left [ \Phi\left( X_\delta + \sqrt{\frac{V}{V  +\delta}} X_\delta + r - r \sqrt{\frac{V}{V + \delta}} + \delta \sqrt{\frac{V}{V + \delta}} 
 - X_\delta\right) \right ]\notag \\
 &\stackrel{(a)}{=} \mathbb{E} \left [ \Phi(X_\delta) + \phi\left( \tilde{X}_\delta \right ) \left(  \sqrt{\frac{V}{V  +\delta}} X_\delta + r - r \sqrt{\frac{V}{V + \delta}} + \delta \sqrt{\frac{V}{V + \delta}} 
 - X_\delta\right) \right ]\notag \\
 &= \mathbb{E} \left [ \Phi(X_\delta) \right ]  + \mathbb{E}\left [ \phi\left( \tilde{X}_\delta \right ) \left(  \sqrt{\frac{V}{V  +\delta}} X_\delta + r - r \sqrt{\frac{V}{V + \delta}} + \delta \sqrt{\frac{V}{V + \delta}} 
 - X_\delta\right) \right ]\notag \\
 &= \mathbb{E} \left [ \Phi(X_\delta) \right ]  + \mathbb{E}\left [ \phi\left( \tilde{X}_\delta \right ) \left( X_\delta \left(  \sqrt{\frac{V}{V  +\delta}} -1\right)  
 \right) \right ] + \left( r - r \sqrt{\frac{V}{V + \delta}} + \delta \sqrt{\frac{V}{V + \delta}} \right)\mathbb{E}\left [ \phi(\tilde{X}_\delta)  \right ]\notag \\
 &= \mathbb{E} \left [ \Phi(X_\delta) \right ]  + \left(  \sqrt{\frac{V}{V  +\delta}} -1\right)\mathbb{E}\left [ \phi\left( \tilde{X}_\delta \right )  X_\delta   
  \right ] + \left( r - r \sqrt{\frac{V}{V + \delta}} + \delta \sqrt{\frac{V}{V + \delta}} \right)\mathbb{E}\left [ \phi(\tilde{X}_\delta)  \right ]\notag \\
  &\stackrel{(b)}{\leq} \mathbb{E} \left [ \Phi(X_\delta) \right ]  + \Bigg |  \sqrt{\frac{V}{V  +\delta}} -1\Bigg | \sqrt{\mathbb{E}\left [ \phi(\tilde{X}_\delta)^2 \right ] \mathbb{E}\left [X_\delta^2 \right ] } + \left( r - r \sqrt{\frac{V}{V + \delta}} + \delta \sqrt{\frac{V}{V + \delta}} \right)\mathbb{E}\left [ \phi(\tilde{X}_\delta)  \right ]\notag \\
  &\leq \mathbb{E} \left [ \Phi(X_\delta) \right ]  + \Bigg |  \sqrt{\frac{V}{V  +\delta}} -1\Bigg | \sqrt{\frac{1}{2 \pi} \left( V + \delta + (r - \delta)^2\right) } + \Bigg |  r - r \sqrt{\frac{V}{V + \delta}} + \delta \sqrt{\frac{V}{V + \delta}} \Bigg | \frac{1}{\sqrt{2 \pi}}\notag \\
  &\stackrel{(c)}{\leq} \mathbb{E} \left [ \Phi(X_\delta) \right ] + \epsilon\notag \\
  &= \mathcal{K}(r - \delta, V + \delta) + \epsilon \notag \\
  &\leq \mathcal{K}(r', V') + \epsilon. \label{cont2}
\end{align}
In equality $(a)$, we used Taylor's Theorem where $\tilde{X}_\delta$ is some random variable which takes values between $X_\delta$ and $\sqrt{\frac{V}{V  +\delta}} X_\delta + r - r \sqrt{\frac{V}{V + \delta}} + \delta \sqrt{\frac{V}{V + \delta}}$, almost surely. In inequality $(b)$, we used Cauchy-Schwarz inequality\footnote{$|\mathbb{E}[XY]|\leq \sqrt{\mathbb{E}[X^2]\mathbb{E}[Y^2] }$ for any two random variables $X$ and $Y$ with finite second-order moments. }. Inequality $(c)$ follows for small enough $\delta$ because 
$$f(\delta) = \Bigg |  \sqrt{\frac{V}{V  +\delta}} -1\Bigg | \sqrt{\frac{1}{2 \pi} \left( V + \delta + (r - \delta)^2\right) } + \Bigg |  r - r \sqrt{\frac{V}{V + \delta}} + \delta \sqrt{\frac{V}{V + \delta}} \Bigg | \frac{1}{\sqrt{2 \pi}},$$
viewed as a function of $\delta$, is continuous in $\delta$; hence $f(\delta) \to 0$ as $\delta \to 0$.

Combining $(\ref{cont1})$ and $(\ref{cont2})$, we have that there exists a small enough $\delta$ such that $|\mathcal{K}(r', V') - \mathcal{K}(r, V)| \leq \epsilon$ for all $(r', V')$ satisfying $(\ref{lilicop})$.

\section{Proof of Lemma \ref{fqq} \label{fqq_proof}}

Let $\Pi$ denote a generic random variable with $\mathbb{E}[\Pi] \geq r'$ and $\text{Var}(\Pi) \leq V$, and let $P_{\Pi}$ denote its probability distribution. Let $\epsilon < \epsilon' < 1$. If $r' =  \Phi^{-1}(\epsilon') + \sqrt{\frac{2V\epsilon'}{\epsilon' - \epsilon}},$ then
    \begin{align*}
        &\mathbb{E}[\Phi(\Pi)]\\
        &= \mathbb{P}\left(\Pi > \Phi^{-1}( \epsilon') \right) \mathbb{E}\left [\Phi(\Pi) | \Pi > \Phi^{-1}( \epsilon') \right ]  + \mathbb{P}\left(\Pi \leq \Phi^{-1}(\epsilon') \right) \mathbb{E}\left [\Phi(\Pi) | \Pi \leq \Phi^{-1}( \epsilon') \right ]  \\
        &> \epsilon' \, \mathbb{P}\left(\Pi > \Phi^{-1}( \epsilon') \right)\\
        &=  \epsilon' \, \mathbb{P}\left(\Pi - \mathbb{E}[\Pi] > \Phi^{-1}( \epsilon') - \mathbb{E}[\Pi] \right)\\
        &= \epsilon' \left(1 -  \mathbb{P}\left(\Pi - \mathbb{E}[\Pi] \leq \Phi^{-1}( \epsilon') - \mathbb{E}[\Pi] \right)\right)\\
        &= \epsilon' \left(1 -  \mathbb{P}\left(\mathbb{E}[\Pi] - \Pi \geq  \mathbb{E}[\Pi] - \Phi^{-1}( \epsilon') \right)\right)\\
        &\geq \epsilon' \left(1 -  \mathbb{P}\left( |\Pi - 
 \mathbb{E}[\Pi] | \geq  \mathbb{E}[\Pi] - \Phi^{-1}( \epsilon') \right) \right)\\
        &\geq \epsilon' \left(1 -  \frac{\text{Var}(\Pi)}{(\mathbb{E}[\Pi] - \Phi^{-1}( \epsilon') )^2}\right)\\
        &> \epsilon.
    \end{align*}
Hence, $\mathcal{K}\left (\Phi^{-1}(\epsilon') + \sqrt{\frac{2V\epsilon'}{\epsilon' - \epsilon}}, V\right ) > \epsilon.$ Furthermore, $K\left(\Phi^{-1}(\epsilon/2 ), V \right) < \epsilon$ which can be seen by choosing the random variable in the infimum in $(\ref{jzz})$ to be equal to $ \Phi^{-1}(\epsilon/2)$ almost surely. Since $\mathcal{K}(r,V)$ is a strictly increasing function of $r$, we showed that the feasible set of the supremum in $(\ref{infr})$ is nonempty and has an upper bound; hence, the supremum, which we now denote by $r^*$, exists. Showing that the supremum is a maximum is straightforward given that the function $\mathcal{K}(r, V)$ is continuous. Lastly, the property $\mathcal{K}(r^*, V) = \epsilon$ follows from the fact that $\mathcal{K}(r, V)$ is continuous and strictly increasing in $r$.

\section{Proof of Lemma \ref{mostgenconv} \label{mostgenconvproof}}

Let $(f, g)$ be a random $(n, R )$ code with average error probability at most $\epsilon$. We will denote this code by $\mathscr{C}$ and its average error probability by $\epsilon_{\mathscr{C}}$. Define 
\begin{align*}
    M_{\mathscr{C}}(n) :=  \lceil \exp(nR) \rceil,
\end{align*}
The distribution of the channel input $X^n$ induced by the code is given by 
\begin{align}
    \overline{P}(x^n) &:= \frac{1}{\lceil \exp(nR) \rceil } \sum_{m=1}^{\lceil \exp(nR) \rceil} \mathbb{P} \left( f(m) = x^n \right),
\end{align}
and we know that $\overline{P} \in \mathcal{P}_{\Gamma,V}(\mathcal{A}^n)$. Let $\overline{P} \circ W$ be the joint distribution over $\mathcal{A}^n \times \mathcal{B}^n$ induced by $\overline{P}$ and the channel $W$. Alternatively, $\overline{P} \circ W$ is the distribution of $(X^n, Y^n)$ induced by the code $\mathscr{C}$, which has the average error probability equal to $\epsilon_{\mathscr{C}}$, and the channel $W$. For any $q \in \mathcal{P}(\mathcal{B}^n)$, consider another channel $\tilde{W} \in \mathcal{P}(\mathcal{B}^n | \mathcal{A}^n)$ such that $\tilde{W}(\cdot | x^n) = q$ for all $x^n$ in the support of $\overline{P}$. Clearly, for the same code $\mathscr{C}$ from before, the average error probability for the channel $\tilde{W}$ is $\epsilon' = 1 - \frac{1}{M_{\mathscr{C}}(n)}$, as the channel output is independent of the channel input and there are $M_{\mathscr{C}}(n)$ messages.

Now consider the
problem of hypothesis testing where a random variable $U$
taking values in $\mathcal{U}$ can have probability measure $P$ or $Q$. Upon
observing $U$, the goal is to declare either $U \sim P$ (hypothesis
$H_1$) or $U \sim Q$ (hypothesis $H_2$). Let $\beta_\alpha(P, Q)$ denote the
minimum attainable error probability under $Q$ when the error probability under $P$ does not exceed $1 - \alpha$. Then the Neyman Pearson lemma guarantees that there exists a (possibly randomized) test $T : \mathcal{U} \to \{0, 1\}$
(where $0$ corresponds to the test selecting $Q$) such that
\begin{align*}
    \sum_{u \in \mathcal{U}} P(u) \mathbb{P}( T(u) = 1) &\geq  \alpha, \sum_{u \in \mathcal{U}} Q(u) \mathbb{P}(T(u) = 1) = \beta_\alpha(P, Q).
\end{align*}
Then for any $\rho > 0$,
\begin{align}
    &\alpha - \rho \beta_\alpha(P, Q) \notag \\
    &= \alpha - \rho  \sum_{u \in \mathcal{U}} Q(u) \mathbb{P}(T(u) = 1) \notag \\
    &\leq \sum_{u \in \mathcal{U}} P(u) \mathbb{P}( T(u) = 1) -  \rho  \sum_{u \in \mathcal{U}} Q(u) \mathbb{P}(T(u) = 1) \notag \\
    &= \sum_{u \in \mathcal{U}} (P(u) - \rho Q(u)) \mathbb{P}( T(u) = 1) \notag \\
    &\leq \sum_{u \in \mathcal{U}} (P(u) - \rho Q(u)) \mathbb{P}( T(u) = 1) \mathds{1}(P(u) > \rho Q(u)) \notag \\
    &= P\left( \frac{P(u)}{Q(u)} > \rho, T(u) =1  \right) - \rho Q\left( \frac{P(u)}{Q(u)} > \rho, T(u) =1 \right) \notag \\
    &\leq P\left( \frac{P(u)}{Q(u)} > \rho 
    \right). \label{yoloheart}
\end{align}
From \cite[Theorem 26]{5452208}, we have 
\begin{align*}
    \beta_{1 - \epsilon_\mathscr{C}}\left(\overline{P} \circ W, \overline{P} \circ q \right) \leq \frac{1}{M_{\mathscr{C}}(n)}.
\end{align*}
Substituting $\alpha = 1 - \epsilon_{\mathscr{C}}$, $P = \overline{P} \circ W$ and $Q = \overline{P} \circ q$ in $(\ref{yoloheart})$, we have  
\begin{align*}
    &\,\,\,\,\, 1 - \epsilon_{\mathscr{C}} \leq (\overline{P} \circ W) \left( \frac{d(\overline{P} \circ W)}{d(\overline{P} \circ q)} > \rho \right) +  \frac{\rho}{M_{\mathscr{C}}(n)} \\
    &\left( 1 - \epsilon_{\mathscr{C}} - (\overline{P} \circ W) \left( \frac{d(\overline{P} \circ W)}{d(\overline{P} \circ q)} > \rho \right)\right)^+ \leq  \frac{\rho}{M_{\mathscr{C}}(n)}
\end{align*}
Hence, 
\begin{align}
    \log M_{\mathscr{C}}(n) &\leq \log \rho - \log \left [ \left( 1 - \epsilon_{\mathscr{C}} -  (\overline{P} \circ W) \left( \frac{d(\overline{P} \circ W)}{d(\overline{P} \circ q)} > \rho \right)\right)^+\right].  \label{fopwr}
\end{align}
Since $(\ref{fopwr})$ holds for any $q$, we have 
\begin{align}
    \log M_{\mathscr{C}}(n) &\leq \sup_{\overline{P} \in \mathcal{P}_{\Gamma, V}(\mathcal{A}^n)}\, \inf_{q \in \mathcal{P}(\mathcal{B}^n)} \log \rho - \log \left [ \left( 1 - \epsilon -  (\overline{P} \circ W) \left( \frac{W(Y^n|X^n)}{q(Y^n)} > \rho \right)\right)^+\right] \notag\\ 
    \implies \log M_{\mathscr{C}}(n) &\leq \log \rho - \log \left [ \left( 1 - \epsilon - \sup_{\overline{P} \in \mathcal{P}_{\Gamma,V}(\mathcal{A}^n)}\, \inf_{q \in \mathcal{P}(\mathcal{B}^n)} (\overline{P} \circ W) \left( \frac{W(Y^n|X^n)}{q(Y^n)} > \rho \right)\right)^+\right].  \label{iadqp}
\end{align}

\section{Proof of Lemma \ref{combinedlemmas} \label{myrefinement}}

$C(\Gamma)$ and $[C(\Gamma)]_n$: For large enough $n$ so that $\text{supp}(P^*) = \text{supp}([P^*]_n)$,  
\begin{align*}
n [C(\Gamma)]_{n} &:= \mathbb{E} \left [ \sum_{i=1}^{n} \log \frac{W(Y_i|X_i)}{Q^*(Y_i)} \right ] \notag  \\
    &= \sum_{a \in \mathcal{A}} n[P^*]_{n}(a) \mathbb{E} \left [  \log \frac{W(Y_{a}|a)}{Q^*(Y_a)} \right] \notag  \\
    &= \sum_{a \in \mathcal{A}} \sum_{i=1}^{n[P^*]_{n}(a)} \sum_{b \in \mathcal{B}} W(b|a)   \log \frac{W(b|a)}{Q^*(b)} \notag \\
    &= \sum_{a \in \mathcal{A}} \sum_{i=1}^{n[P^*]_{n}(a)}  \left(C(\Gamma) - C'(\Gamma)(\Gamma - c(a)) \right) \notag  \\
    &= nC(\Gamma) - n C'(\Gamma)  \left(\Gamma - \sum_{a \in \mathcal{A}} [P^*]_{n}(a) c(a) \right) \notag \\
    &\in  \left [ n C(\Gamma) - 2C'(\Gamma) J c_{\max}, n C(\Gamma)   \right ]. \notag
\end{align*}
$V(\Gamma)$ and $[V(\Gamma)]_n$:
\begin{align*}
    n\,[V(\Gamma))]_{n} &=\text{Var}\left(\sum_{i=1}^{n} \log \frac{W(Y_i|X_i)}{Q^*(Y_i)} \right) \notag \\
    &= n  \sum_{j=1}^J [P^*]_{n}(a_j) \nu_{a_j} \notag \\
    &\in \left [ n V(\Gamma) - 2J \nu_{\max}, n V(\Gamma) + 2J \nu_{\max}  \right ].
\end{align*}

$Q^{cc}$ and $Q^*$: Recall that $Q^{\text{cc}}$ is the distribution of $Y^{n}$ when the channel input $X^{n}$ is uniformly distributed over the type class $T^{n}_{\mathcal{A}}([P^*]_{n})$.
\begin{align*}
    Q^{\text{cc}}(y^n) &= \sum_{x^n \in T^n_{\mathcal{A}}([P^*]_n)} \frac{1}{|T^n_\mathcal{A}([P^*]_n)|} \prod_{i=1}^n W(y_i|x_i)\\
    &\leq \sum_{x^n \in T^n_{\mathcal{A}}([P^*]_n)} \kappa\,\exp\left(-n H([P^*]_n) + \frac{s([P^*]_n) - 1}{2} \log n  \right) \prod_{i=1}^n W(y_i|x_i),
\end{align*}
where the last inequality above follows from the well-known formula for the size of the type class, e.g., \cite[Problem 2.2]{korner1}, where $\kappa > 0$ is a constant. On the other hand, 
\begin{align*}
    Q^*(y^n) &= \sum_{x^n \in \mathcal{A}^n} P^*(x^n) \prod_{i=1}^n W(y_i|x_i)
\end{align*}
For some constant $\tau > 0$, we have 
\begin{align}
    &\,\frac{Q^*(y^n)}{Q^{\text{cc}}(y^n)} \notag \\
    &\geq \frac{\sum_{x^n \in \mathcal{A}^n} P^*(x^n) \prod_{i=1}^n W(y_i|x_i)}{\sum_{x^n \in T^n_{\mathcal{A}}([P^*]_n)} \kappa\,\exp\left(-n H([P^*]_n) + \frac{s([P^*]_n) - 1}{2} \log n  \right) \prod_{i=1}^n W(y_i|x_i)} \notag\\
    &\geq \frac{\sum_{x^n \in T^n_{\mathcal{A}}([P^*]_n)} P^*(x^n) \prod_{i=1}^n W(y_i|x_i)}{\sum_{x^n \in T^n_{\mathcal{A}}([P^*]_n)} \kappa\,\exp\left(-n H([P^*]_n) + \frac{s([P^*]_n) - 1}{2} \log n  \right) \prod_{i=1}^n W(y_i|x_i)} \notag\\
    &= \frac{\sum_{x^n \in T^n_{\mathcal{A}}([P^*]_n)} \exp\left( -n H([P^*]_n) - n D([P^*]_n || P^*) \right) \prod_{i=1}^n W(y_i|x_i)}{\sum_{x^n \in T^n_{\mathcal{A}}([P^*]_n)} \kappa\,\exp\left(-n H([P^*]_n) + \frac{s([P^*]_n) - 1}{2} \log n  \right) \prod_{i=1}^n W(y_i|x_i)} \notag\\
    &\stackrel{(a)}{\geq} \frac{\sum_{x^n \in T^n_{\mathcal{A}}([P^*]_n)} \exp\left( -n H([P^*]_n) - \tau/n \right) \prod_{i=1}^n W(y_i|x_i)}{\sum_{x^n \in T^n_{\mathcal{A}}([P^*]_n)} \kappa\,\exp\left(-n H([P^*]_n) + \frac{s([P^*]_n) - 1}{2} \log n  \right) \prod_{i=1}^n W(y_i|x_i)} \notag\\
    &= \frac{\exp\left( -n H([P^*]_n) - \tau/n \right)\sum_{x^n \in T^n_{\mathcal{A}}([P^*]_n)} \prod_{i=1}^n W(y_i|x_i)  }{ \kappa\,\exp\left(-n H([P^*]_n) + \frac{s([P^*]_n) - 1}{2} \log n  \right)\sum_{x^n \in T^n_{\mathcal{A}}([P^*]_n)}\prod_{i=1}^n W(y_i|x_i) } \notag \\
    &= \frac{ \exp\left( -n H([P^*]_n) - \tau/n \right) }{ \kappa\,\exp\left(-n H([P^*]_n) + \frac{s([P^*]_n) - 1}{2} \log n  \right) } \notag \\
    &= \frac{1}{\kappa} \exp \left(-\frac{\tau}{n} - \frac{s([P^*]_n) - 1}{2} \log n  \right). \label{ie}
\end{align}
In inequality $(a)$ above, we used a 'reverse Pinsker's inequality' \cite[Lemma 6.3]{1603768}, where the constant $\tau$ depends on $P^*$.
From $(\ref{ie})$, we obtain 
\begin{align*}
    &\log \frac{Q^*(y^n)}{Q_{\text{cc}}(y^n)}\\
    &\geq - \frac{s([P^*]_n) - 1}{2} \log (n) - \kappa 
\end{align*}
for sufficiently large $n$, where $\kappa$ is an appropriately redefined constant in the last inequality above.  
\section*{Acknowledgment}

This research was supported by the US National Science
Foundation under grant CCF-1956192.


\ifCLASSOPTIONcaptionsoff
  \newpage
\fi



\bibliographystyle{IEEEtran}
\end{document}